\documentclass[]{jfm}

\usepackage{graphicx}
\usepackage{epstopdf,epsfig}
\usepackage{newtxtext}
\usepackage{newtxmath}
\usepackage{natbib}
\usepackage{hyperref}
\usepackage{caption}
\usepackage{subcaption}
\usepackage{mathtools}

\hypersetup{
    colorlinks = true,
    urlcolor   = blue,
    citecolor  = black,
}

\newcommand{\RomanNumeralCaps}[1]
\shorttitle{Dynamic Slip wall model.}
\shortauthor{P. Raje, and K. Duraisamy}

\newcommand{\comment}[1]{}

\title{A new dynamic slip approach for wall-modeled Large Eddy Simulations in a Consistent Discontinuous Galerkin Framework}

\author{Pratikkumar Raje\aff{1}
  \corresp{\email{praje@umich.edu}},
  Karthik Duraisamy\aff{1}}

\affiliation{\aff{1}Department of Aerospace Engineering, University of Michigan, Ann Arbor, MI 48109}

\begin{document}

\maketitle
 
\begin{abstract}

A  wall-modeled large eddy simulation approach is proposed in a Discontinuous Galerkin (DG) setting, building on the slip-wall concept of Bae et al. (JFM'19) and the universal scaling relationship by Pradhan and Duraisamy (JFM'23). The effect of the order of the DG approximation is introduced via the length scales in the formulation. The level of under-resolution is represented by a slip Reynolds number and the model attempts to incorporate the effects of the numerical discretization and the subgrid-scale model. The dynamic part of the new model is based on a modified form of Germano identity --performed on the universal scaling parameter-- and is coupled with the dynamic Smagorinsky model. A sharp modal cut-off filter is used as the test filter for the dynamic procedure, and the dynamic model can be easily integrated into any DG solver. Numerical experiments on channel flows show that grid independence of the statistics is achievable and predictions for the mean velocity and Reynolds stress profiles agree well with the direct numerical simulation, even with significant under-resolution. When applied to flows with separation and reattachment, the model also consistently predicts one-point statistics in the reverse flow and post-reattachment regions in good agreement with experiments. The performance of the model in accurately predicting equilibrium and separated flows using significantly under-resolved meshes can be attributed to several aspects that work synergistically: the optimal finite-element projection framework; the interplay of the scale-separation and numerical discretization within the DG framework; and the consistent dynamic procedures for subgrid and wall modeling.

\end{abstract}


\section{Introduction}

%%%%%%%%%%%%%%%%%%%%%%%%%%%%%%%%%%%%%%%%%%%%%%%%%%%%%%%%%%%%%%%%%%%%%%%%%%%%%%%%%%%%%%%%%%%%%%%

Wall-bounded turbulent flows are of particular relevance to many engineering applications. Computational costs of large eddy simulations (LES) increase significantly with the increase in Reynolds number. This is especially true for flows at friction velocity ($u_\tau$) based Reynolds number $Re_\tau > 10^3$ which is the range of Reynolds numbers relevant to industrial applications~\citep{smits2013wall}. A Direct Numerical Simulation (DNS) resolves all the relevant scales of motion and offers the highest possible fidelity \citep{moin1998direct}. However, substantial grid requirements along with time step limitations at high Reynolds numbers make DNS infeasible for computing flows of practical relevance. On the other hand, Reynolds-averaged Navier-Stokes (RANS) equations model all the relevant scales of motion and places less restrictive demands on computational costs but offers a lower fidelity \citep{wilcox1998turbulence}. It may not be a reliable tool for computing flows for which the turbulence models are not calibrated.

%%%%%%%%%%%%%%%%%%%%%%%%%%%%%%%%%%%%%%%%%%%%%%%%%%%%%%%%%%%%%%%%%%%%%%%%%%%%%%%%%%%%%%%%%%%%%%%

A wall-resolved (WR) LES resolves dynamically important energy-carrying eddies and models the nearly universal and nearly isotropic small i.e. subgrid-scales (SGS) \citep{sagaut2005large}. For a WRLES of a turbulent boundary layer at high Reynolds number, however, a vast majority of the computational resources have to be spent on the viscous and logarithmic layers since the grid point requirement for each of these layers scale as O(Re$_{\tau}^{2}$) \citep{larsson2016large}. To alleviate this `near wall problem of LES', wall modeled (WM) LES offers a practical solution, which aims to bypass the resolution of the inner layer of the turbulent boundary layers. In WMLES, turbulent motions in the inner layer are modeled, whereas outer layer turbulent motions are resolved as in a conventional LES \citep{piomelli2002wall}. Wall-stress models and hybrid LES/RANS are the two different approaches to model the inner layer and perform WMLES. As these approaches still resolve the outer layer of the turbulent boundary layer, they can - in principle - offer better fidelity than RANS techniques.

%%%%%%%%%%%%%%%%%%%%%%%%%%%%%%%%%%%%%%%%%%%%%%%%%%%%%%%%%%%%%%%%%%%%%%%%%%%%%%%%%%%%%%%%%%%%%%%

A hybrid RANS/LES technique, including the detached eddy simulation (DES) in a WMLES set-up, uses RANS equations in the inner layer to estimate the wall stress and switches to the LES mode in the outer layer \citep{heinz2020review}. The LES solution is used to feed information to a RANS model at some distance away from the wall. The predictions, however, depend on the choice of the RANS model and the modeling of the RANS/LES interface \citep{piomelli2003inner,davidson2005hybrid,davidson2006hybrid,keating2006dynamic,shur2008hybrid,choi2009compressible}. On the other hand, a wall-stress model computes wall shear stress using a log-law or the solution of some form of thin boundary-layer equations (TBLE) on an embedded grid between the first grid point and the wall \citep{larsson2016large,bose2018wall}. In this approach, the filtered wall-shear stress is estimated at each time step using the LES information from an off-wall grid point on an LES mesh. The wall shear stress is then passed onto the LES grid as a Neumann boundary condition. Several wall-stress modeling strategies with varying complexities have been developed and studied over the years \citep{deardorff1970numerical,schumann1975subgrid,piomelli1989new,piomelli1999large,cabot2000approximate,piomelli2002wall,sagaut2005large,piomelli2008wall,larsson2016large,bose2018wall}.

A traditional algebraic equilibrium wall-stress model using a log-law, e.g. Reichardt's profile \citep{reichardt1951vollstandige}, has the advantage of low computational cost. However, it generally performs poorly in non-equilibrium conditions and complex geometries, in particular, involving flows with boundary layer separation \citep{park2017wall,goc2020wall,whitmore2024slip}. Moreover, non-monotonic grid convergence in the prediction of the size of turbulent separation bubbles is also observed for the equilibrium wall stress models \citep{goc2020wall,whitmore2021large,agrawal2022non}. On the other hand, several non-equilibrium wall models have shown promise in predicting separated flows \citep{balaras1996two,wang2002dynamic,hickel2013parametrized,park2014improved,park2016numerical}. However, the two-layer zonal wall models employing embedded grids require some effort in grid generation and domain decomposition in the pre-processing step, especially on unstructured meshes \citep{bodart2011wall,park2016numerical,park2016wall}. The model predictions obtained using a wall-stress model depend on the choice of the exchange location (EL) i.e., the location from which the instantaneous LES solution is sampled to feed into the wall model, even for the simple geometry of a turbulent flow in a channel \citep{kawai2012wall,frere2017application}. Consequently, the EL becomes a parameter of the simulation, which needs to be adjusted according to the flow field characteristics. Setting the EL requires knowledge of the boundary layer thickness, which is a property of the solution. Moreover, many of the wall-stress models introduce complexities with accompanying empiricism to treat complex flows, e.g. sensors to turn off the wall-stress models at the separation point \citep{bodart2011wall,bose2018wall,agrawal2024non}.

Accurate and reliable prediction of separated flows at high Reynolds numbers remains a pacing research issue within the Computational Fluid Dynamics (CFD) community. Several efforts to validate the state-of-the-art WMLES techniques in predicting separated flows at appropriate Reynolds numbers in a realistic external aerodynamics configuration have been undertaken recently. NASA CFD Vision 2030 report \citep{slotnick2014cfd}) has identified WMLES for complex 3D flows of practical relevance as one of the key milestones along the CFD technology development roadmap. \cite{park2016wall,lehmkuhl2018large,goc2020wall,goc2021large} have investigated predictive capabilities of equilibrium and non-equilibrium models within the WMLES framework in the characterization of the flow around an aircraft by considering the JAXA Standard Model and NASA Common Research model with wing/body/tail configuration, showing promise in practical applications, yet identifying several areas of improvement.

Since the solution may not be accurately computed near the wall on coarser near-wall LES grids due to the presence of steep wall normal gradients, one promising alternative to the wall-stress models is the virtual-wall model in which the LES domain is terminated at some finite distance above the wall \citep{chung2009large,inoue2011large,cheng2015large}. Instantaneous slip velocities obtained using a reduced form of the TBLE are then provided at this virtual boundary or the `virtual-wall', which corresponds to the location of the bottom boundary in the LES. The offset of the virtual wall is set to be proportional to the mesh size. This treatment of the wall-slip boundary condition has been shown to capture the quantitative features of a separation-reattachment turbulent boundary-layer flow at low to moderately large Reynolds numbers. However, identification of the lifted virtual-wall can be challenging for complex practical engineering geometric configurations. 

\cite{bose2014dynamic} propose a wall modeling strategy in which formal boundary conditions for the filtered Navier-Stokes equations are derived instead of relying on the true (unfiltered) boundary conditions for the filtered fields. Unlike traditional wall-stress models and hybrid RANS/LES approaches, the method does not use a wall-stress model or a RANS model in the inner layer to estimate the wall-stress. As a result, sampling of the LES solution at the off-wall grid points is not required. The model is derived using the properties of a modified form of the differential filter \citep{germano1986differential}, and it does not make any assumptions about the local state of the boundary layer or any RANS/LES hybridization. Instantaneous wall slip velocities can be estimated using this slip-wall model when the near-wall solution is under-resolved in the case of coarse grid resolutions. The model is compatible with an arbitrary LES filter and can be motivated using the RANS-type momentum equation \citep{yang2017physical}. It offers a promising alternative to the wall-stress models to predict high Reynolds number flows with complex geometries involving separation and reattachment.

The slip wall model relates the velocity field at the wall to the wall-normal derivative of the velocity field via a wall-adjacent length scale called slip length. The slip length depends on a model coefficient $C_w$ and the near-wall grid resolution $\Delta$. The model recovers the no-slip condition as the near wall grid is refined and in the limit $\Delta \rightarrow 0$, and smoothly admits a wall slip velocity as the near wall grid resolution is coarsened and the flow is no longer fully resolved. The slip wall model is a general boundary condition applicable to any geometrically complex surface, including two orthogonally or non-orthogonally intersecting walls. Moreover, it is naturally suited to handle boundary layer separation as it will smoothly revert to a no-slip condition at the separation point without additional sensors or damping functions.

%%%%%%%%%%%%%%%%%%%%%%%%%%%%%%%%%%%%%%%%%%%%%%%%%%%%%%%%%%%%%%%%%%%%%%%%%%%%%%%%%%%%%%%%%%%%%%%

The value of the model parameter i.e. the slip length is found to depend on the Reynolds number $Re_\tau$ of the flow, grid resolution, SGS model, and the numerical discretization \citep{carton2017assessment,bae2019dynamic,pradhan2023unified}. \cite{bose2014dynamic} proposed a dynamic procedure to calculate the slip length based on the Germano's identity. However, attempts to reproduce the results for a high $Re_\tau$ channel flow were unsuccessful \citep{bae2019dynamic}. The wall-stress invariant dynamic wall model (WSIM) of \cite{bae2019dynamic} provides an alternate dynamic procedure to estimate the slip length. The model predictions for the channel flow at the high $Re_\tau$ cases are found to depend on the grid resolution and grid convergence studies were not carried out. Numerical experiments with arbitrary constant values of slip length using NASA's discontinuous-Galerkin (DG) solver \textit{eddy} in the implicit LES set-up failed to yield stable computations when applied to a channel flow at $Re_\tau \gtrsim 1000$ with a high order polynomial basis ($p = 3$ and $p = 7$) \citep{carton2017assessment}.

%%%%%%%%%%%%%%%%%%%%%%%%%%%%%%%%%%%%%%%%%%%%%%%%%%%%%%%%%%%%%%%%%%%%%%%%%%%%%%%%%%%%%%%

Existing dynamic versions of the slip-wall model of \cite{bose2014dynamic,bae2019dynamic} are found to be sensitive to the numerical implementation details, including the numerical discretization and the choice of SGS model in a finite volume framework. Moreover, some models \citep{bae2019dynamic}  show a significant log-layer mismatch with the DNS. Given the limitations of the existing dynamic slip-wall models, recent works use a Prandtl mixing length-based model to estimate slip lengths \citep{whitmore2024slip}. Other strategies to estimate optimal slip-lengths include an optimization procedure to reproduce a known wall shear stress distribution for turbulent channel flows at a range of Reynolds numbers and grid resolutions and model the behavior of the slip lengths using a curve fit \citep{whitmore2023evaluation}. \emph{Ad hoc} sensor-based modeling strategies to change the model forms for the slip-lengths that switch between the mixing-length-based and parameterized forms in the separation regions have also been studied \citep{whitmore2024slip}. Application of the static slip-length models to separated flows in the Boeing speed bump and the JAXA Standard Model configurations suggests that robust separation predictions require the development of an accurate method for computing slip lengths.

%%%%%%%%%%%%%%%%%%%%%%%%%%%%%%%%%%%%%%%%%%%%%%%%%%%%%%%%%%%%%%%%%%%%%%%%%%%%%%%%%%%%%%%%%%%%%%%

\cite{pradhan2023unified} employed an optimal finite-element projection framework to obtain \textit{a priori} estimates of the wall slip velocity for a typical WMLES using DNS data for a channel flow \citep{lee2015direct} and propose improvements to the slip wall model of \cite{bose2014dynamic}. The optimal projection framework is used to modify the slip length, and it is represented as a function of the Reynolds number based on local slip velocity magnitude and near-wall local grid resolution $Re_{slip}$. A new model parameter $\lambda$ is introduced to represent the effect of the numerical method or the order of projection $p$ in the DG set-up and SGS model. Using an \textit{a priori} estimate for $\lambda$, the $Re_{slip}$ model for the modified slip length is shown to give good predictions for a range of high $Re_\tau$ channel flow cases with the constant coefficient Smagorinsky SGS model using a DG solver with orders of projection up to $p = 3$.

%%%%%%%%%%%%%%%%%%%%%%%%%%%%%%%%%%%%%%%%%%%%%%%%%%%%%%%%%%%%%%%%%%%%%%%%%%%%%%%%%%%%%%%%%%%%%%%

The present study begins with the modified form of the slip wall model proposed by \cite{pradhan2023unified} which uses the $Re_{slip}$ model for the modified model coefficient. The main objective is to establish a dynamic modeling procedure for the model parameters. We use the dynamic Smagorinsky model \citep{DSM} as the SGS model. The choice of the SGS model is found to be critical to obtain the correct slope of the velocity profile in the log layer \citep{bae2019dynamic}. On the other hand, values of the slip length are found to be responsible for a shift in the mean velocity profile relative to the DNS data, and they do not affect the shape of the mean velocity profile.

%%%%%%%%%%%%%%%%%%%%%%%%%%%%%%%%%%%%%%%%%%%%%%%%%%%%%%%%%%%%%%%%%%%%%%%%%%%%%%%%%%%%%%%%%%%%%%

The manuscript is organized as follows. Section \ref{sec2} presents the discontinuous Galerkin discretization framework used in the present work. Section \ref{sec:sec3} provides an overview of the original slip wall model formulation by \cite{bose2014dynamic} along with the modification introduced by \cite{pradhan2023unified}. The dynamic modeling procedures of \cite{bose2014dynamic} and \cite{bae2019dynamic} to calculate the slip length are discussed briefly in Section \ref{sec:sec4}. The proposed dynamic modeling strategy to estimate the model parameter $\lambda$ using a modified form of Germano identity and the $Re_{slip}$ model is presented in Section \ref{sec:sec5}. The key assumptions to arrive at the final form of the dynamic model are also discussed. The proposed dynamic slip wall model is tested on a range of channel and periodic hill flows in Section \ref{sec:sec6} and results are compared with the available DNS and experimental data along with an equilibrium wall-stress model. Finally, conclusions are drawn in Section \ref{sec:sec7}.

%%%%%%%%%%%%%%%%%%%%%%%%%%%%%%%%%%%%%%%%%%%%%%%%%%%%%%%%%%%%%%%%%%%%%%%%%%%%%%%%%%%%%%%%%%%%%%%

%%%%%%%%%%%%%%%%%%%%%%%%%%%%%%%%%%%%%%%%%%%%%%%%%%%%%%%%%%%%
\section{Discontinuous Galerkin discretization}\label{sec2}
The governing equations in this work are the  compressible Navier-Stokes equations in their conservative form written as
\begin{equation} \label{eqn:navier_stokes}
  \frac{\partial \mathbf{U}}{\partial t} + \nabla \cdot \mathbf{F}(\mathbf{U})
  - \nabla \cdot \mathbf{G}(\mathbf{U}, \nabla \mathbf{U}) = \mathbf{0},
\end{equation}
where $\mathbf{U} \in \mathbb{R}^s  $ is the conservative state vector of rank $s$, consisting of density, momentum, and total energy components, $\mathbf{F}$ is the inviscid flux, and $\mathbf{G}$ is the viscous flux. We note that boldface denotes a state vector. We use the Discontinuous Galerkin (DG) method for the spatial discretization. The DG method combines the concepts of finite element and finite volume methods and allows for high-order approximations, geometric flexibility, and natural parallelization. The computational domain $\Omega$ is divided into non-overlapping elements $K$, each having a sub-domain $\Omega_K$ and boundary $\partial \Omega_K$. These elements can have arbitrary shapes and sizes, allowing for efficient representation of complex geometries. A polynomial approximation is typically used to represent the solution using a ${L}_2$-projection within each element. The degree of the polynomial $p$ can vary, and higher-degree polynomials enable higher-order accuracy. The DG space ${\mathcal{V}_h}$ is defined as
\begin{equation}
{\mathcal{V}_h} \triangleq \left\{\phi\in{L_2}(\Omega):\phi_h\equiv\phi|_{\Omega_K}\in P^p,\forall\Omega_K\in\Omega \right\},
\end{equation}
where the space of polynomials up to degree $p$ is denoted as $P^p$, and $\phi_h$ is the basis function defined on $\Omega_K$. Defining ${\mathcal{V}_h}$ in this manner allows for discontinuities in the solution across element boundaries. The element-wise solution $\mathbf{U}_h$ that approximates $\mathbf{U}$ in $\Omega_k$ takes the form
 \begin{equation}
  \mathbf{U}(\mathbf{x},t) \approx \mathbf{U}_h(\mathbf{x},t) = \sum_{j=1}^{n_p} \mathbf{W}_{k,j}(t) \phi_{k,j}(\mathbf{x}), \hspace{0.1in} \mathbf{x} \in \Omega_k, 
  \label{Eqn:DG_soln_approx}
 \end{equation}
where $\mathbf{U}_{k,j}$ represents the coefficients associated with the $j^{th}$ basis function $\phi_{k,j}$ and $n_p$ represents the total number of degrees of freedom within the element $k$ of order $p$. 

The DG method employs a weak formulation of the governing equations which is obtained by multiplying Eqn.~(\ref{eqn:navier_stokes}) by test functions, which are the same as the basis functions, integrating by parts, and coupling the elements via numerical fluxes,
\begin{equation} \label{eqn:weak_form}
\begin{split}
  \int_{\Omega_K} \mathbf{\phi}_h^T \frac{\partial \mathbf{U}_h}{\partial t} \, \mathrm{d}\Omega \; -
  \int_{\Omega_K} \nabla \mathbf{\phi}_h^T \cdot \left[ \mathbf{F} \left(\mathbf{U}_h \right) - \mathbf{G} \left( \mathbf{U}_h, \nabla \mathbf{U}_h \right) \right] \, \mathrm{d}\Omega \; + &\\
  \int_{\partial \Omega_K} \mathbf{\phi}_h^T \left[ \widehat{\mathbf{F}} \left( \mathbf{U}_h^+, \mathbf{U}_h^- \right) - \widehat{\mathbf{G}} \left( \mathbf{U}_h^+,\mathbf{U}_h^-,\nabla \mathbf{U}_h^+, \nabla \mathbf{U}_h^- \right) \right] \cdot \mathbf{n} \, \mathrm{d}S \; &\\ -
  \int_{\partial \Omega_K} \left( \mathbf{U}_h^+- \{ \mathbf{U}_h \} \right)^T \mathbf{G} \left( \mathbf{U}_h^+, \nabla \mathbf{\phi}_h^+ \right) \cdot \mathbf{n} \, \mathrm{d}S &=
  \mathbf{0}, \; \forall \mathbf{\phi}_h \in \boldsymbol{\mathcal{V}}_h.
\end{split}
\end{equation}
$\partial \Omega_K$ represents the element boundary, and on that boundary, $(\cdot)^+$ and $(\cdot)^-$ represent quantities taken from the current and neighboring element, respectively. Approximate numerical fluxes are denoted by $\widehat{(\cdot)}$, $\{\cdot\}$ represents a face average or boundary value, and $\mathbf{n}$ is the outward pointing normal vector. The boundary conditions are set through the numerical fluxes. 

Substituting Eqn.~(\ref{Eqn:DG_soln_approx}) into Eqn.~(\ref{eqn:weak_form}), we get the final update equation which can be written as,
\begin{equation}
    \mathbf{M} \dfrac{d \mathbf{W}}{dt} = -\mathbf{RHS},
\end{equation}
where $\mathbf{M}$ is the spatial mass matrix and $\mathbf{RHS}$ consists of the volume and surface integrals. Then, the spatial residual vector can be defined as
\begin{equation}
    \mathbf{R} \equiv \dfrac{d \mathbf{W}}{dt} = -\mathbf{M}^{-1} \mathbf{RHS},
\end{equation}
We solve for the expansion coefficients $\mathbf{W}$ which then provide an approximation of the solution to the governing equations over the entire computational domain. The solver used in the present study is discussed in Section \ref{sec:sec6} and Appendix \ref{appendixA}.

%%%%%%%%%%%%%%%%%%%%%%%%%%%%%%%%%%%%%%%%%%%%%%%%%%%%%%%%%%%%%%%%%%%%%%%%%%%%%%%%%%%%%%%%%%

\section{Slip-wall modeling}\label{sec:sec3}

The slip wall model is essentially a wall boundary condition. The main idea is that the slip velocity is a natural consequence of the near wall under-resolution of the LES mesh. This has also been shown by \cite{pradhan2023unified} using the optimal finite-element projection framework wherein the $L_2$ projection of channel flow DNS data onto grids suitable for a WMLES results in slip velocities at the wall. Also, the magnitude of the slip velocity is shown to increase with an increase in near-wall grid under-resolution. This shows that the near equivalence in the boundary conditions for the unfiltered and filtered variables does not hold in the case of a coarse LES when wall modeling becomes necessary. A slip wall model is an alternative to the traditional wall-stress modeling approach wherein the wall stress is not estimated directly but is indirectly affected through the non-vanishing filtered velocities at the wall. It provides estimates of the slip velocities at the wall when the LES grid resolution is insufficient to accurately resolve the near-wall region and the no-slip condition is not satisfied.

\cite{bose2014dynamic} use the properties of a modified differential filter to derive a slip velocity boundary condition given as
\begin{equation}
\overline{u}_i = \overline{C}_w \overline{\Delta}_w  \dfrac{\partial \overline{u}_i}{\partial n}, 
\label{eqn:slip-model}
\end{equation}
where $n$ is the wall-normal direction, $\overline{C}_w$ is a tunable model coefficient, whereas $\overline{\Delta}_w$ is related to the near-wall grid resolution. In Eqn.~(\ref{eqn:slip-model}), the slip velocity only depends on the wall-normal derivative of the velocity field and is a direct consequence of the constraint placed on the differential filter that the slip length vanishes at the boundaries. The magnitude of the slip length i.e. $\overline{C}_w \overline{\Delta}_w$ imposes a filter length scale at the wall; if it vanishes at the wall, then the filtered velocity field will exactly satisfy a no-slip boundary condition. The slip wall boundary condition smoothly admits a wall slip velocity as the near-wall LES resolution is coarsened and the flow is no longer fully resolved. It is pertinent to note that while Eqn.~(\ref{eqn:slip-model}) is derived from a specific choice of the form of the filter kernel, previous studies \citep{carton2017assessment,pradhan2023unified} show that the slip wall model can still perform well even without using the specified filter explicitly.

\cite{pradhan2023unified} characterize the wall slip velocity in a WMLES in terms of a Reynolds number based on slip velocity magnitude and near-wall under-resolution using the optimal finite-element projection framework and propose a modified form of the slip wall model given by 
\begin{equation}
\overline{u}_i = \dfrac{\overline{C}_{w,\lambda}}{\lambda} \overline{\Delta}_w  \dfrac{\partial \overline{u}_i}{\partial n} , \hspace{0.1in} \text{where} \hspace{0.1in} \overline{C}_w = \dfrac{\overline{C}_{w,\lambda}}{\lambda} \hspace{0.02in} .
\label{eqn:slip-model-new}
\end{equation}
The model coefficient $ \overline{C}_{w,\lambda}$ is a function of the slip-velocity based Reynolds number $\overline{Re}_{slip}$ and $ \lambda$, where $\overline{Re}_{slip} = \overline{u}_s (\overline{\Delta}^{e}_{w} / p)/ \overline{\nu} $. Here, $\overline{u}_s$ is the magnitude of the wall slip velocity, and $\overline{\nu}$ is the kinematic viscosity of the fluid. Note that, $p$ denotes the order of polynomial basis used in the DG solver with $\overline{\Delta}^{e}_{w}$ being the element size adjacent to the wall, and their ratio represents the effective grid size. The model parameter $\lambda$ contains the effect of the order of projection $p$ and hence the numerical method along with the SGS model. Using the above form of $\overline{C}_{w,\lambda}$, it is found that given an SGS model $ \overline{C}_{w,\lambda} / \lambda$ admits a universal scaling relationship for a particular value of $\lambda$ for a wide range of the parameter space. As a result, the model incorporates the effect of Reynolds number, near-wall grid under-resolution, SGS model, and numerical discretization.

%%%%%%%%%%%%%%%%%%%%%%%%%%%%%%%%%%%%%%%%%%%%%%%%%%%%%%%%%%%%%%%%%%%%%%%%%%%%%%%%%%%%%%

\section{Previous dynamic slip-wall models}\label{sec:sec4}

\cite{bose2014dynamic} presented a dynamic procedure to compute the slip length $(\overline{C}_w \overline{\Delta}_w)$ in the slip wall model given by Eqn.~(\ref{eqn:slip-model}). It uses a modified form of Germano's identity, which represents the invariance of the total Reynolds stress at the test-filtered level. The model coefficient $(\overline{C}_w \overline{\Delta}_w)$ is computed as 
\begin{equation}
\left( \overline{C}_w \overline{\Delta}_w \right)^2 \Delta_{R}^{2}  \dfrac{\partial \widehat{\overline{u}}_i}{\partial n} \dfrac{\partial \widehat{\overline{u}}_j}{\partial n} + T_{ij} - \widehat{\tau}_{ij}  = \widehat{\overline{u}_i \overline{u}_j } - \widehat{\overline{u}}_i \widehat{\overline{u}}_j, 
\label{eqn:bose_slip}
\end{equation}
where $ \Delta_R = ( \widehat{\overline{\Delta}}_{w} / \overline{\Delta}_w )  $ is the ratio of the test filter width to the grid filter width at the wall, and a value of $\Delta_R = 1.4$ is recommended. Here, $\overline{(\cdot)}$ represents a grid filtered quantity, a hat, i.e., $\widehat{(\cdot)}$ denotes the test filtering operation, $T_{ij}$ and $\tau_{ij}$ depict the SGS stress tensors at the test and grid filter levels, respectively. The slip length is assumed to be equal for the three spatial directions. Eqn.~(\ref{eqn:bose_slip}) is solved for $\left( \overline{C}_w \overline{\Delta}_w \right) $  using a least squares method. The model was tested on a series of high Reynolds number channel flows and NACA $4412$ airfoil at near-stall conditions.

\cite{bae2019dynamic} proposed an alternate dynamic modeling strategy for the slip length $\left( \overline{C}_w \overline{\Delta}_w \right) $ as an improvement over the \cite{bose2014dynamic} dynamic model. The dynamic model is based on a combination of the invariance of wall stress condition under test filtering and a modified form of Germano's identity and is referred to as the wall-stress invariant model (WSIM). The proposed dynamic modeling approach, however, is not unique, and different modeling choices are possible. The dynamic model is given by
\begin{equation}
    \left( \overline{C}_w \overline{\Delta}_w \right)^2 =  \dfrac{L_{ij} M_{ij} + F_{ij} M_{ij}}{M_{kl} M_{kl}},
    \label{eqn:dynamic_bae}
\end{equation}
where
\begin{equation}
    L_{ij} = \overline{u}_i \overline{u}_j - \widehat{\overline{u}}_i \widehat{\overline{u}}_j,
\end{equation}
\begin{equation}
    M_{ij} = \left[ \dfrac{\partial \overline{u}_i}{\partial n} \dfrac{\partial \overline{u}_j}{\partial n} - \Delta_{R}^{2} \dfrac{\partial \widehat{\overline{u}}_i}{\partial n} \dfrac{\partial \widehat{\overline{u}}_j}{\partial n} \right],
\end{equation}
and $F_{ij}$ contains different wall stresses, namely Reynolds stress, subgrid stress, viscous stress, and pressure tensors computed from the specified velocity field. The model was tested on a statistically stationary plane turbulent channel, a non-equilibrium three-dimensional transient channel, and a zero-pressure-gradient flat-plate turbulent boundary layer.

%%%%%%%%%%%%%%%%%%%%%%%%%%%%%%%%%%%%%%%%%%%%%%%%%%%%%%%%%%%%%%%%%%%%%%%%%%%%%%%%%%%%%%%

\section{A new dynamic slip-wall model}\label{sec:sec5}

We propose a dynamic procedure to compute the model coefficient $\lambda$ in the modified slip wall model given by Eqn.~(\ref{eqn:slip-model-new}) rather than the slip length in Eqn.~(\ref{eqn:slip-model}) following insights from  \cite{pradhan2023unified}. We start with the Germano identity \citep{germano1990averaging}, which can be written as
\begin{equation}
     T_{ij} - \widehat{\tau}_{ij} = \widehat{\overline{u}_i \overline{u}}_j - \widehat{\overline{u}}_i \widehat{\overline{u}}_j ,
     \label{eqn:Germano}
\end{equation}
where the SGS stresses at the grid and test filtered levels are given by
%%\noindent \textit{Dynamic procedure for the model coefficient : $ \lambda $ }
\begin{equation}
    \tau_{ij} = \overline{u_i u_j} - \overline{u}_i \overline{u}_j, \hspace{0.1in} \text{and} \hspace{0.1in}  T_{ij} = \widehat{\overline{u_i u_j}} - \widehat{\overline{u}}_i \widehat{\overline{u}}_j.
\end{equation}
Eqn.~(\ref{eqn:Germano}) represents an exact identity and does not involve any assumptions. Subtracting ($\widehat{\overline{u}_i \overline{u}_j } - \overline{u}_i \overline{u}_j$) from both sides of Eqn.~(\ref{eqn:Germano}), we get
\begin{equation}
    T_{ij} - \widehat{\tau}_{ij} - ( \widehat{\overline{u}_i \overline{u}_j } - \overline{u}_i \overline{u}_j ) = \overline{u}_i \overline{u}_j - \widehat{\overline{u}}_i \widehat{\overline{u}}_j 
    \label{eqn:mod_Germano}
\end{equation}
%%\textbf{Assumption :}
We assume that the slip velocity at the test-filtered level takes a form similar to that for the grid-filtered level, and it is given by  
\begin{equation}
\widehat{\overline{u}}_i = \widehat{\overline{C}}_w \widehat{\overline{\Delta}}_{w}  \dfrac{\partial \widehat{\overline{u}}_i}{\partial n},
\label{eqn:slip_vel_test}
\end{equation}
where the model coefficient $ \widehat{\overline{C}}_w = \widehat{\overline{C}}_{w,\lambda} / \lambda$ has a form similar to that of coefficient at the grid filtered level $\overline{C}_{w} = \overline{C}_{w,\lambda}/ \lambda$. Next, we assume that $\lambda$ is constant between the grid and test filtered levels. This assumption is based on the findings of \cite{pradhan2023unified} using the optimal finite element projection framework. The universality of the model coefficient $\lambda$ depends on the choice of the length scale $\Delta_w$ used in the slip-wall model. If the cube root of the cell volume is used as the length scale, $\lambda$ is found to remain fairly constant across different resolutions and Reynolds numbers for a given SGS model. Next, we assume the slip length to be equal in the streamwise and spanwise directions, while zero in the wall-normal direction i.e. no transpiration based on the findings of \cite{pradhan2023unified}. In general, it can be different in the streamwise, spanwise, and wall-normal directions as observed in the \emph{a priori} studies on the slip-wall model using the optimal finite element projection framework by \cite{pradhan2023unified}. However, numerical experimentation using arbitrary constant values of the slip lengths in the streamwise and spanwise directions for turbulent channel flows did not significantly affect the model predictions. Similar observations are made in the numerical experiments of \cite{bae2019dynamic}. Notably, in the previous works on dynamic slip-wall models \citep{bose2014dynamic,bae2019dynamic}, slip length is assumed to be the same in the three spatial directions.

The model coefficient $ \widehat{\overline{C}}_{w,\lambda}$ is assumed to be a function of Reynolds number based on slip velocity magnitude and the near-wall grid resolution at the test filtered level along with $\lambda$. Substituting for the slip velocities at the grid filtered level and test filtered level using Eqn.~(\ref{eqn:slip-model}) and Eqn.~(\ref{eqn:slip_vel_test}) in the right-hand-side of Eqn.~(\ref{eqn:mod_Germano}), we get
\begin{equation}
\left( \overline{C}_w \overline{\Delta}_w \right)^2 \dfrac{\partial \overline{u}_i}{\partial n} \dfrac{\partial \overline{u}_j}{\partial n} - \left( \widehat{\overline{C}}_w \widehat{\overline{\Delta}}_{w} \right)^2 \dfrac{\partial \widehat{\overline{u}}_i}{\partial n} \dfrac{\partial \widehat{\overline{u}}_j}{\partial n} = \overline{u}_i \overline{u}_j - \widehat{\overline{u}}_i \widehat{\overline{u}}_j \hspace{0.02in} .
\label{eqn:dyn_process_1}
\end{equation}

Now, $\overline{\Delta}_w$ and $\widehat{\overline{\Delta}}_{w}$ depend on the grid resolution, $p$, and the filter used. On the other hand, the model coefficients $ \overline{C}_w $ and $  \widehat{\overline{C}}_w $ depend on $Re_{slip}$, $p$ at the grid and test filtered levels, respectively and the model coefficient $\lambda$. In principle, we can use the above equation to find $\lambda$ for a given model for $ \overline{C}_w $ and hence $  \widehat{\overline{C}}_w $. However, this would result in a significantly complex non-linear equation in $\lambda$. We choose an alternate approach to simplify the process with an aim to keep a balanced mixture of physical content and mathematical simplicity and rewrite Eqn.~(\ref{eqn:dyn_process_1}) as
\begin{equation}
\left( \overline{C}_w \overline{\Delta}_w \right)^2 \left[ \dfrac{\partial \overline{u}_i}{\partial n} \dfrac{\partial \overline{u}_j}{\partial n} - C_{wR}^{2} \Delta_{R}^{2}  \dfrac{\partial \widehat{\overline{u}}_i}{\partial n} \dfrac{\partial \widehat{\overline{u}}_j}{\partial n} \right] = \overline{u}_i \overline{u}_j - \widehat{\overline{u}}_i \widehat{\overline{u}}_j, 
\label{eqn:dyn_proc_2}
\end{equation}
%\left( \dfrac{\widehat{\overline{C}}_w}{\overline{C}_w} \right)^2 \left( \dfrac{\widehat{\overline{\Delta}}_{w}}{\overline{\Delta}_w} \right)^2
where $ C_{wR} = \widehat{\overline{C}}_w / \overline{C}_w $.  In this work, we use the value for $ \Delta_{w} $ as per \cite{pradhan2023unified} and $\Delta_{R}$ is given by
\begin{equation}
    \Delta_{R} = \dfrac{\widehat{\overline{\Delta}}_{w}}{\overline{\Delta}_w} = \dfrac{p}{p^\star},
    \label{eqn:deltaR}
\end{equation}
where $p^\star$ is the sharp modal cut-off filter order, as discussed in Appendix \ref{appendixA}. On the other hand, \cite{pradhan2023unified} show that the model coefficient $C_{w, \lambda}$ is a function of grid resolution, and its value increases when the grid resolution is changed from $\Delta^+$ to $2 \Delta^+$. In other words, given that the test filter width is coarser than the grid filter width,  $ \widehat{\overline{C}}_w $ can be expected to be greater than $\overline{C}_w $, thereby resulting in the ratio $ C_{wR} $ to be greater than one. We use a value of $ C_{wR} = 2$ in this work. Sensitivity studies using different plausible values of $ C_{wR} $ are shown in Appendix \ref{appendixC}. 

Let 
\begin{equation}
M_{ij} = \left[ \dfrac{\partial \overline{u}_i}{\partial n} \dfrac{\partial \overline{u}_j}{\partial n} - C_{wR}^{2} \Delta_{R}^{2} \dfrac{\partial \widehat{\overline{u}}_i}{\partial n} \dfrac{\partial \widehat{\overline{u}}_j}{\partial n} \right] \hspace{0.1in}  \text{and} \hspace{0.1in}  L_{ij} = \overline{u}_i \overline{u}_j - \widehat{\overline{u}}_i \widehat{\overline{u}}_j
\end{equation}
for notational convenience. Eqn.~(\ref{eqn:dyn_proc_2}) can then be equivalently written as
\begin{equation}
\left( \dfrac{\overline{C}_{w,\lambda}}{\lambda} \overline{\Delta}_w \right)^2 M_{ij} = L_{ij} \hspace{0.02in} .
\label{eqn:dyn_proc_4}
\end{equation}
Eqn.~(\ref{eqn:dyn_proc_4}) represents six independent equations in space for a single unknown $\lambda$ given the model for $\overline{C}_{w,\lambda}$. Thus, the system is overdetermined, and we use the method of least squares to obtain $\lambda$, which is then given by
\begin{equation}
    \left( \dfrac{\overline{C}_{w,\lambda}}{\lambda} \overline{\Delta}_w \right)^2 = max \left( \dfrac{<L_{ij} M_{ij}>}{<M_{kl} M_{kl}>}, 0 \right), 
    \label{eqn:dynamic_slip_0}
\end{equation}
where $< >$ indicates that the numerator and denominator are first averaged over an element followed by an averaging over the directions of homogeneity i.e. stream-wise and span-wise in case of channel flows and span-wise in case of periodic flows, and the ratio is clipped to have a maximum value of zero. We can rewrite the above equation as
\begin{equation}
\dfrac{\overline{C}_{w,\lambda}}{\lambda} \overline{\Delta}_w  - l_s = 0,
\label{eqn:dynamic_slip_1}
\end{equation}
where
\begin{equation}
   l_s = 	\sqrt{max \left( \dfrac{<L_{ij} M_{ij}>}{<M_{kl} M_{kl}>}, 0 \right)}  \hspace{0.02in} .
\end{equation}
%\textbf{Pradhan and Duraisamy 2023 model fit for $C_{w,\lambda}$}
\cite{pradhan2023unified} provide a model fit for $\overline{C}_{w,\lambda}$ based on $L_2$ projected channel flow DNS data, which is given by
\begin{equation}
    \overline{C}_{w,\lambda} = \begin{dcases}
        0.725 \hspace{0.05in} log_{10} \left(\overline{Re}_{slip}/\lambda \right) - 0.925, & \text{if} \hspace{0.1in} \log_{10} \left(\overline{Re}_{slip}/\lambda \right) > 4.25  \\
        0.6 \hspace{0.05in} log_{10} \left(\overline{Re}_{slip}/\lambda \right) - 0.41, & \text{if} \hspace{0.1in} \log_{10} \left(\overline{Re}_{slip}/\lambda \right) > 3.18  \\
        0.475 \hspace{0.05in} log_{10} \left(\overline{Re}_{slip}/\lambda \right) - 0.003, & \text{if} \hspace{0.1in} \log_{10} \left(\overline{Re}_{slip}/\lambda \right) > 0.7  \\
        0.33, & \text{otherwise.} 
    \end{dcases}
    \label{eqn:Cw_model}
\end{equation}
Substituting for $\overline{C}_{w,\lambda}$ in Eqn.~(\ref{eqn:dynamic_slip_1}), we get a non-linear equation with $\lambda$ as the only unknown, which can be found dynamically using a numerical method. The Secant method is used to find $\lambda$ using Eqns.~(\ref{eqn:dynamic_slip_1}) and~(\ref{eqn:Cw_model}). The parameter $\lambda$, thus obtained, can reach unrealistically high values, especially at high Reynolds numbers on coarse near-wall LES meshes. We prescribe an upper limit to $\lambda$ as
   \begin{equation}
       \lambda_f = max(\lambda,3 \lambda_{CCSM}), \text{where} \hspace{0.05in} \lambda_{CCSM} = 1.125.
   \end{equation}    
Here, $\lambda_{CCSM}$ is the value of $\lambda$ for the constant coefficient Smagorinsky model (CCSM) obtained in the \emph{a priori} analysis of channel flow DNS data and traditional wall model predictions using the optimal finite element projection framework in \cite{pradhan2023unified}. The limiter value is based on numerical experimentation across the range of Reynolds numbers and flow configurations considered in this work. Finally, we use $\lambda_f$ in Eqn.~(\ref{eqn:slip-model-new}). 

The dynamic model in Eqn.~(\ref{eqn:dynamic_slip_1}) is essentially identical to the dynamic slip-wall model of \cite{bose2014dynamic}. The model given by Eqn.~(\ref{eqn:dynamic_slip_1}) can be transformed to Eqn.~(\ref{eqn:bose_slip}) using the Germano identity. However, the model form in Eqn.~(\ref{eqn:bose_slip}) is found to be sensitive to the implementation details including the numerical discretization and the choice of the SGS model in a finite-volume framework in previous studies \citep{park2016wall,bae2019dynamic}. Our attempts to use this model form in the DG framework with $p \geq 2$ resulted in unstable simulations with the constant coefficient and the dynamic Smagorinsky SGS models.

We remark that the dynamic modeling procedure to obtain $\lambda$ as discussed above is not unique. The model coefficient can be obtained using a number of modeling choices e.g. modified form of the Germano identity used in \cite{bae2019dynamic}. Eqn.~(\ref{eqn:dynamic_slip_0}) has a form similar to that of the dynamic model of \cite{bae2019dynamic}, but it does not contain the additional wall-stress terms in the numerator, which originates from the invariance of wall-stress condition under test filtering. Those additional wall-stress terms are expected to predict the same wall stress regardless of the grid resolution (or filter) and act as an effective self-regulating mechanism to control the changes in slip length to predict the correct wall stress. In our case, a similar effect is obtained by enforcing the \cite{pradhan2023unified} model for $\overline{C}_{w,\lambda}$ which is a function of the slip wall velocity and near-wall grid resolution via the slip-based Reynolds number. The parameter $\overline{C}_{w,\lambda}$ and hence the slip length appropriately changes when the near-wall grid is refined or coarsened and/or when the Reynolds number is increased or decreased, resulting in an appropriate change in the wall stress. Note that the wall stress can be related to the slip length $ (\overline{C}_{w} \overline{\Delta}_w)$ and slip velocity $U_{slip}$ using the slip wall model as : $\tau_w = (\nu + \nu_{SGS}) \left( U_{slip} /(\overline{C}_{w} \overline{\Delta}_w) \right) $, where $\nu_{SGS}$ is the SGS or eddy viscosity.  

\subsection{Implementation of the wall boundary condition}
We assume that there is no transpiration and slip is only allowed in the wall-parallel directions. It is useful to note that, the slip-wall model allows for transpiration as considered in the previous studies of \cite{carton2017assessment}, \cite{bose2014dynamic}, and \cite{bae2019dynamic}. However, using the optimal finite-element projection framework, \cite{pradhan2023unified} show that the slip length associated with the wall-normal velocity is approximately zero. Hence, it can be set to zero without significant loss of generalizability. In the current implementation of the dynamic slip-wall model, we compute the wall-normal derivatives of the slip-velocity components at the wall using Eqn.~(\ref{eqn:slip-model-new}). The wall-parallel slip-velocity components at the wall are computed using the solution inside the element adjacent to the wall. This is then used to compute the wall stress and it is applied as a Neumann boundary condition complemented by a slip boundary condition for the velocity. The numerical implementation is done using the following steps:
\begin{itemize}
    \item At every integration point, a ghost value is created, where the wall-parallel slip-velocity components at the wall are obtained from the element interior state $\mathbf{U}^{+}_{h}$ as
\begin{equation}
  \overline{u}^{b}_{h,i} = \overline{u}^{+}_{h,i} - \overline{u}^{+}_{h,j} n_j n_i.  
\end{equation} 
    Wall-normal velocity gradients are also calculated using the interior solutions.
    \item Slip-wall parameters $C_{w,\lambda}$ and $\lambda$ are computed using the dynamic slip-wall model.
    \item The wall-normal derivatives of the slip-velocity components at the wall are then computed using the slip-wall model given by Eqn.~(\ref{eqn:slip-model-new}) as follows: 
\begin{equation}
    \left.\frac{\partial \overline{u}_{h,i}}{\partial n}\right|_{w} = \frac{\overline{u}^{b}_{h,i}}{\Delta (C_{w,\lambda}/\lambda)},
\end{equation}
   \item Finally, wall stress components at the quadrature points of the boundary faces are computed using the following formula:
\begin{equation}
\tau'_{w}= \left( \nu + \nu_{SGS} \right) \left.\frac{\partial \overline{u}_{h,i}}{\partial n}\right|_{w},
\end{equation}
where we consider the contribution of the mean wall stress from the viscous and the sub-grid stresses for the wall stress. 
\item The projected wall stress $\tau_{w,i}= \tau'_{w} n_i$ is applied as a Neumann boundary condition. 
\end{itemize}

%%%%%%%%%%%%%%%%%%%%%%%%%%%%%%%%%%%%%%%%%%%%%%%%%%%%%%%%%%%%%%%%%%%%%%%

\section{Numerical experiments.}\label{sec:sec6}

%%%%%%%%%%%%%%%%%%%%%%%%%%%%%%%%%%%%%%%%%%%%%%%%%%%%%%%%%%%%%%%%%%%%%%%%%%%%%%%%%

In this work, we use CaslabDG, an in-house Discontinuous-Galerkin (DG) solver for the computations. The governing equations are the filtered compressible Navier-Stokes equations in conservative form. The solver was successfully used previously to compute statistically stationary channel flows at high $Re_\tau$ \citep{pradhan2023unified} using a constant coefficient Smagorinsky model with up to $3$ orders of the polynomial basis $p$. The solver is parallelized using Message Passing Interface (MPI). Inviscid fluxes are approximated using Roe approximate Riemann solver \citep{roe1981approximate}. An SGS model is used for the unresolved SGS stresses in the filtered Navier-Stokes equations. The SGS viscosity is added to the molecular viscosity and the viscous flux contains both molecular and turbulence contributions. The second form of Bassi and Rebay \citep{bassi2000gmres}, popularly known as the BR-2 scheme, is used for the viscous fluxes. The governing equations are marched in time using an explicit third-order Runge-Kutta TVD (RK3-TVD) scheme. 

The solver uses the Lagrange nodal basis evaluated at the Gauss-Legendre quadrature points and the number of quadrature points $ngp$ in each of the three directions is related to the polynomial degree of approximation by $ngp = (p + 2)$. The integrals are approximated using the Gauss quadrature rule. The basis and test functions are created using a tensor product of the one-dimensional Lagrange interpolating polynomials which forms a non-hierarchical nodal basis. The corresponding number of degrees of freedom in each element is $(p + 1)^3$. We use a Lagrange polynomial basis of degree $p = 2$ for all the wall-modeled LES computations. It is to be noted that we do not use an explicit filter for the spatial filtering operation, but rely on implicit filtering through the numerical discretization and grid resolution. Also, the polynomial basis degree of $p = 2$ used in this work does not warrant for polynomial dealiasing, which can be achieved by explicitly filtering the solution at every time step \citep{diosady2013design,gassner2013accuracy,brazell2015using}.

The results obtained using the dynamic slip-wall model are compared with those obtained using an equilibrium wall-stress model (EQWM). For the EQWM, we compute the wall friction $\tau_w$ from the instantaneous velocity taken at the furthest distance from the wall inside the first element. The computed wall friction is then used as the Neumann boundary condition applied at the quadrature points of the boundary faces. Our implementation is similar to the work of \cite{carton2017assessment}. We use the Reichardt function of the form
\begin{equation}
    u^+ = \dfrac{1}{\kappa} ln(1 + \kappa z^+) + 6.646 \left[ 1 - exp\left( \dfrac{-z^+}{11} \right) - \dfrac{z^+}{11}  exp\left( \dfrac{-z^+}{3} \right)  \right], 
    \label{Eqn:Reichardt}
\end{equation}
where $ \kappa = 0.38$, as the equilibrium wall-stress model, and use the Newton-Raphson method to iterate on the values of $u^+$ and $y^+$. This wall function supports the theoretical velocity profile down to the wall. We have chosen this approach for its simplicity and efficiency, and it is shown to give excellent results for statistically steady channel flows at high Reynolds numbers in \cite{carton2017assessment}. Note that for the simulations of separated flows in periodic hill configuration, we do not use any adjustments to the implementation of the EQWM like turning off the model in the separation region.

%%%%%%%%%%%%%%%%%%%%%%%%%%%%%%%%%%%%%%%%%%%%%%%%%%%%%%%%%%%%%%%%%%%%%%%%%%%%%%%%%%

%%%%%%%%%%%%%%%%%%%%%%%%%%%%%%%%%%%%%%%%%%%%%%%%%%%%%%%%%%%%%%%%%%%%%%%%

\subsection{Sharp modal cut-off filter as a test filter }\label{subsec:filter}

The dynamic modeling procedure requires filtering at two different levels i.e. grid filter and test filter to calculate the value of the model coefficient $\lambda$. In a DG framework, this is equivalent to using two different orders of polynomial basis for approximating the solution. The Lagrange interpolation polynomials, which are used as the basis functions within our work, are not hierarchical i.e. every basis function contains high-order solution content. As a result, unlike a spectral method, we cannot directly use a sharp cut-off filter to remove the higher-order modes. To reduce the order of projection which would result in a coarser filtering operation, the solution coefficients need to be transformed to a modal representation the hierarchical form of which allows for a classification of solution modes based on polynomial degree. The solution can then be coarsely filtered by setting the higher-order modes to zero or by scaling the higher-order coefficients by a factor $ \alpha \in [0, 1]$. In this work, we use a cut-off filter order $p^\star = 1$ and set the modes of degree greater than one to zero for the test filter. This is equivalent to assuming a test filter to be about twice the width of the grid filter, which is generally followed in finite difference or finite volume methods \citep{pope2000turbulent}. Once the filtered forms of modal solution coefficients are obtained, an inverse transformation is performed to get the filtered nodal solution coefficients thereby obtaining the coarse-filtered solution. In this work, we follow the procedure outlined by \cite{brazell2015using} to implement the sharp modal cut-off filter in our solver, and it is discussed in detail in Appendix \ref{appendixA}.

%%%%%%%%%%%%%%%%%%%%%%%%%%%%%%%%%%%%%%%%%%%%%%%%%%%%%%%%%%%%%%%%%%%%%%%%%%%%%%

\subsection{Dynamic Smagorinsky model}\label{sec:rules_submission}

The dynamic Smagorinsky Model (DSM) is a simple eddy viscosity model that relates the unresolved SGS stresses to the resolved strain-rate $\overline{S}_{ij}$ via a turbulent viscosity $\nu_{SGS}$ as
\begin{align}
 \tau_{ij}^{SGS} = 2 \overline{\rho} \nu_{SGS} \overline{S}_{ij} \hspace{0.05in}, \hspace{0.1in} \text{where} \hspace{0.1in}  \overline{S}_{ij} = \dfrac{1}{2} \left( \dfrac{\partial \overline{u}_i }{\partial x_j } + \dfrac{\partial \overline{u}_j }{\partial x_i } \right) \hspace{0.02in} .
\end{align}
The SGS eddy viscosity is related to a characteristic velocity and a length scale on dimensional grounds, and it is given by 
\begin{equation}
    \nu_{SGS} = ( C_S \Delta )^2 | \overline{S} | , 
\end{equation}
where $C_S$ is the Smagorinsky coefficient, $ | \overline{S} | = \sqrt{2 \overline{S}_{ij} \overline{S}_{ij} } $ is the strain-rate magnitude, and $\Delta $ is the filter width or a representative grid size. The DSM improves upon the original Smagorinsky model by dynamically adjusting the model coefficient $(C_S \Delta)$ based on local flow properties. The idea is to seek a more accurate representation of the turbulence, especially in regions with varying flow conditions. DSM also provides a near-wall correction that can lead to proper near-wall behavior of the SGS viscosity without the use of wall-damping functions. The dynamic calculation of the coefficient is based on an explicitly performed second-level filter operation called the test filter that is applied to the grid-filtered variables. As mentioned before, we denote the test filter operation by a hat, and we use the sharp modal cut-off filter as the test filter as discussed in \ref{subsec:filter}. The model coefficient $(C_S \Delta)$ is calculated as

\begin{equation}
(C_s \Delta)^2  = \dfrac{1}{2} \dfrac{ L_{ij}^{d} M_{ij} }{M_{kl} M_{kl}},
\label{eqn:dyn-smag}
\end{equation}
where the Leonard stress tensor $L_{ij}$ and its deviatoric part $ L_{ij}^{d} $ are given by
\begin{align}
    L_{ij} = \widehat{ \overline{u}_i \overline{u}_j} - \widehat{ \overline{u}}_i  \widehat{ \overline{u}}_j , \hspace{0.1in}  L_{ij}^{d} = L_{ij} - \dfrac{1}{3} L_{kk} \delta_{ij},
\end{align}
and 
\begin{equation}
    M_{ij} =  \widehat{ | \overline{S} | \overline{S}_{ij} }  - \Delta^{2}_{R} | \widehat{\overline{S}} | \widehat{\overline{S}}_{ij} \hspace{0.02in} .
\end{equation}
The derivative and the test filter operations do not commute for the sharp modal cut-off filter. We follow \cite{brazell2015using} to determine the second term of $M_{ij}$ by computing the test-filtered velocity followed by the derivatives of the test-filtered velocity to form the strain-rate tensor. Another possible choice is to use the test filter operation on the grid-filtered strain-rate, but this approach does not have any advantages over the method used here as shown by \cite{brazell2015using}. The parameter $\Delta_R$ is calculated as per the recommendation of \cite{brazell2015using} and it is given by
\begin{equation}
    \Delta_{R} = \dfrac{\widehat{\overline{\Delta}}_{w}}{\overline{\Delta}_w} = \dfrac{p + 1}{p^\star + 1}.
    \label{eqn:deltaR}
\end{equation}

%%%%%%%%%%%%%%%%%%%%%%%%%%%%%%%%%%%%%%%%%%%%%%%%%%%%%%%%%%%%%%%%%%%%%%%%%

The numerator in Eqn.~(\ref{eqn:dyn-smag}) can assume local negative values and this is physically consistent as it corresponds to energy backscatter i.e. energy from the SGS scales is transferred back to the resolved scales. However, negative SGS viscosity values can numerically destabilize the simulation, especially when the sum $(\nu + \nu_{SGS})$ becomes negative. Thus, it is customary to perform some type of averaging of the numerator and denominator, generally in the directions of homogeneity. In this work, we perform two-step averaging. First, the numerator and denominator are averaged over an element to get their representative single values in each element. After this, the numerator and denominator are averaged over the homogeneous directions to get the final averaged numerator and denominator as $ < L_{ij}^{d} M_{ij} > $ and $<M_{kl} M_{kl}>$, respectively. Finally, the ratio $[0.5 (<L_{ij}^{d} M_{ij}>/<M_{kl} M_{kl}>)]$ is clipped to  get non-negative values.

%%%%%%%%%%%%%%%%%%%%%%%%%%%%%%%%%%%%%%%%%%%%%%%%%%%%%%%%%%%%%%%%%%%%%%%%%

\subsection{Application to statistically stationary channel flows}\label{sec:types_paper}

The new dynamic slip wall model is applied to a series of statistically stationary turbulent channel flows that are homogeneous in directions parallel to the wall. The fully developed turbulent flow between the two parallel walls is separated by a distance $2 \delta$ in the $z$-direction, where $\delta$ is the half-channel height. The flow is assumed to be periodic in the stream-wise ($x$) and span-wise ($y$) directions. The simplicity of geometry and boundary conditions makes this canonical flow configuration an appealing test case, and it has been used to validate the performance of previous dynamic slip wall models \citep{bose2014dynamic,bae2019dynamic}. 

The friction Reynolds number is set as $Re_\tau = \overline{\rho} u_\tau \delta / \mu$, with $u_\tau = \sqrt{\tau_w/ \overline{\rho}}$, $u_\tau$ being the friction velocity based on the wall shear stress $\tau_w$ taken from the available DNS. The friction Reynolds number is imposed through a constant forcing in the x-momentum equation using the pressure gradient $ \langle dp/dx \rangle = - \tau_w / \delta$. A number of WMLES studies use the constant pressure gradient as the forcing strategy in finite-volume \citep{yang2017physical,bae2019dynamic} and discontinuous Galerkin frameworks \citep{carton2017assessment,frere2017application,lv2021discontinuous}. For a DNS, \cite{quadrio2016does} show that the specific choice of the forcing term, i.e. constant mass flow or constant pressure gradient, does not produce important statistical consequences and one-point statistics do not show any appreciable difference between the simulations.

The size of the computational domain is $2 \pi \delta$ in the $x$-direction and $\pi \delta$ in the $y$-direction. The degree of polynomial $p$ used for all the simulations presented here is $2$ and the sharp modal cut-off order $p^\star$ is $1$. For all the cases considered, the flow is initially evolved for at least time $20 \delta / u_\tau$ units and statistics are sampled for an additional $10 \delta / u_\tau$ time units. One-point statistics including the mean velocity and Reynolds shear and normal stresses are compared with the  DNS of \cite{lee2015direct} and \cite{hoyas2022wall}. 

The performance of the proposed dynamic slip-wall model is validated using the cases listed in Table~\ref{tab:grid}, which shows the simulated Reynolds numbers and the grid resolutions in inner and outer layer units. The meshes are uniform in the stream-wise, span-wise, and wall-normal directions. The first element size in the wall-normal direction for all the considered cases is significantly coarser than a conventional LES mesh and the resolution is insufficient to resolve  near-wall turbulent structures. As a result, none of the simulated Reynolds numbers with the grids given in Table~\ref{tab:grid} are wall-resolved.

%%%%%%%%%%%%%%%%%%%%%%%%%%%%%%%%%%%%%%%%%%%%%%%%%%%%%%%%%%%%%%%%%%%%%%%%

\begin{table}
  \begin{center}
\def~{\hphantom{0}}
  \begin{tabular}{lccccccccccccc} 
Case & $Re_\tau$ & $N_x \times N_y \times N_z$  & $\Delta_x^+$ & ${\Delta_x}/{\delta}$  &  $\Delta_y^+$ & ${\Delta_y}/{\delta}$  & $\Delta_{z}^+$ & ${\Delta_{z}}/{\delta}$ \\[3pt]
DSW-2000-G1 & 1994.756  & $8 \times 8  \times  8$  &  720.68  & 0.393 & 360.34 & 0.196  & 229.4  & 0.125 \\
DSW-2000-G2 & 1994.756  & $16 \times 16  \times  16$  & 360.34  & 0.196 & 180.17 & 0.098  & 114.7  & 0.0625 \\
DSW-2000-G3 & 1994.756  & $32 \times 32  \times  32$  & 180.17  & 0.098 & 90.1 & 0.049  & 57.35  & 0.03125 \\
DSW-5200-G1 & 5185.897  & $8 \times 8  \times  8$  &  1627.35  & 0.393 & 813.67 & 0.196  & 518  & 0.125 \\
DSW-5200-G2 & 5185.897  & $16 \times 16  \times  16$  & 813.67  & 0.196 & 406.84 & 0.098  & 259  & 0.0625 \\
DSW-5200-G3 & 5185.897  & $32 \times 32  \times  32$  & 406.84  & 0.098 & 203.42 & 0.049  & 129.5  & 0.03125 \\
DSW-10000-G1 & 10049  & $8 \times 8  \times  8$  &  3935  & 0.393 & 1967.52 & 0.196  & 1252.56  & 0.125 \\
DSW-10000-G2 & 10049  & $16 \times 16  \times  16$  & 1967.52  & 0.196 & 983.76 & 0.098  & 626.28  & 0.0625 \\
DSW-10000-G3 & 10049  & $32 \times 32  \times  32$  & 983.76  & 0.098 & 491.88 & 0.049  & 313.14  & 0.03125 \\
\end{tabular}
\caption{Summary of mesh parameters for the different simulated Reynolds numbers. Here, $\Delta_x$, $\Delta_y$, and $\Delta_z$ are the effective grid sizes in the stream-wise ($x$), span-wise ($y$), and wall-normal ($z$) directions, respectively, $\delta$ is the half channel height, and $\Delta_x^+$,  $\Delta_y^+$, and $\Delta_z^+$ are normalized with wall units. $N_x$, $N_y$, and $N_z$ represent the number of elements in the stream-wise, span-wise, and wall-normal directions, respectively. The number of degrees of freedom in each direction is given by $(p+1) N_x$, $(p+1) N_y$, and $(p+1) N_z$, where $p$ is the degree of the polynomial basis. Note that the numerical experiments are labeled following the convention
[Dynamic slip-wall model (DSW)]-[$Re_\tau$]-[grid resolution]. }
  \label{tab:grid}
  \end{center}
\end{table}

%%%%%%%%%%%%%%%%%%%%%%%%%%%%%%%%%%%%%%%%%%%%%%%%%%%%%%%%%%%%%%%%%%%%%%%%%%%%%

%%%%%%%%%%%%%%%%%%   Re 10000 grid converence %%%%%%%%%%%%%%%%%%%%%%%%%%%%%%%%%%%%%%%%%%%%%%

\begin{figure}
\centering
\begin{subfigure}{0.49\textwidth}
\centering
\includegraphics[width=0.99\linewidth, trim = 0mm 0mm 0mm 0mm]{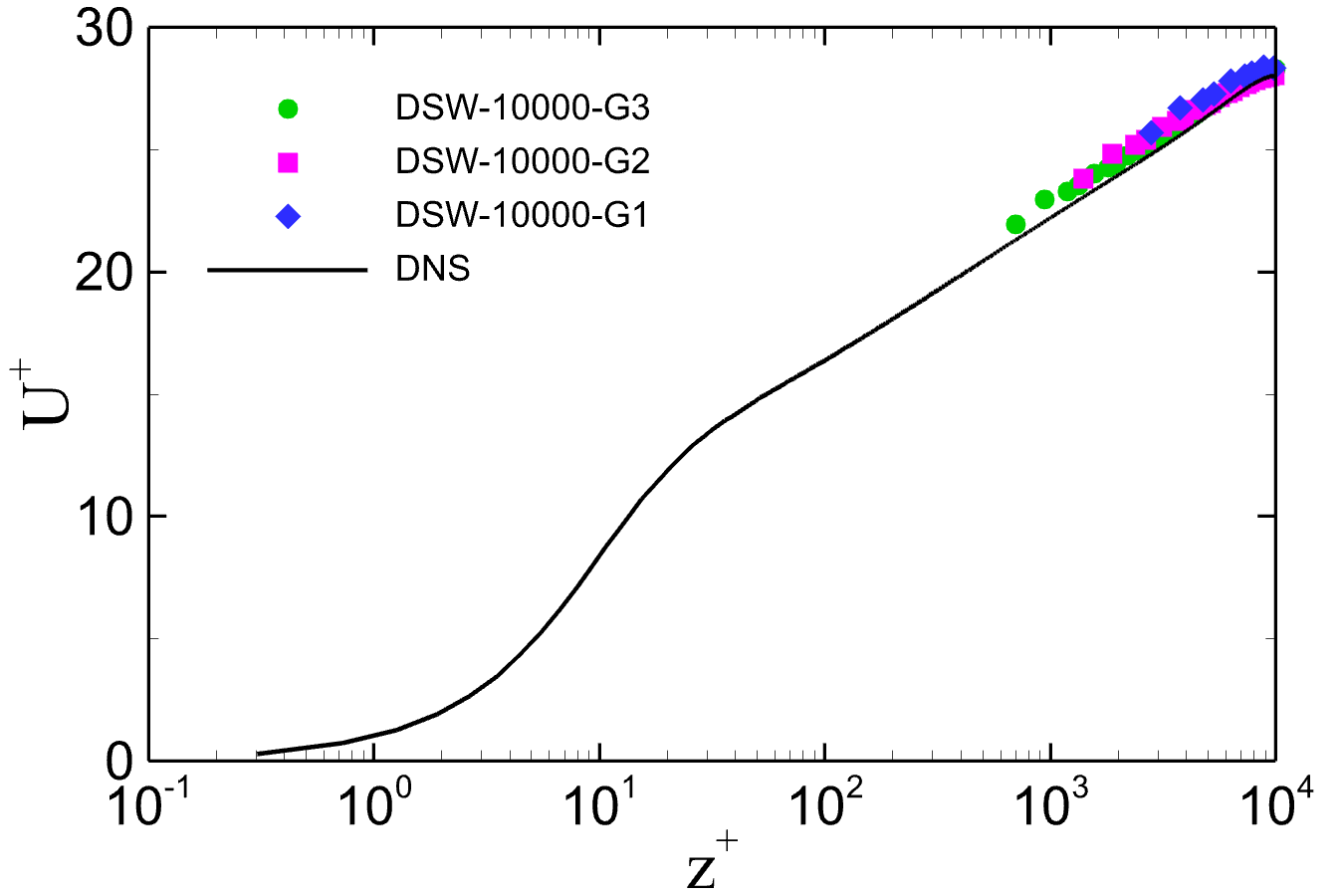}
\caption{}
\end{subfigure} 
\begin{subfigure}{0.49\textwidth}
\centering
\includegraphics[width=0.99\linewidth, trim = 0mm 0mm 0mm 0mm]{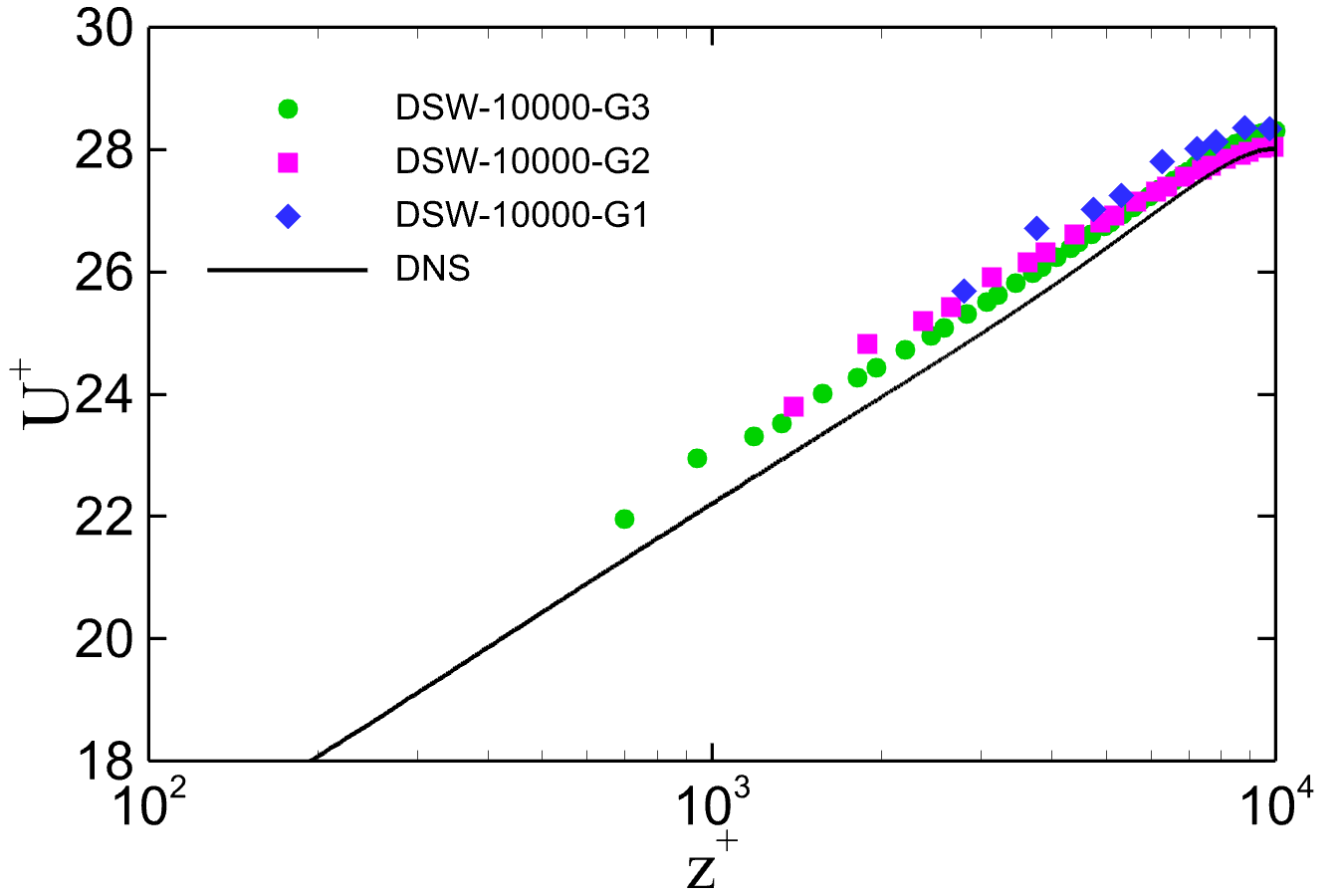}
\caption{}
\end{subfigure}	
\caption{Grid refinement study for the proposed dynamic wall model showing comparisons between model predictions and DNS for the stream-wise mean velocity at $Re_\tau \approx 10000$. (a) Classical visualization and (b) Visualization focused on the bulk profile and the top interface of the first element.}
\label{fig:DSW-Re10000}
\end{figure}

A grid sensitivity study for channel flow at $Re_\tau \approx 10000$ is shown in Fig.~\ref{fig:DSW-Re10000}. Starting with a coarse mesh with $8 \times 8 \times 8$ elements, the number of elements in each of the three directions is doubled at each level of refinement. This corresponds to three different coarse near-wall resolutions of $\Delta z = 0.125 \delta, 0.0625 \delta,$ and $0.03125 \delta$ i.e. $\Delta_{z_w}^{+} \approx 1253$, $626$, and $313$. The details of the mesh parameters are given in Table~\ref{tab:grid}. For all the cases, the first off-wall grid point lies in the log-layer. The results are plotted starting from the second off-wall element at $(p+1)$ quadrature points in each element. It can be seen that the mean velocity for the coarsest mesh G1 has a slight positive log-layer mismatch which reduces upon grid refinement. The difference between the model predictions at each of the successive grid refinement levels is less than $1 \%$. Grid refinement study for the other two Reynolds number cases shows a similar trend.

%%%%%%%%%%%%%%%%%%%%  $Re_\tau = 10000 grid refinement %%%%%%%%%%%%%%%%%%%%%%%%%
	
%%%%%%%%%%%%%%%%%%%%%%%%%%%%%%%%%%%%%%%%%%%%%%%%%%%%%%%%%%%%%%%%%%%%%%%%%%

\begin{figure}
\centering
\begin{subfigure}{0.49\textwidth}
\centering
\includegraphics[width=0.99\linewidth, trim = 0mm 0mm 0mm 0mm]{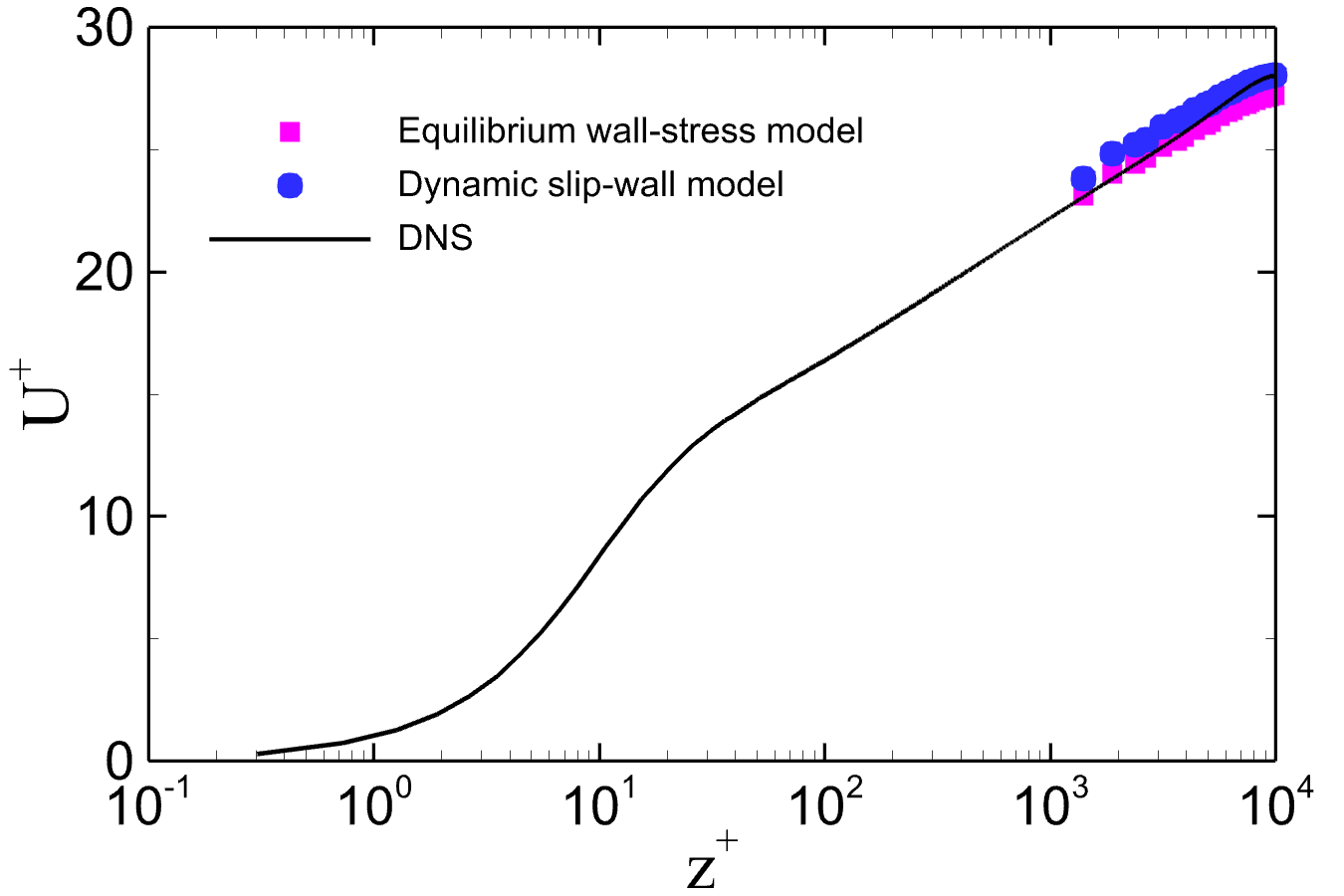}
\caption{}
\end{subfigure} 
\centering
\begin{subfigure}{0.49\textwidth}
\centering
\includegraphics[width=0.99\linewidth, trim = 0mm 0mm 0mm 0mm]{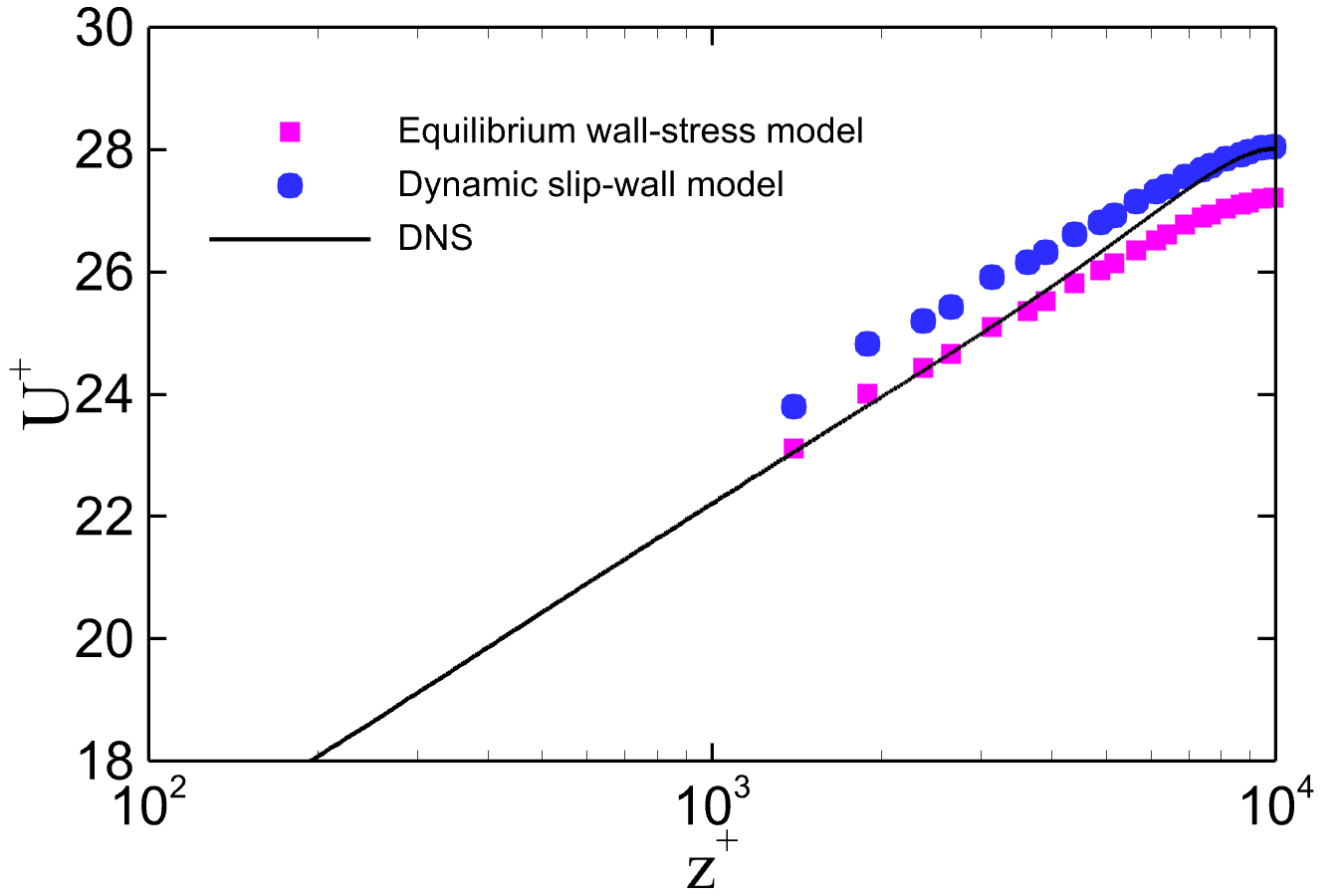}
\caption{}
\end{subfigure} \\
\begin{subfigure}{0.49\textwidth}
\centering
\includegraphics[width=0.99\linewidth, trim = 0mm 0mm 0mm 0mm]{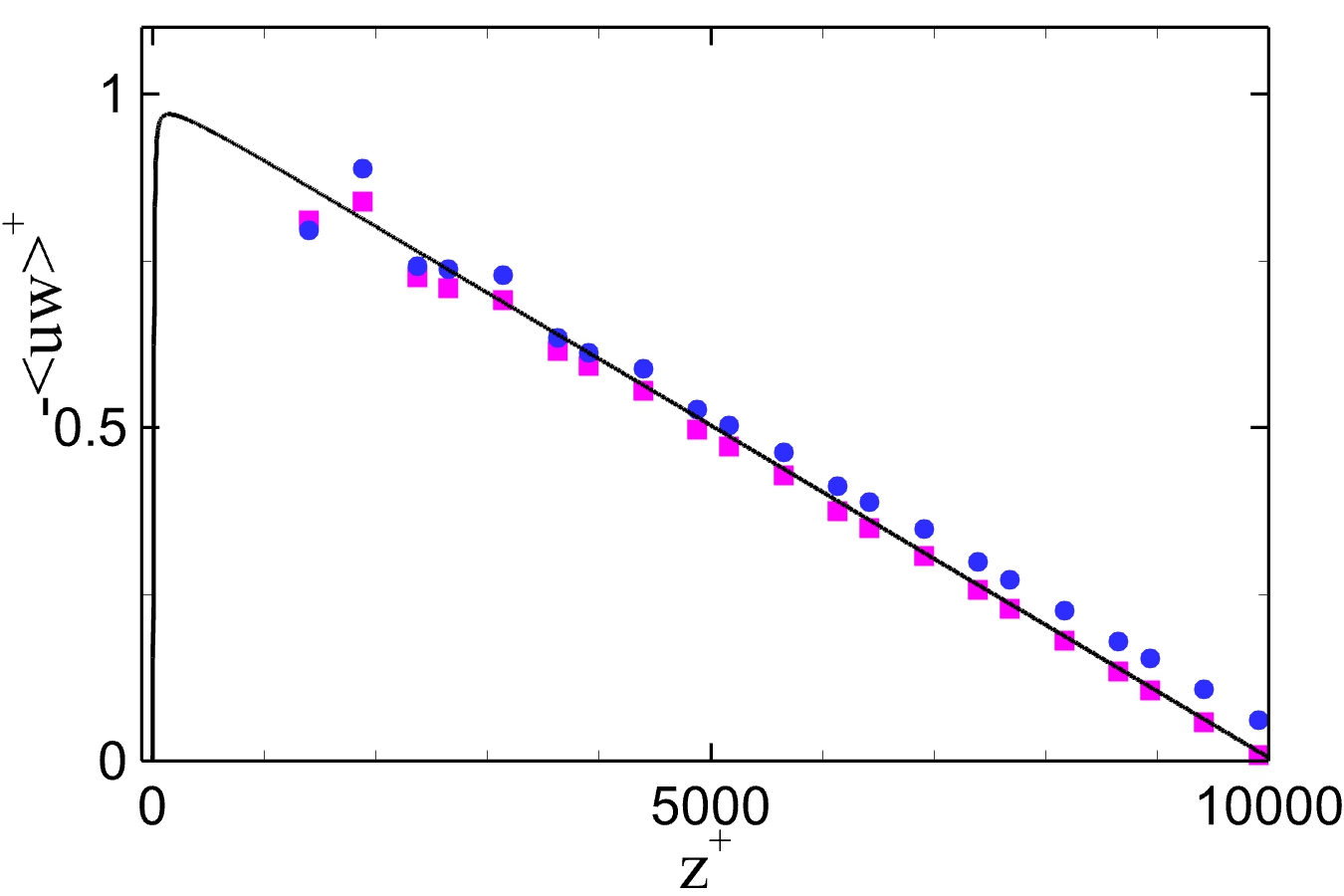}
\caption{}
\end{subfigure} 
\begin{subfigure}{0.49\textwidth}
\centering
\includegraphics[width=0.99\linewidth, trim = 0mm 0mm 0mm 0mm]{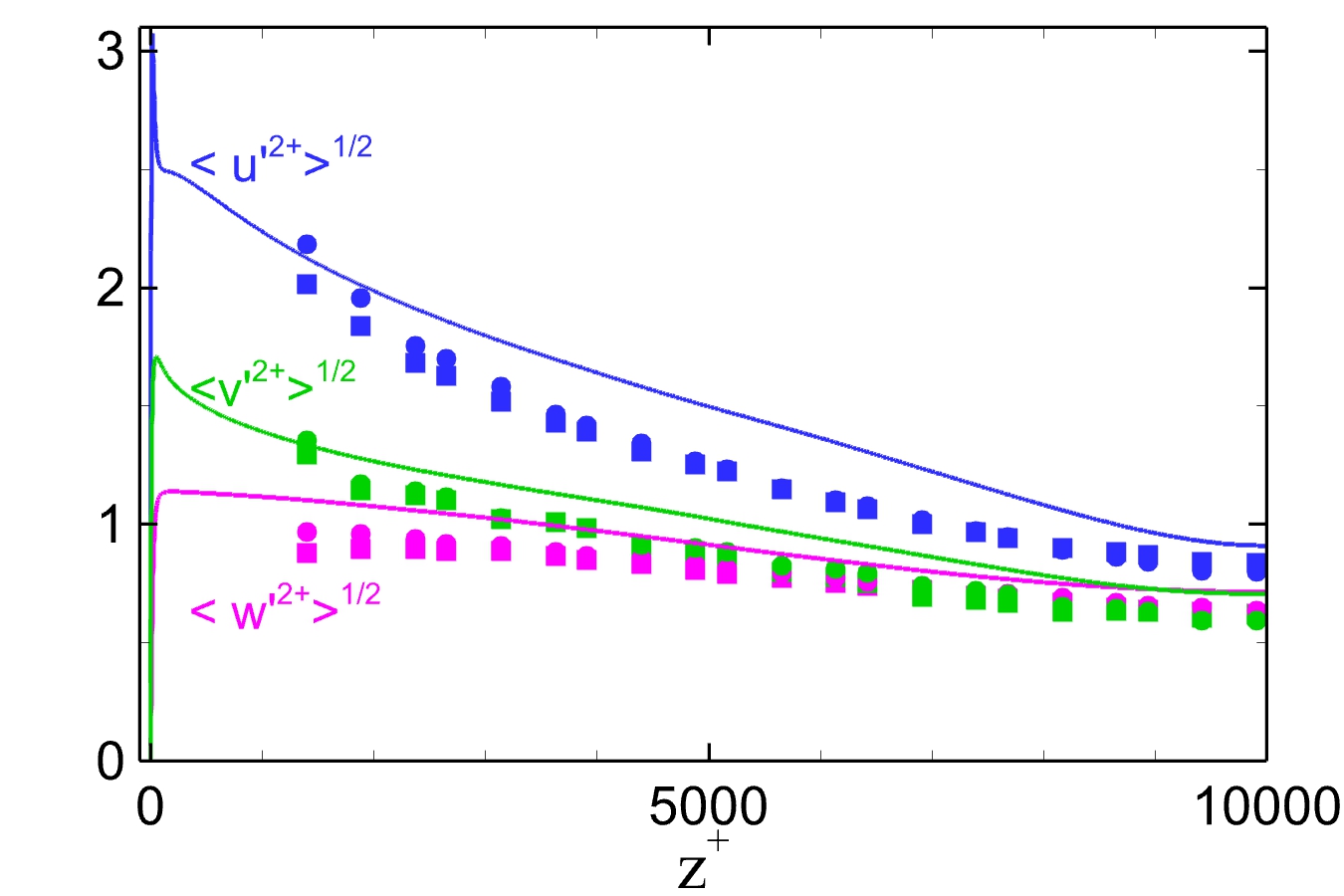}
\caption{}
\end{subfigure}	
\caption{Comparison between the proposed dynamic slip wall model and equilibrium wall-stress model predictions using grid G2 with the DNS for (a) full mean velocity profile, (b) mean velocity in the log region, (c) Reynolds shear stress, (d) r.m.s. velocity fluctuations at $Re_\tau \approx 10000$.}
\label{fig:Re10000-TradVsDynSlip}
\end{figure}

A comparison between the dynamic slip-wall model and the equilibrium wall-stress model predictions with the DNS is shown in Fig.~\ref{fig:Re10000-TradVsDynSlip}. The model predictions are obtained on Grid G2 with $16 \times 16 \times 16$ elements. The slope of the mean velocity profiles obtained using the two models is similar, however there is a slight shift between them. Both the mean velocity profile predictions match well with the DNS. The two model predictions for the Reynolds shear stress profiles also match the DNS well. The stream-wise Reynolds stress predicted by the two models is also similar but there is a slight mismatch with the DNS. On the other hand, the span-wise and wall-normal Reynolds stress profiles obtained using the two models closely agree with the DNS.

%%%%%%%%%%%%%%%%%%%%  $Re_\tau = 10000 model compars %%%%%%%%%%%%%%%%%%%%%%%%%

%%%%%%%%%%%%%%%%%%%%%%%%%%%%%%%%%%%%%%%%%%%%%%%%%%%%%%%%%%%%%%%%%%%%%%%%%%

%%%%%%%%%%%%%%%%%%%%%%%%%%%%%%%%%%%%%%%%%%%%%%%%%%%%%%%%%%%%%%%%%%%%%%%%%%%%%

\begin{figure}
\centering
\begin{subfigure}{0.49\textwidth}
\centering
\includegraphics[width=0.99\linewidth, trim = 0mm 0mm 0mm 0mm]{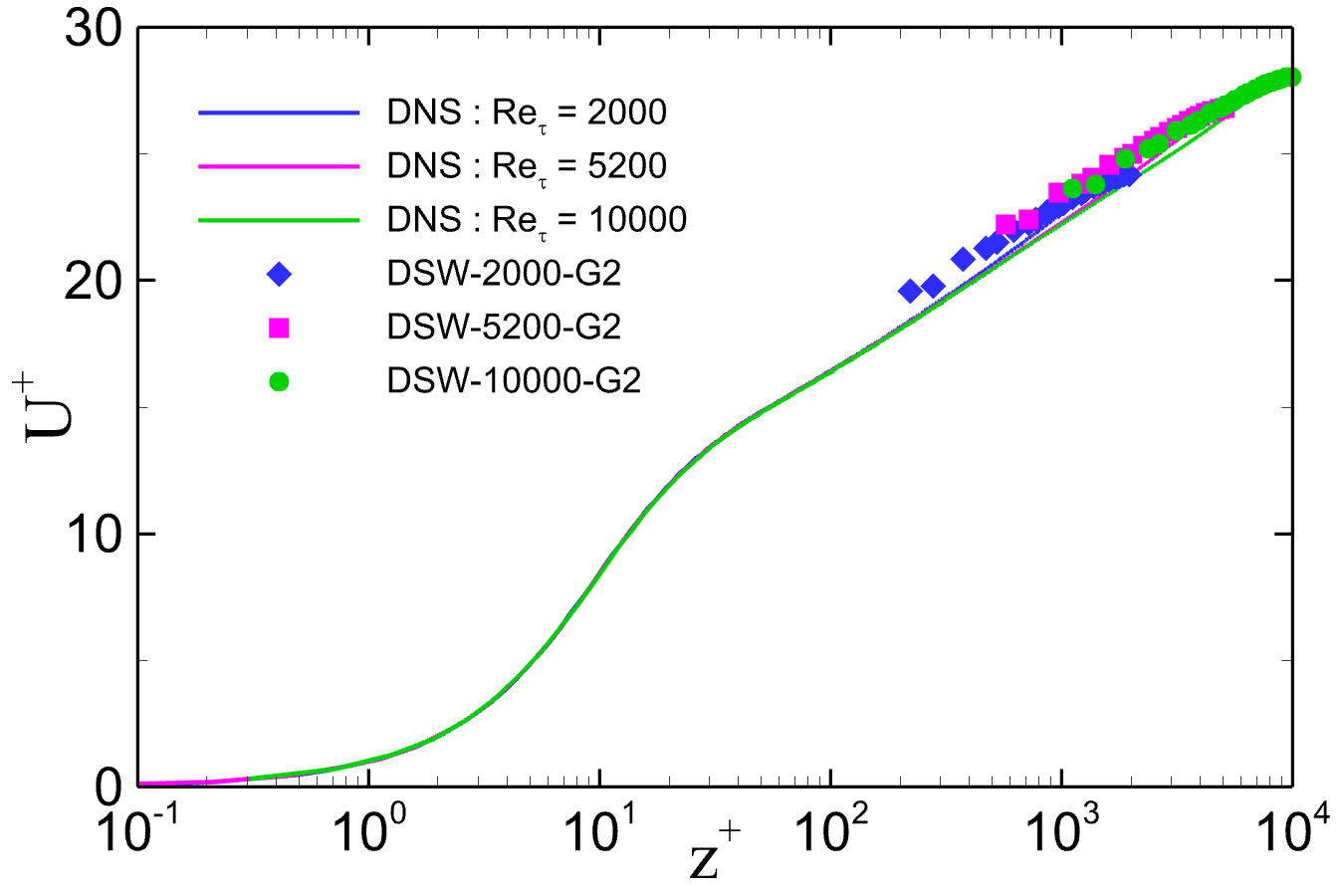}
\caption{}
\end{subfigure} 
\begin{subfigure}{0.49\textwidth}
\centering
\includegraphics[width=0.99\linewidth, trim = 0mm 0mm 0mm 0mm]{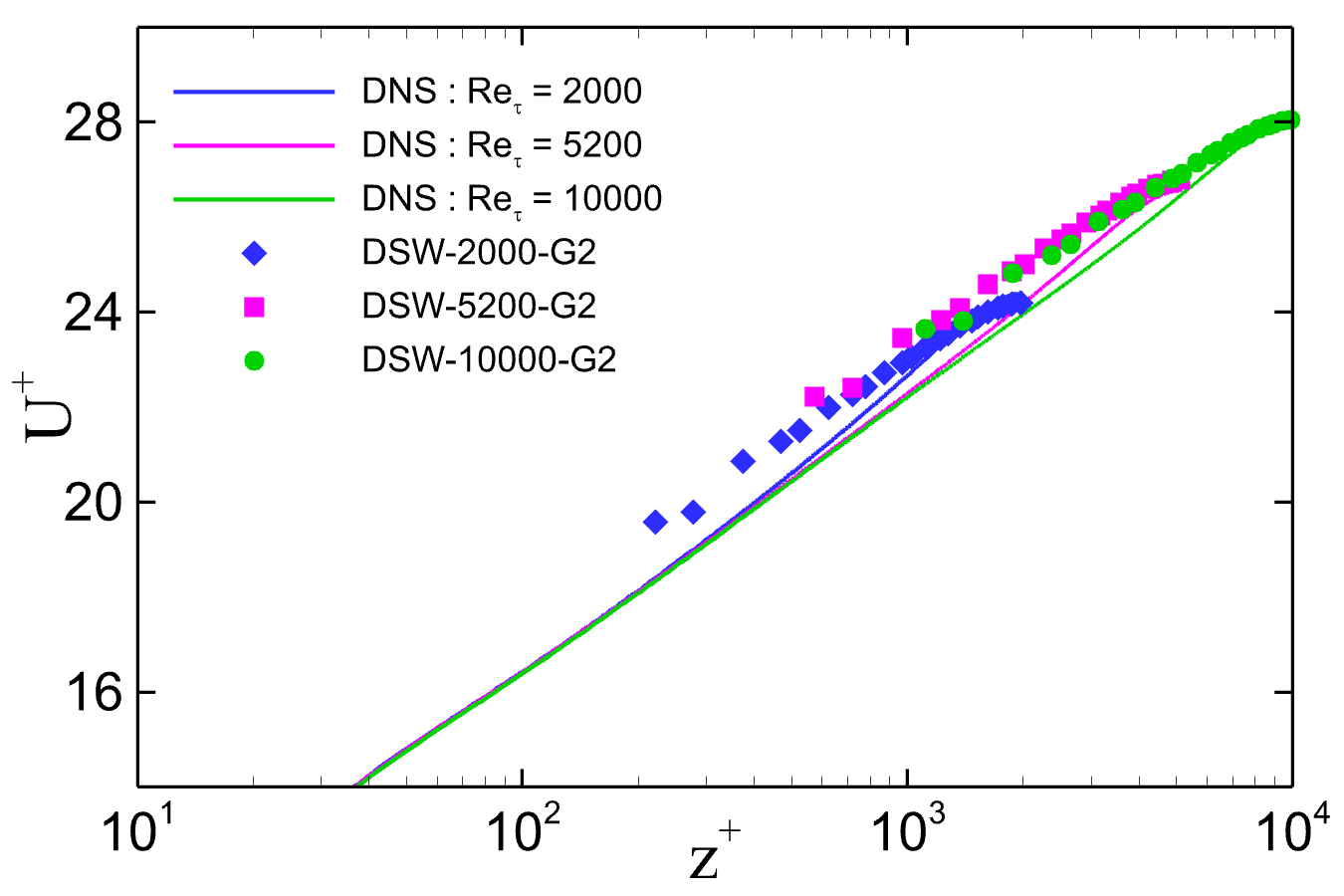}
\caption{}
\end{subfigure} \\
\begin{subfigure}{0.49\textwidth}
\centering
\includegraphics[width=0.99\linewidth, trim = 0mm 0mm 0mm 0mm]{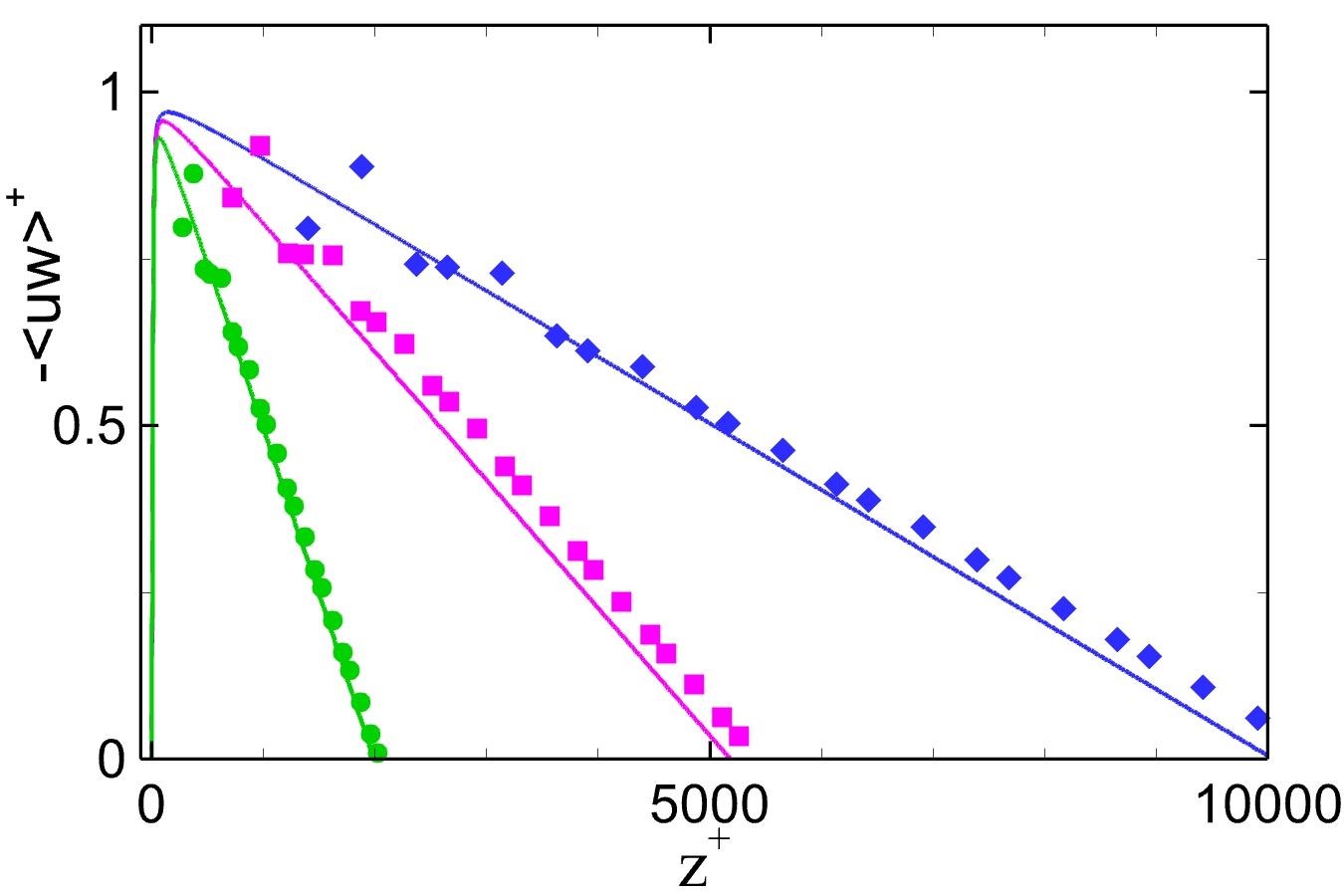}
\caption{}
\end{subfigure} \\
\begin{subfigure}{1.0\textwidth}
\centering
\includegraphics[width=1.0\linewidth, trim = 0mm 0mm 0mm 0mm]{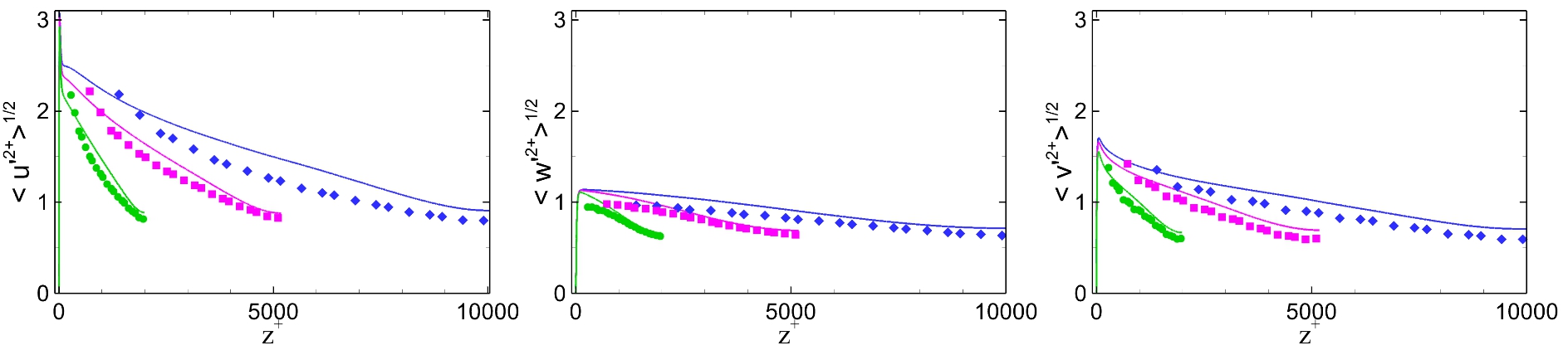}
\caption{}
\end{subfigure}	
\caption{Comparison between DNS and proposed dynamic slip wall model predictions using grid G2 for (a) full mean velocity, (b) mean velocity profile in the log region, (c) Reynolds shear stress, (d) r.m.s. velocity fluctuations at $Re_\tau \approx 2000, 5200,$ and $10000$. }
\label{fig:Re_vary}
\end{figure}

\begin{figure}
\centering
\includegraphics[width=0.45\linewidth, trim = 0mm 0mm 0mm 0mm]{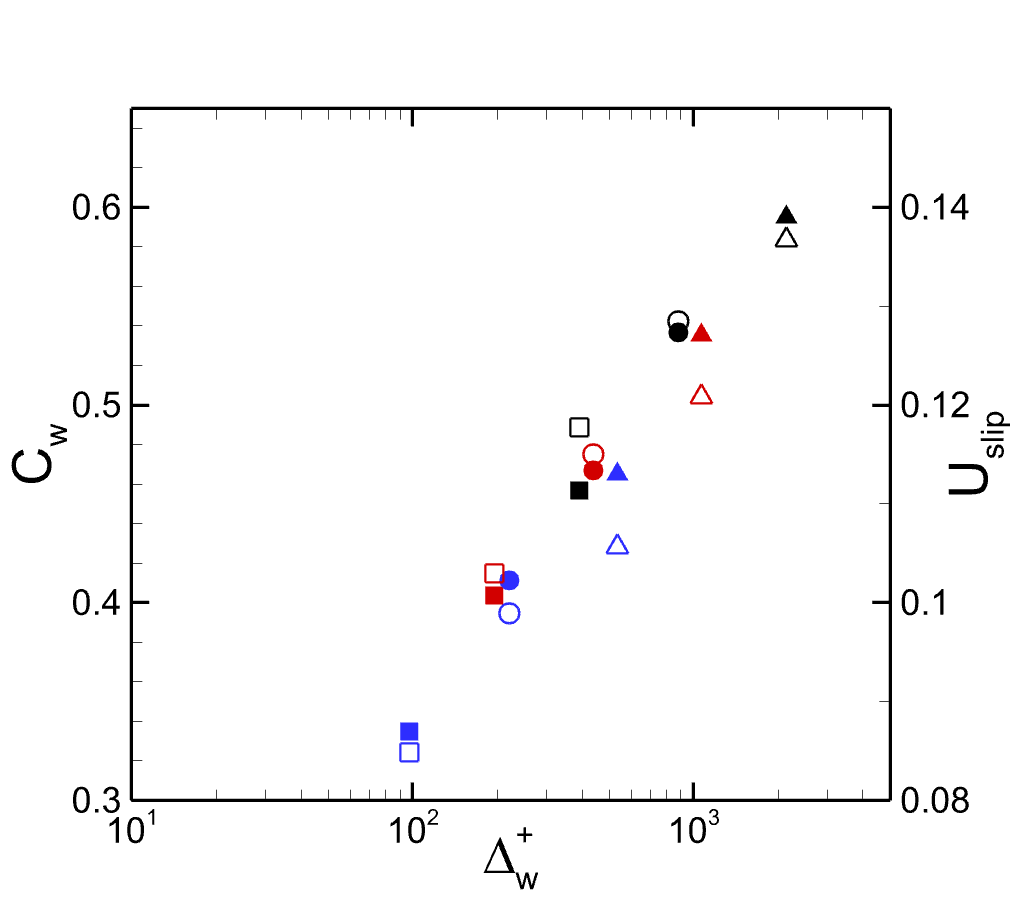}
\caption{Slip parameter $C_w$ (filled symbols) and stream-wise slip-velocity $U_{slip}$ (unfilled symbols) as a function of near-wall grid resolution $\Delta_{w}^{+}$ at $Re_\tau \approx 2000$ (squares), $5200$ (circles), and $10000$ (deltas). Color code : black -- grid G1; red -- grid G2; blue -- grid G3.}	
\label{fig:cw-uslip-channel}
\end{figure}

One-point statistics on grid G2 at $Re_\tau \approx 2000, 5200,$ and $10000$ are presented in Fig.~\ref{fig:Re_vary} and compared with the available DNS. The first and second moments agree well with the DNS at the three Reynolds numbers. The model parameter $C_w$ and mean streamwise slip velocity $U_{slip}$ at the three Reynolds numbers and grid resolutions are plotted in Fig.~\ref{fig:cw-uslip-channel}. It is important to note that the slip-wall model is sensitive to the numerical implementation details, including the numerical discretization and the choice of the SGS model. Consequently, different optimal slip lengths are required for the correct prediction of the wall stress depending on the numerical setup. As a result, the quantitative assessment of model parameters like $C_w$ or $\lambda$ with DNS is difficult. However, $C_w$ variation with the near-wall grid resolution $\Delta_{w}^{+}$ follows an expected trend that is qualitatively similar to that observed by \cite{whitmore2023evaluation} for the optimal slip length estimates, i.e. larger slip lengths on coarser near-wall grid resolutions and decay of the slip length in the limit $\Delta_{w}^{+} \to 0 $. Moreover, the slip wall velocity trend is also consistent with the \textit{a priori} filtering tests of \cite{pradhan2023unified} using the optimal finite-element projection framework i.e. an increase in the slip velocity with an increase in the Reynolds number on identical grids, and an increase in slip velocity at the same Reynolds number upon coarsening the grid.

%%%%%%%%%%%%%%%%%%%%  $Re_\tau vary on Grid 2 %%%%%%%%%%%%%%%%%%%%%%%%%

%%%%%%%%%%%%%%%%%%%%%%%%%%%%%%%%%%%%%%%%%%%%%%%%%%%%%%%%%%%%%%%%%%%%%%%%%%

%%%%%%%%%%%%%  Cw-Uslip Error %%%%%%%%%%%%%%%%%%%%%%%%%%%%%%%%%%%

%%%%%%%%%%%%%%%%%%%%%%%%%%%%%%%%%%%%%%%%%%%%%%%%%%%%%%%%%%%%%%%%%%%%%%%%%%

%%%%%%%%%%%%%  mass flow rate %%%%%%%%%%%%%%%%%%%%%%%%%%%%%%%%%%%

\begin{figure}
\centering
\includegraphics[width=0.45\linewidth, trim = 0mm 0mm 0mm 0mm]{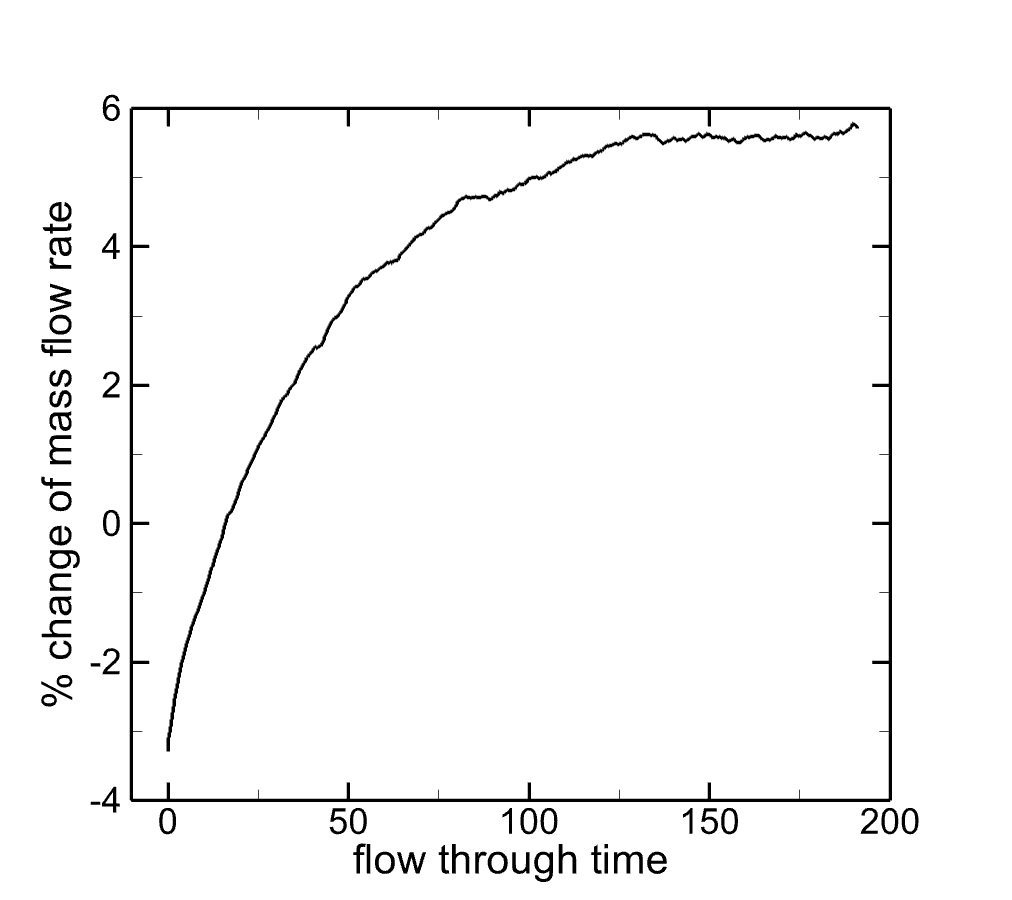}
\caption{Change in the mass flow rate relative to the nominal value plotted as a function of the flow through time obtained at $Re_\tau \approx 2000$ on Grid G2.}	
\label{fig:mass-flow}
\end{figure} 

%%%%%%%%%%%%%%%%%%%%%%%%%%%%%%%%%%%%%%%%%%%%%%%%%%%%%%%%%%%%%%%%%%%%%%%%%%

The model parameter $C_w$ and hence the slip length is found to increase as the near-wall grid is coarsened at the considered Reynolds numbers (see Fig~\ref{fig:cw-uslip-channel}). The slip length modifies how momentum is transferred between the wall and the fluid. It effectively tunes the wall stress to account for near-wall turbulence effects without fully resolving the boundary layer. A larger slip length increases the velocity slip at the wall as seen in Fig.~\ref{fig:cw-uslip-channel} leading to a higher velocity near the boundaries. For the considered cases, this results in the reduction of wall shear stress. For the same driving pressure gradient, this implies that more of the pressure gradient contributes to accelerating the bulk flow instead of being dissipated near the walls. As a result, the velocity profile is modified such that the mass flow rate $U_b = (1/2 \delta) \int_{-\delta}^{\delta} \overline{u(z)} dz $ and hence the bulk Reynolds number $Re_b =  U_b \delta / \nu $ is found to increase for these cases as evident from the shift in mean velocity profiles relative to the DNS in Figs.~\ref{fig:DSW-Re10000} and \ref{fig:Re_vary}. The maximum increase in the mass flow rate relative to the nominal value is about $6$ \% on the coarsest grid G1 and the increase reduces upon grid refinement. Figure~\ref{fig:mass-flow} shows the change in the mass flow rate variation relative to its nominal value with the flow through time ($L_x/U_b$) at $Re_\tau \approx 2000$ on grid G2, and similar trends are observed for the other cases.

It is important to note that, the underlying assumptions for the slip condition become invalid for significantly larger near-wall grid resolutions commensurate to larger filter sizes. The shift in the mean velocity relative to the DNS is expected to increase as the near-wall grid is coarsened, as seen in Fig~\ref{fig:DSW-Re10000}. Similar observations are made in the study of \cite{bae2019dynamic} using the constant pressure gradient forcing term. Note that the increase in mass flow upon coarsening the grid is also observed in LES with no-slip boundary condition albeit on relatively finer grids than the WMLES grids.

\begin{figure}
\centering
\includegraphics[width=0.45\linewidth, trim = 0mm 0mm 0mm 0mm]{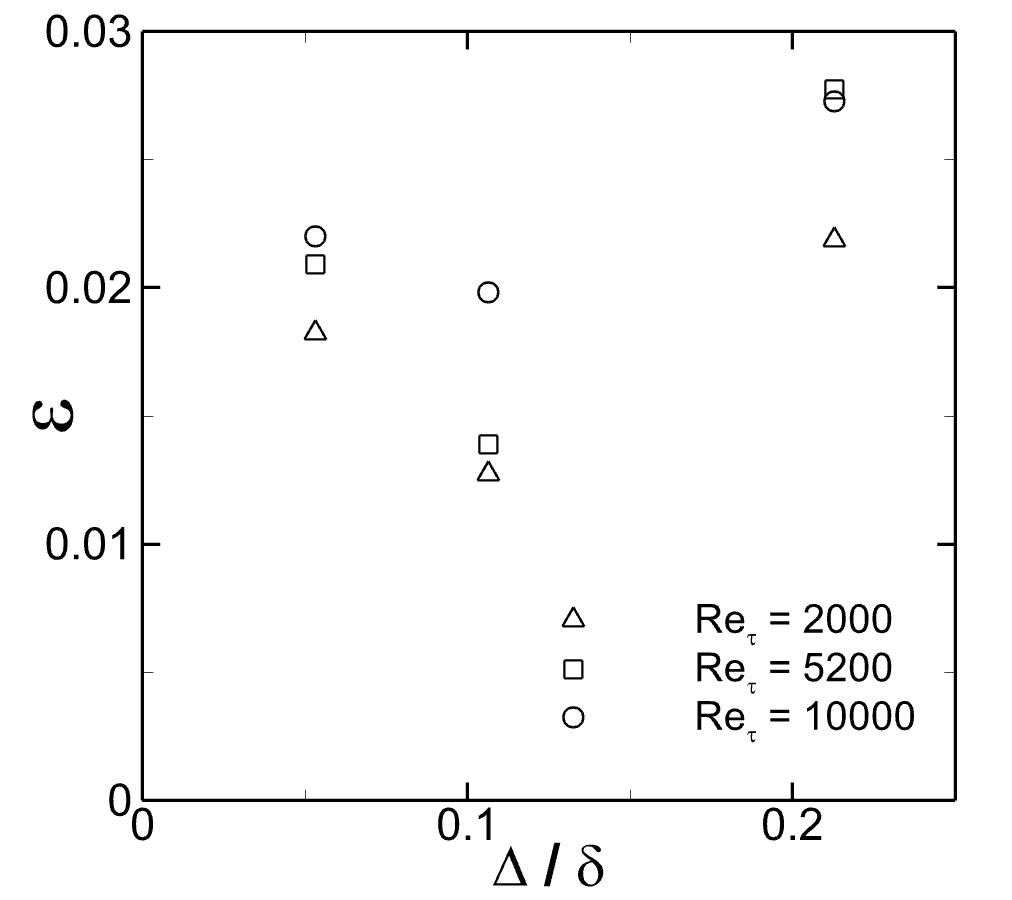}
\caption{Normalized $L_2$ error, $\mathcal{E}$, in stream-wise mean velocity $U^+$ as a function of grid resolution $\Delta$ at $Re_\tau \approx 2000, 5200$, and $10000$. }	
\label{fig:channel_flow_Error}
\end{figure}

A quantitative assessment of the dynamic slip-wall model is performed in terms of the normalized $L_2$ error in the stream-wise mean velocity $U^+$ predictions w.r.t the DNS for all the cases presented in Table~\ref{tab:grid}. The calculations exclude the first near-wall element. The normalised $L_2$ error is determined between the second off-wall element $\Delta^{+}_{2e}$ and the half channel height $\delta^+$ as
\begin{equation}
\mathcal{E} = \left[ \dfrac{\int_{\Delta^{+}_{2e}}^{\delta^{+}} (U^{+}_{DSW} - U^{+}_{DNS})^2  dz^+}{\int_{\Delta^{+}_{2e}}^{\delta^{+}} (U^{+}_{DNS})^2  dz^+} \right]^{1/2} \hspace{0.02in} .
\end{equation}
Here, $U^{+}_{DSW}$ and $U^{+}_{DNS}$ represent the mean velocity obtained using the proposed dynamic slip-wall model and DNS, respectively. The error is evaluated at $(p+1)$ quadrature points within each element in the wall-normal direction $z^+$, and the integration for each element is performed using quadrature. The error $\mathcal{E}$ is plotted as a function of the representative grid size $\Delta$ in Fig.~\ref{fig:channel_flow_Error}. We consider $\Delta$ based on element volume, $\Delta = (\Delta_x \Delta_y \Delta_z)^{1/3}$ with $\Delta_x$, $\Delta_y$, and $\Delta_z$ taken as the effective grid sizes in the stream-wise, span-wise, and wall-normal directions, respectively. The $L_2$ error slightly increases with an increase in Reynolds number on an identical grid. However, the maximum error is less than $3 \%$ for all the cases considered here demonstrating the performance of the model at practically relevant Reynolds numbers on significantly under-resolved near-wall LES mesh resolutions.

%%%%%%%%%%%%%  L2 Error %%%%%%%%%%%%%%%%%%%%%%%%%%%%%%%%%%%

%%%%%%%%%%%%%%%%%%%%%%%%%%%%%%%%%%%%%%%%%%%%%%%%%%%%%%%%%%%%%%%%%%%%%%%%%%

\begin{figure}
		\centering
		\begin{subfigure}{0.49\textwidth}
			\centering
			\includegraphics[width=1.\linewidth, trim = 0mm 0mm 0mm 0mm]{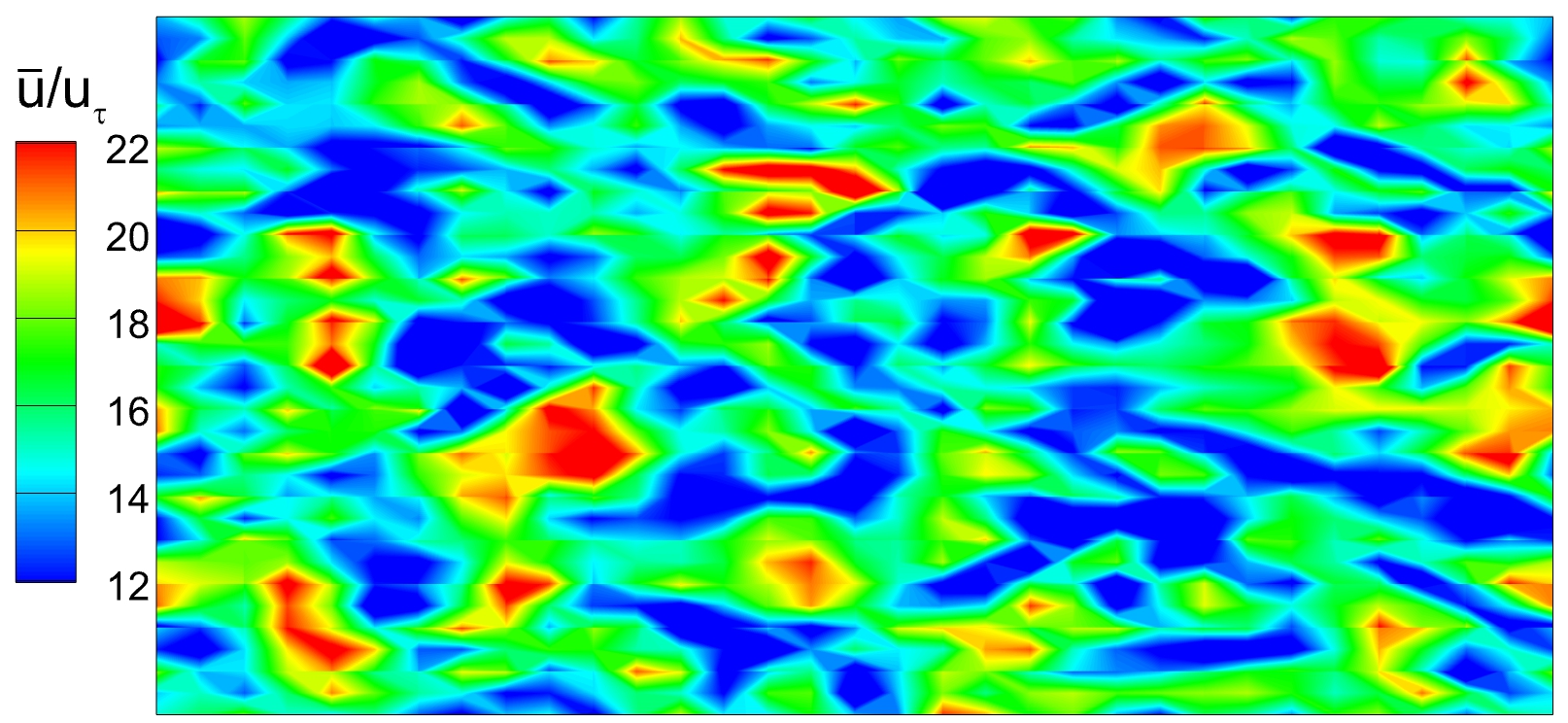}
			\caption{}
		\end{subfigure} 
		\begin{subfigure}{0.49\textwidth}
			\centering
			\includegraphics[width=1.\linewidth, trim = 0mm 0mm 0mm 0mm]{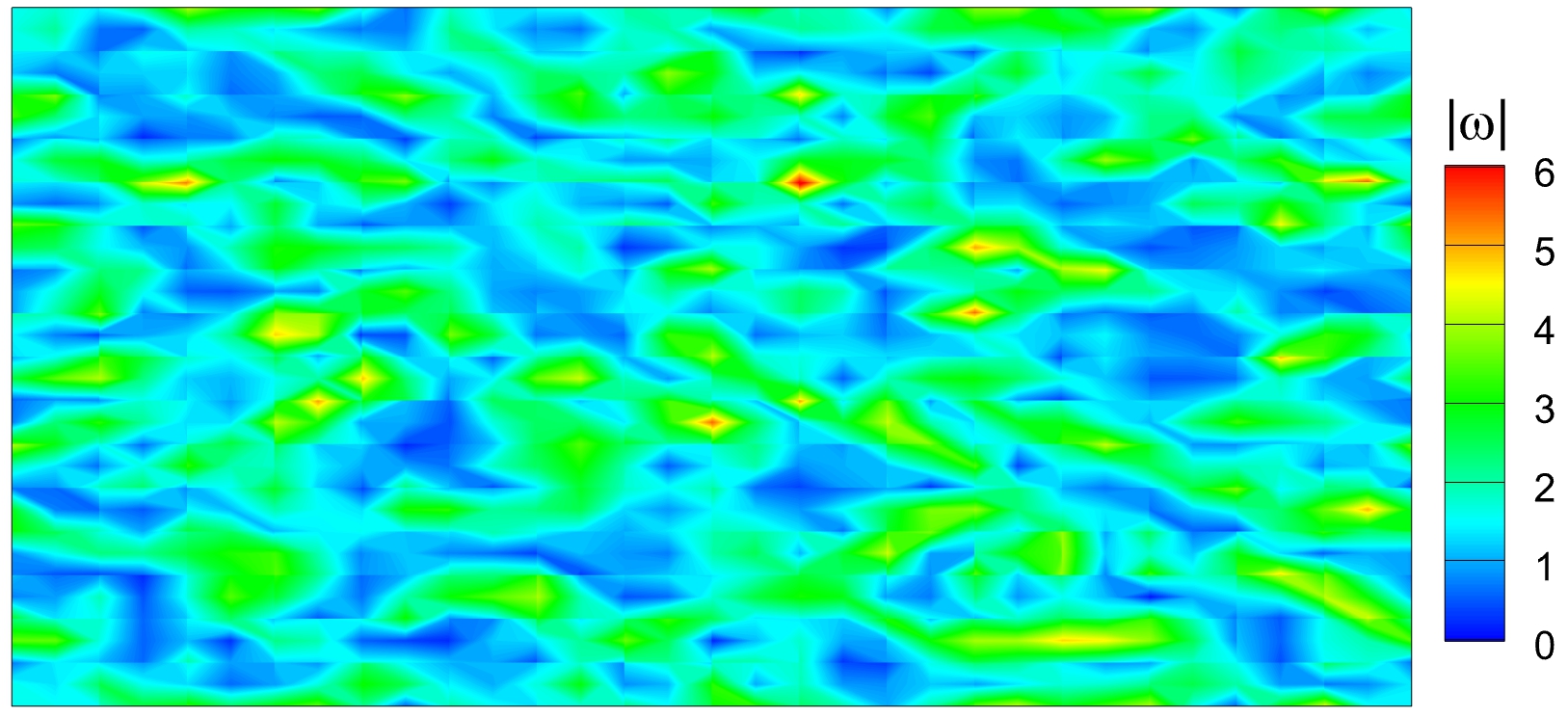}
			\caption{}
		\end{subfigure}	
  \caption{Snapshots of (a) normalized stream-wise slip velocity and (b) vorticity magnitude on the bottom wall obtained using the new dynamic slip-wall model at $Re_\tau \approx 10000$ using grid G2.}
  \label{fig:DSW-u}
	\end{figure}

\begin{figure}
		\centering
		\includegraphics[width=0.75\linewidth, trim = 0mm 0mm 0mm 0mm]{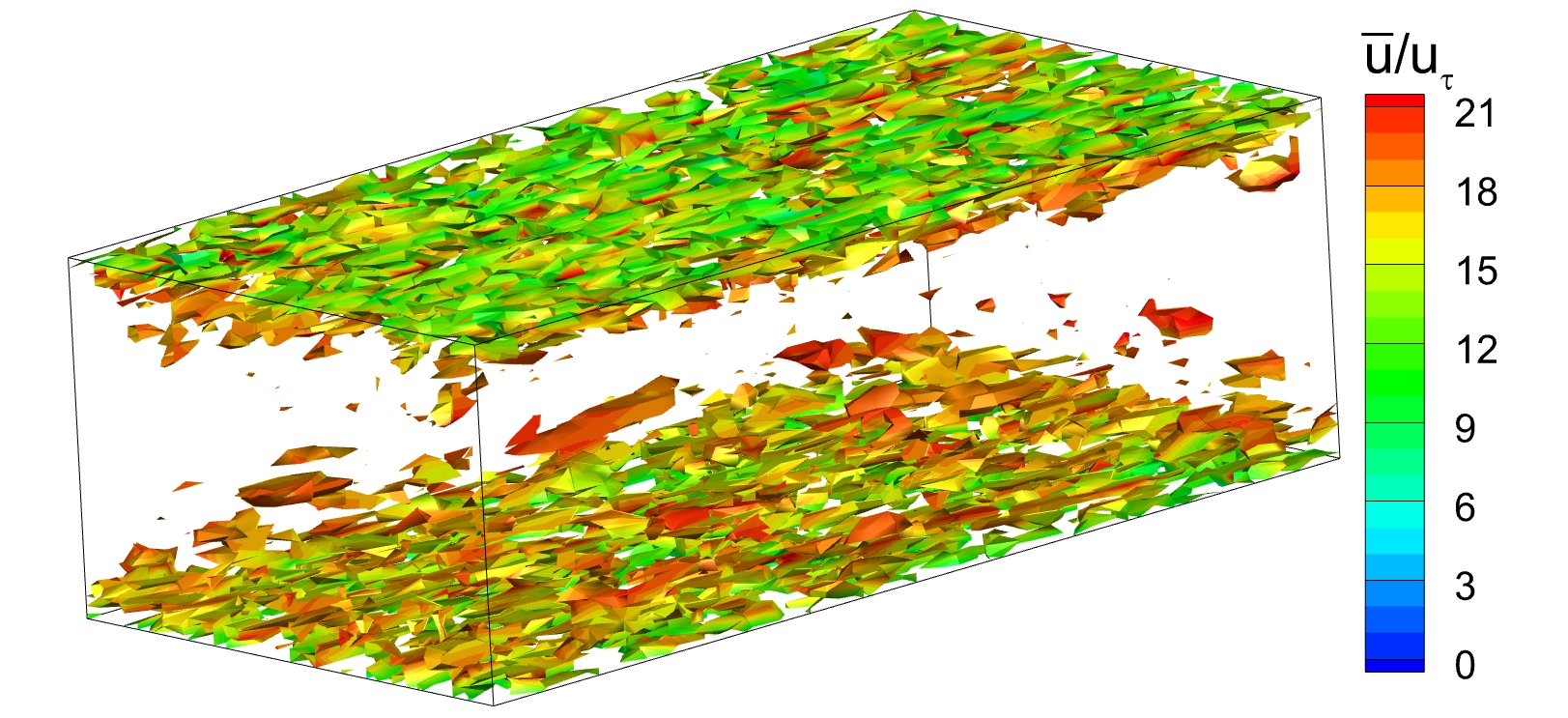}
  \caption{Iso-surafces of Q-criterion colored with normalized stream-wise velocity $\overline{u}/u_\tau$ obtained using the new dynamic slip wall model at $Re_\tau \approx 10000$ case using grid G2.}
  \label{fig:DSW-Qcri}
	\end{figure}

Instantaneous snapshots of the stream-wise slip velocity normalized by the friction velocity $u_\tau$ on the bottom wall at $Re_\tau \approx 10000$ employing G2 grid is shown in Fig.~\ref{fig:DSW-u}(a). The mean slip velocity at the wall increases as the Reynolds number increases and the simulation resolves a smaller fraction of the inner layer of the boundary layer. The mean stream-wise slip velocities at the wall are approximately $10.9 u_\tau$, $13.9 u_\tau$, and $15.35 u_\tau$ for $Re_\tau \approx 2000, 5200,$ and $10000$, respectively on grid G2; the centerline velocity is approximately $28 u_\tau$. This behavior is consistent with the \textit{a priori} filtering tests of \cite{pradhan2023unified} using the optimal finite-element projection framework. The snapshot of vorticity magnitude levels on the bottom wall is also shown. The visualization of the near-wall eddies is shown in Fig.~\ref{fig:DSW-Qcri} using the Q-criterion.

%%%%%%%%%%%%%%%%%%%%%%%%%%%%%%%%%%%%%%%%%%%%%%%%%%%%%%%%%%%
	
	%%%%%%%%%%%%%%%%%%%%%%%%%%%%%%%%%%%%%%%%%%%%%%%%%%%%%%%%%%%%%%%%

%%%%%%%%%%%%%%%%%%%%%%%%%%%%%%%%%%%%%%%%%%%%%%%%%%%%%%%%%%%
	
	%%%%%%%%%%%%%%%%%%%%%%%%%%%%%%%%%%%%%%%%%%%%%%%%%%%%%%%%%%%%%%%%

%%%%%%%%%%%%%%%%%%%%%%%%%%%%%%%%%%%%%%%%%%%%%%%%%%%%%%%%%%%%%%%%%%%%%%%

\subsection{Application to separated flows}\label{sec:periodic_hill}

We next apply the new dynamic slip-wall model to periodic hill flows at different Reynolds numbers. The flow configuration consists of a channel flow with constrictions and forms a generic case of an internal flow separating from a curved surface. Periodic boundary conditions are applied in the stream-wise ($x$) and span-wise ($y$) directions. The flow separates at the hill crest resulting in a large recirculation bubble, and it reattaches further downstream. A Reynolds number based on the bulk velocity at the crest $Re_b = \overline{\rho} U_b h/\mu$ determines the flow conditions for this case with $U_b$ being the bulk velocity and $h$ the hill height. The constant mass flow rate is ensured by adding a source term in the $x$-momentum equation. This forcing term is dynamically adjusted to provide the correct mass flow rate at the hill crest, and therefore the correct bulk Reynolds number. The mass flux is constant to five digits of accuracy for the present computations after an initial transient.

The periodic hill case has been extensively studied over the past 15 years, both experimentally and numerically. \cite{rapp2011flow} performed experiments in a water channel at $Re_b$ ranging from $5600$ to $37000$. Several DNS and LES \citep{breuer2009flow,diosady2014dns,balakumar2014dns,gloerfelt2015investigation,krank2018direct} studies have also been conducted. Many studies have also been performed to test the performance of WMLES \citep{balakumar2014dns,carton2017assessment}. The availability of high-quality data from experiments, DNS, and LES makes this a good benchmark test case to evaluate the performance of the slip-wall model in the presence of separation and reattachment processes. \cite{de2018validation} place two-dimensional periodic hill cases at Level 4 complexity in their list of benchmark test cases, which grow in complexity from Level 1 to Level 5. This test case is well-documented and well-posed with consistent DNS/LES predictions between different codes that match well with the experiments. As pointed out in \cite{gloerfelt2019large}, this benchmark test case has been a choice in several European projects and workshops to investigate the reliability of RANS/LES strategies. 

%%%%%%%%%%%%%%%%%%%%%%%%%%%%%%%%%%%%%%%%%%%%%%%%%%%%%%%%%%%%%%%%%%%%%%%%%%%
The periodic hill flows involve massive separation on the hill's leeward sides, the length of which is about $50 \%$ of that of the periodic segment. The principal challenge of this flow arises from the separation on the curved hill surface and the fact that the reattachment point, and hence the whole flow, are highly sensitive to the separation process. The flow exhibits complex dynamics, including separation, reattachment, unsteady shear layer, large recirculation bubble, and strong acceleration on the windward wall. Resolving these delicate flow details, especially on significantly coarser grids is a challenging task. For example, in a recent work, \cite{zhou2021wall} developed a data-driven wall model using the feedforward neural network and training data from the wall-resolved LES of the periodic hill flows with different Reynolds numbers and hill geometries. The model shows a good performance for the turbulent channel flows at $Re_\tau$ as high as $5200$. However, significant discrepancies in the mean velocity and Reynolds stress predictions are observed between their data-driven wall model and the WRLES for the periodic hill case with validation studies limited to $Re_b \approx 10600$.

%%%%%%%%%%%%%%%%%%%%%%%%%%%%%%%%%%%%%%%%%%%%%%%%%%%%%%%%%%%%%%%%%%%%%%%%%%%%%%

The size of the computational domain is $L_x = 9h$, $L_y =  4.5h$, and $L_z = 3.035h$ in the streamwise ($x$), spanwise ($y$), and wall-normal  ($z$) directions, respectively. Piecewise third-order polynomial functions give the coordinates of the curved hill, and the second hill geometry is described by the same equations with a horizontal translation \cite{rapp2011flow}. We use two grids; a coarse grid with $50 \times 24 \times 9$ elements i.e. $150 \times 72 \times 27 $ ($ = 0.2916$ million) degrees of freedom and a fine grid with $75 \times 36 \times 15$ elements i.e. $225 \times 108 \times 45$ ($ \approx 1.1$ million) degrees of freedom. In comparison to our grids, a DNS of $Re_b \approx 10600$ performed by \cite{krank2018direct} using a $7^{th}$ order DG solver, used $128 \times 64 \times 64$ elements i.e. $896 \times 448 \times 448$ ($ \approx 180$ million) degrees of freedom whereas to perform an implicit LES, a mesh with $448 \times 224 \times 224$ ($ \approx 22.5$ million) degrees of freedom was used.

\begin{figure}
		\centering
		\begin{subfigure}{0.49\textwidth}
			\centering
			\includegraphics[width=1.\linewidth, trim = 0mm 0mm 0mm 0mm]{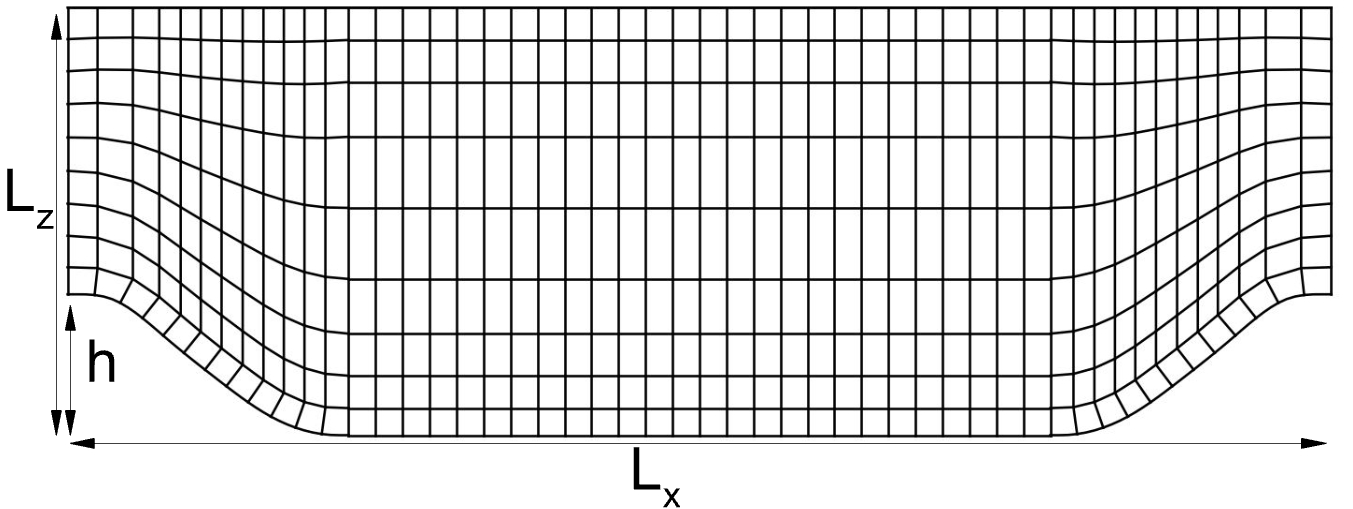}
			\caption{Grid G1 : $50 \times 24 \times 9$}
		\end{subfigure} 
		\begin{subfigure}{0.49\textwidth}
			\centering
			\includegraphics[width=1.\linewidth, trim = 0mm 0mm 0mm 0mm]{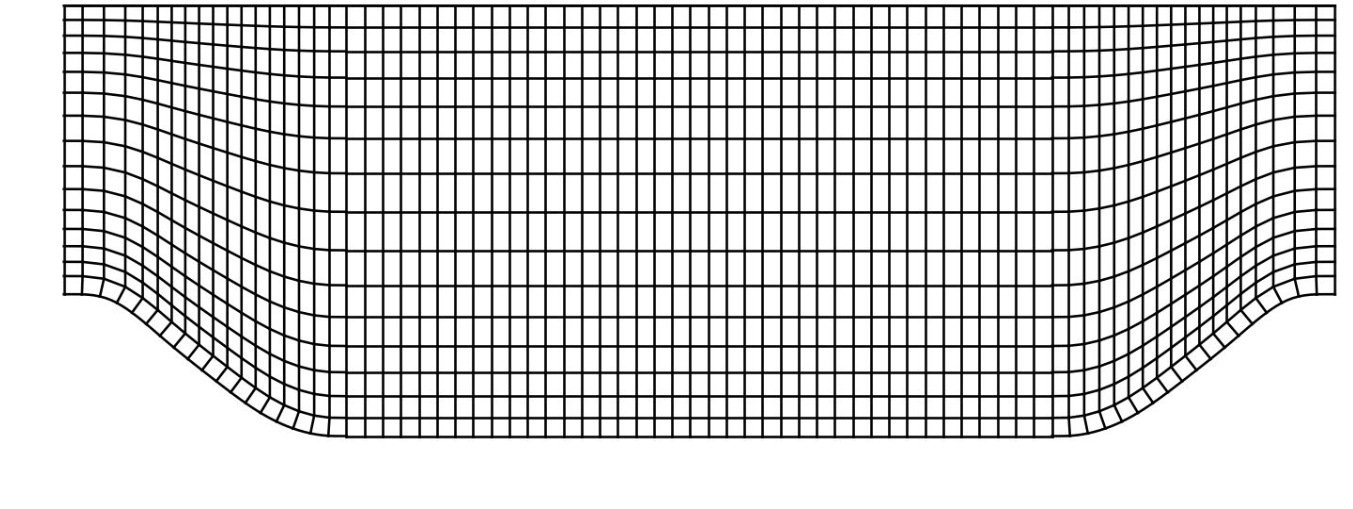}
			\caption{Grid G2 : $75 \times 36 \times 15$}
		\end{subfigure}	
  \caption{Coarse and fine grids used to compute the periodic hill flows.}
  \label{fig:hill_grids}
	\end{figure}

The grids are approximately uniform in the stream-wise and span-wise directions and a mild stretching is used in the wall-normal direction. The mesh is perpendicular to the wall in the first cell away from the wall. The effective element sizes at the hill crest, a key region for the periodic hill flow, are $\Delta_{x} (= \Delta^{e}_{x}/p) \approx 0.105h$ and $\Delta_{z} (= \Delta^{e}_{z}/p) \approx 0.093h$ for the coarse grid and $\Delta_{x} \approx 0.065h$ and $\Delta_{z} \approx 0.064h$ for the fine grid. Figure~\ref{fig:hill_grids} shows the two grids used in the computations. We consider two high Reynolds numbers cases of $ Re_b \approx 10600$ and $37000$ for which high-quality experimental data is available.

%%%%%%%%%%%%%%%%%%%%%%%%%%%%%%%%%%%%%%%%%%%%%%%%%%%%%%%%%%%%%%%%

%%%%%%%%%%%%%  Grid %%%%%%%%%%%%%%%%%%%%%%%%%%%%%%%%%%%
	
	%%%%%%%%%%%%%%%%%%%%%%%%%%%%%%%%%%%%%%%%%%%%%%%%%%%%%%%%%%%%%%%%

%% GRID REFINEMENT : Reb = 10595 :Mean wall-normal velocity

%%%%%%%%%%%%%%%%%%%%%%%%%%%%%%%%%%%%%%%%%%%%%%%%%%%%%%%%%%%
	
	\begin{figure}
		\centering
		\begin{subfigure}{1.0\textwidth}
			\centering
			\includegraphics[width=1.0\linewidth, trim = 0mm 0mm 0mm 0mm]{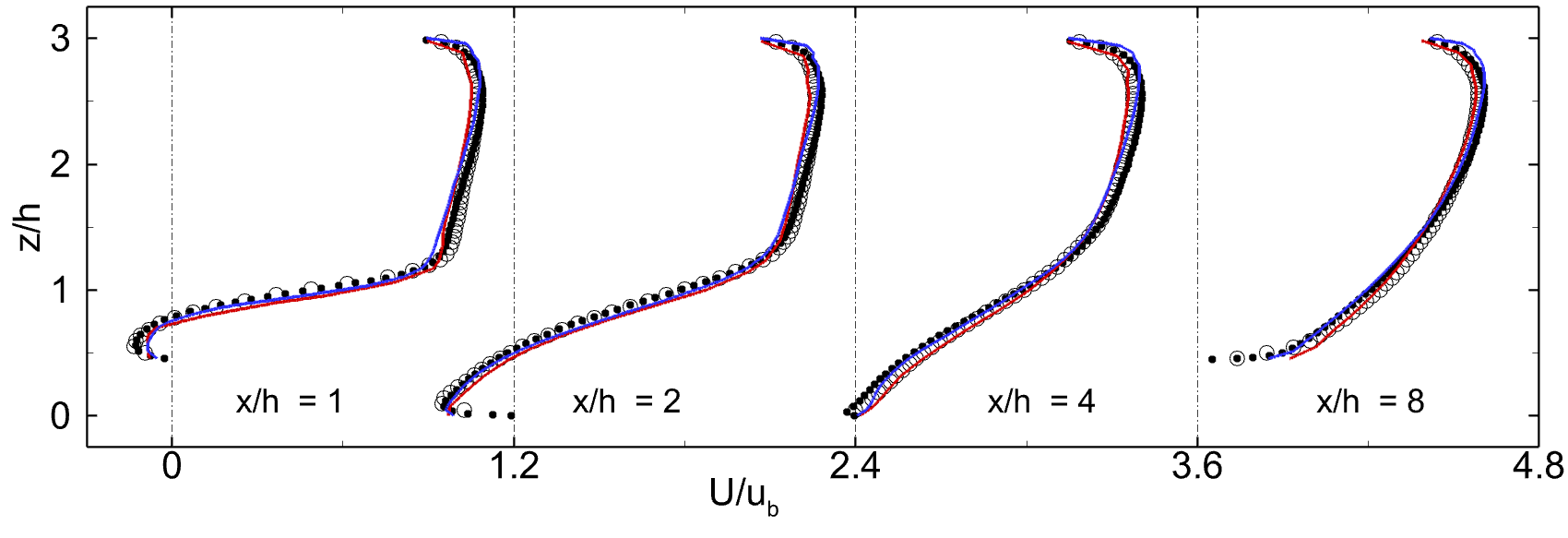}
			\caption{}
		\end{subfigure} \\ 
		\begin{subfigure}{1.0\textwidth}
			\centering
			\includegraphics[width=1.0\linewidth, trim = 0mm 0mm 0mm 0mm]{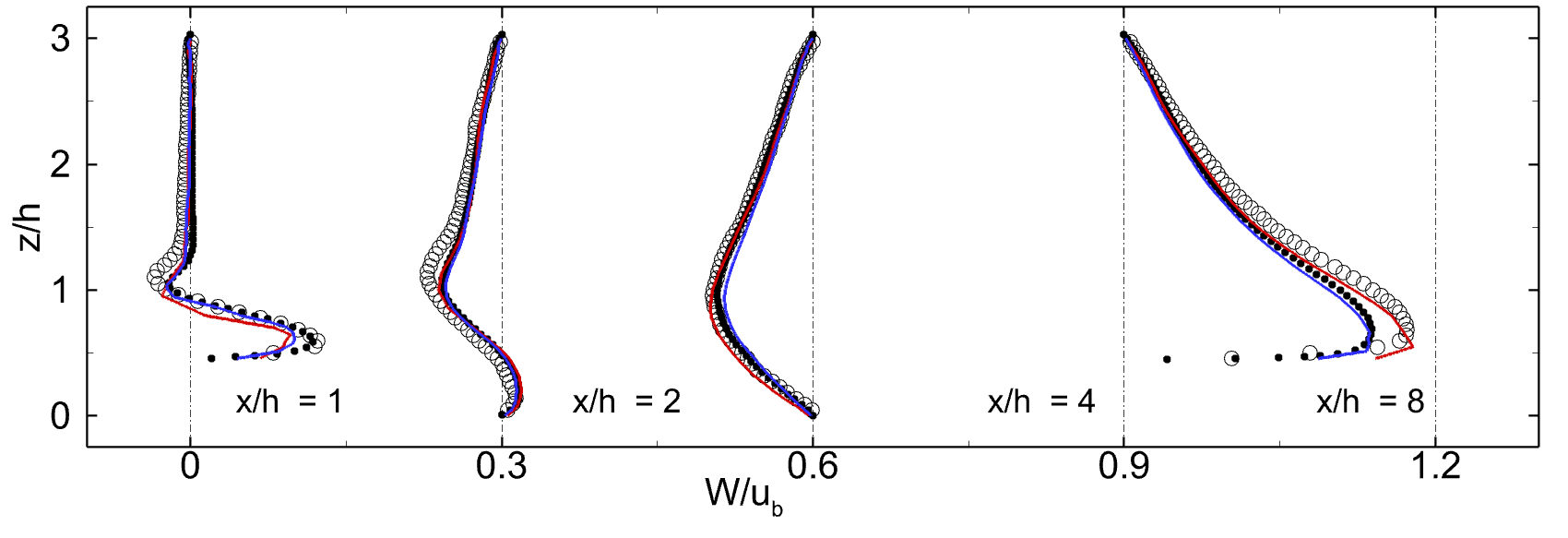}
			\caption{}
		\end{subfigure}	
  \caption{Effect of grid refinement on the mean velocity profiles in the streamwise (U) and vertical (W) directions at different stations for the $Re_b \approx 10600$ case. Red solid lines, Grid G1; blue solid lines, Grid G2; unfilled circles, Experiment \cite{rapp2011flow}; filled circles, WRLES by \cite{breuer2009flow}.}
  \label{fig:Re10k-grid-refine}
	\end{figure} 
	
	%%%%%%%%%%%%%%%%%%%%%%%%%%%%%%%%%%%%%%%%%%%%%%%%%%%%%%%%%%%%%%%%

\begin{figure}
		\centering
		\begin{subfigure}{1.0\textwidth}
			\centering
			\includegraphics[width=1.0\linewidth, trim = 0mm 0mm 0mm 0mm]{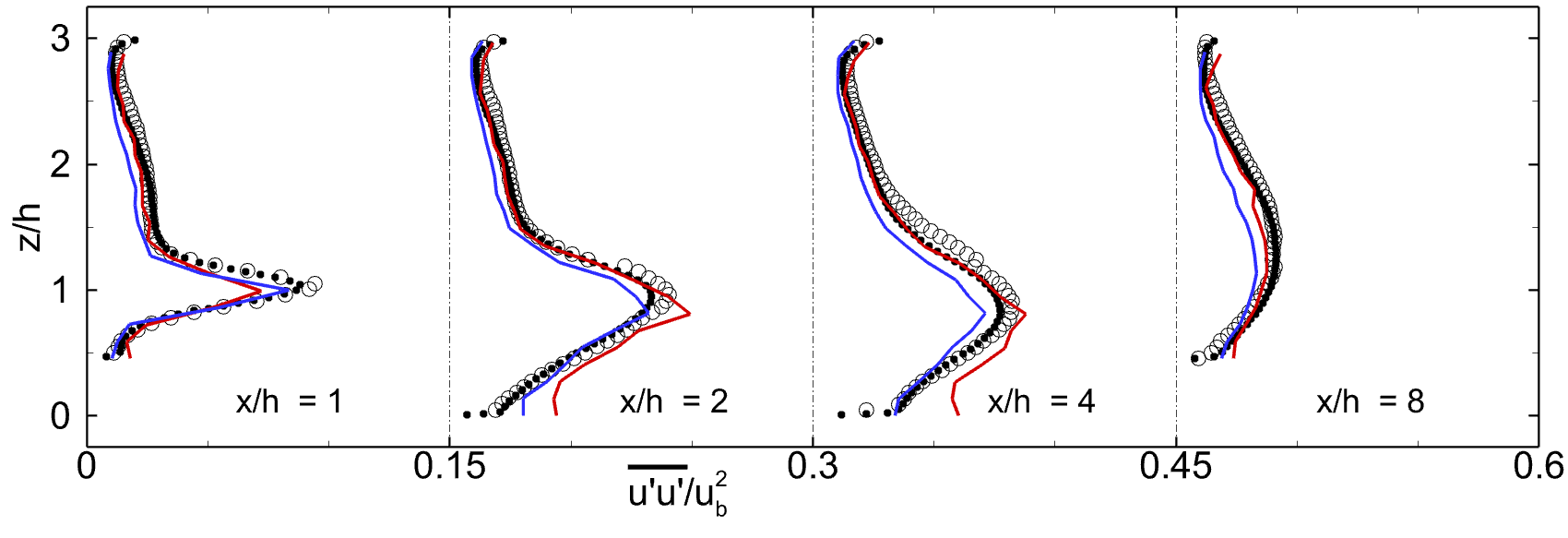}
			\caption{}
		\end{subfigure} \\ 
		\begin{subfigure}{1.0\textwidth}
			\centering
			\includegraphics[width=1.0\linewidth, trim = 0mm 0mm 0mm 0mm]{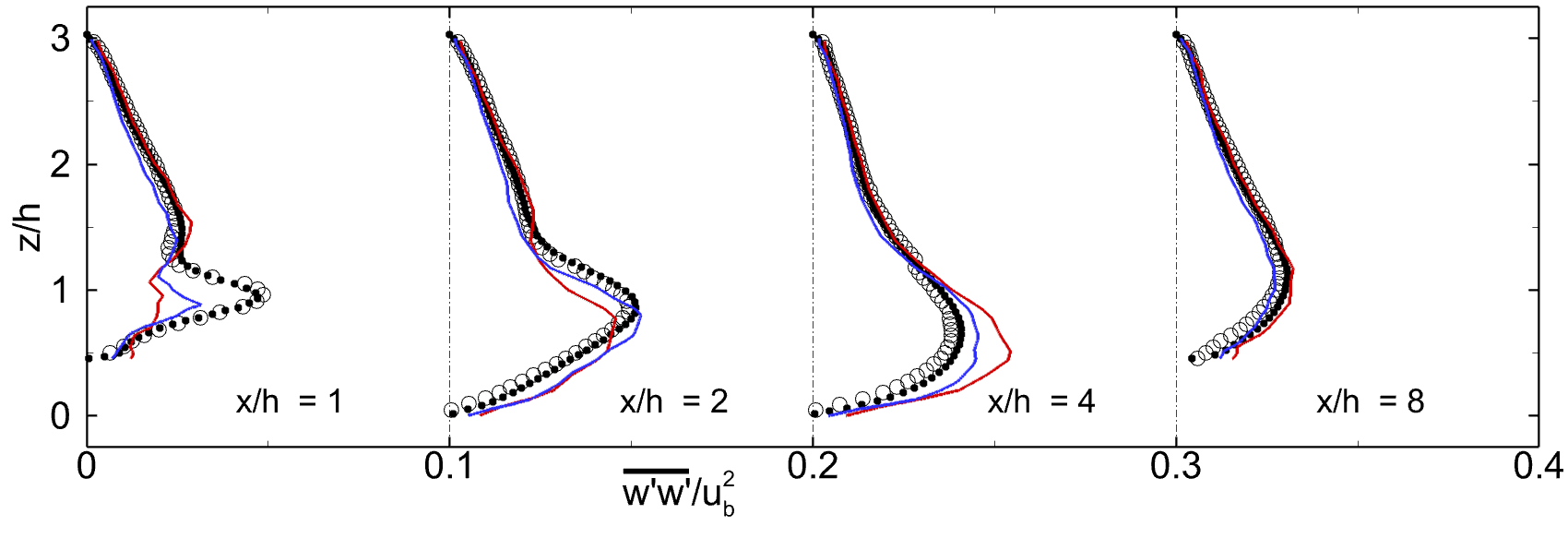}
			\caption{}
		\end{subfigure}	\\
          \begin{subfigure}{1.0\textwidth}
			\centering
			\includegraphics[width=1.0\linewidth, trim = 0mm 0mm 0mm 0mm]{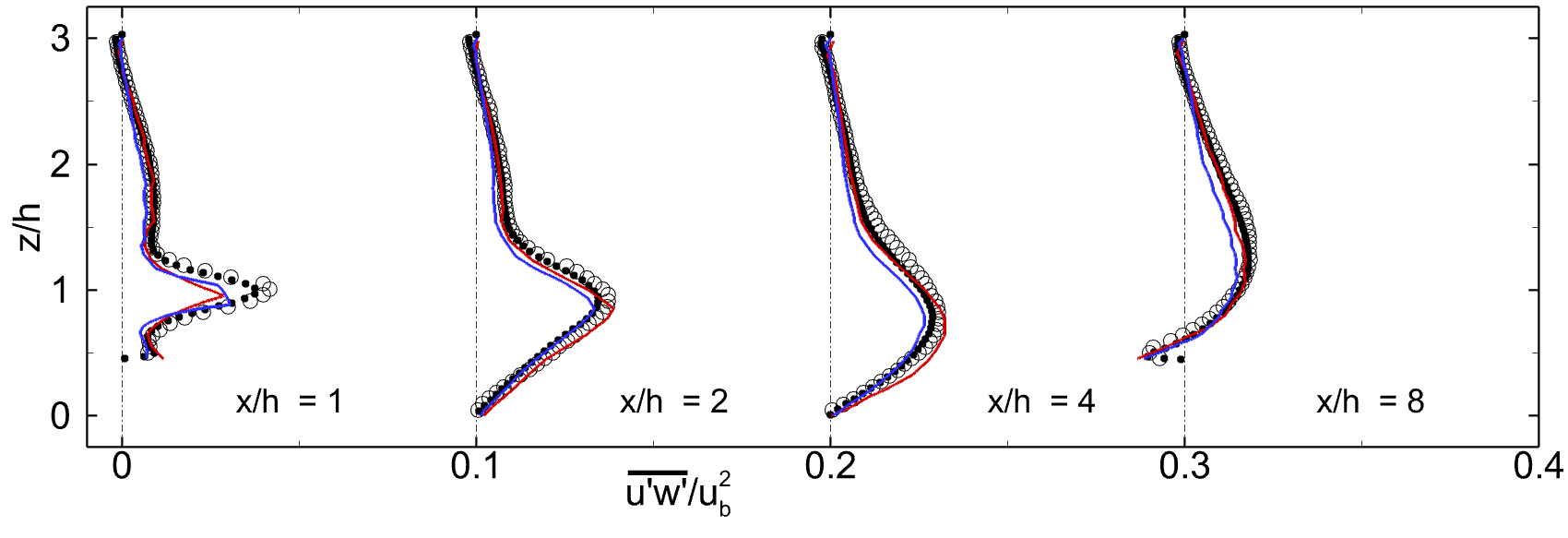}
			\caption{}
		\end{subfigure}
  \caption{Effect of grid refinement on the Profiles of Reynolds stresses in the streamwise (U) and vertical (W) directions at different stations for the $Re_b \approx 10600$ case. Red solid lines, equilibrium wall-stress model; Red solid lines, Grid G1; blue solid lines, Grid G2; unfilled circles, Experiment \cite{rapp2011flow}; filled circles, WRLES by \cite{breuer2009flow}.}
  \label{fig:Re10k-Re_stresses}
	\end{figure}

We first study the effect of mesh resolution on the dynamic slip-wall model predictions at $Re_b \approx 10600$. Wall-normal variation of the mean stream-wise and wall-normal velocity profiles obtained on the coarse and fine grids is shown in Fig.~\ref{fig:Re10k-grid-refine} while Reynolds stream-wise, wall-normal, and shear stress profiles are shown in Fig.~\ref{fig:Re10k-Re_stresses}. The model predictions are compared with the experimental data and the WRLES results of \cite{breuer2009flow} at four stream-wise locations of $x = 1h, 2h, 4h$, and $8h$ which covers the separated as well as post-reattachment regions. The mean stream-wise and wall-normal velocity profile predictions at these locations on the two grids closely match each other and they compare well with the experiments and WRLES. Note that the wall-normal velocity prediction at $x/h = 8$ is particularly sensitive to the grid refinement. This has also been observed in previous LES studies of \cite{gloerfelt2019large}. Notably, the WRLES results for wall-normal velocity, which has a lower amplitude than the mean streamwise velocity, show discrepancies with the experiments at $x/h = 8$. Interestingly, WRLES results match well with the dynamic slip-wall predictions at this location.

The reverse flow velocities are captured well on the two grids. The Reynolds shear stress profiles on the two grids are also similar to each other and they show a good match with the experiment at the four locations. On the other hand, stream-wise and wall-normal Reynolds stress profile predictions on the coarse grid follow the qualitative trend well and the predictions improve on the fine grid and get closer to the experimental data. It is to be noted that, despite the very coarse grid resolution, the agreement between the resolved part of the Reynolds shear and normal stresses with the measurement is reasonably good. 

\begin{figure}
		\centering
		\begin{subfigure}{1.0\textwidth}
			\centering
			\includegraphics[width=1.0\linewidth, trim = 0mm 0mm 0mm 0mm]{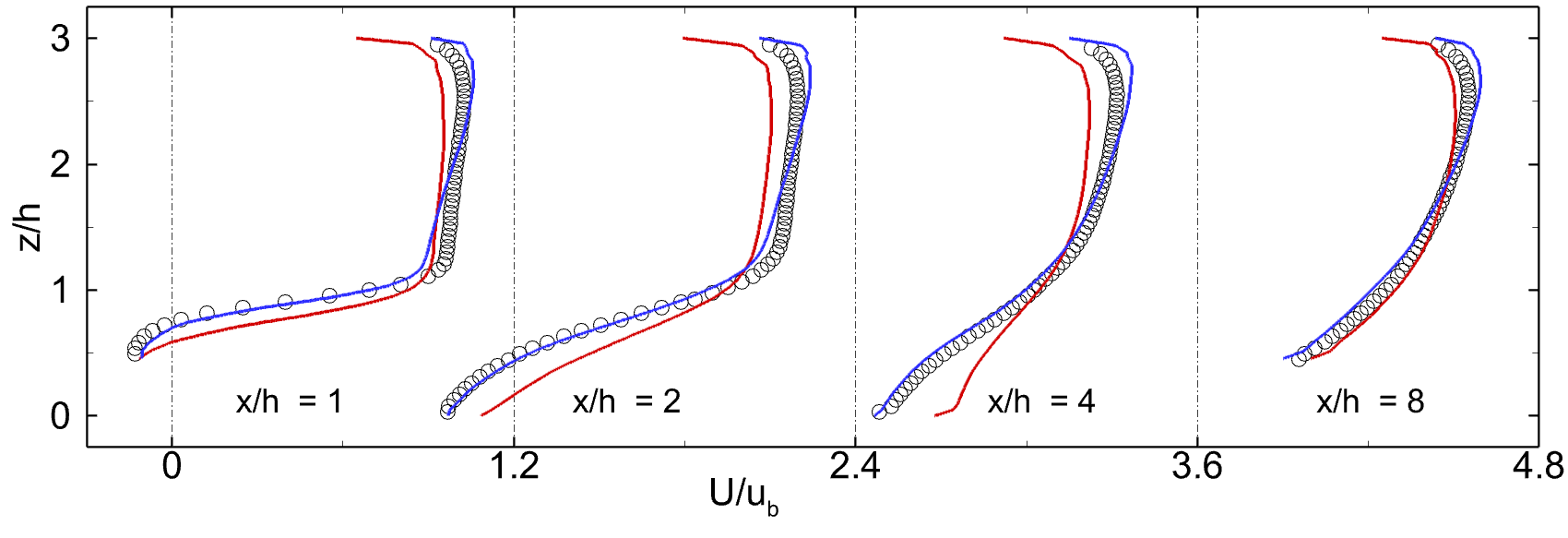}
			\caption{}
		\end{subfigure} \\ 
		\begin{subfigure}{1.0\textwidth}
			\centering
			\includegraphics[width=1.0\linewidth, trim = 0mm 0mm 0mm 0mm]{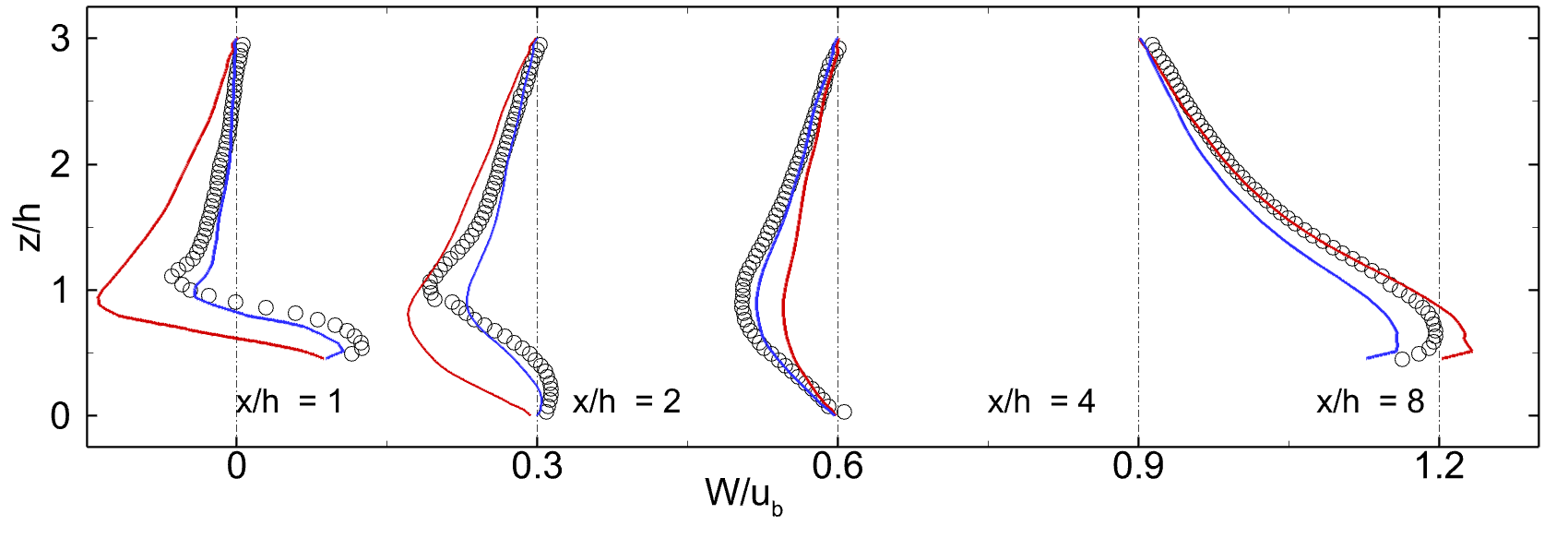}
			\caption{}
		\end{subfigure}	
  \caption{Mean velocity profiles in the streamwise (U) and vertical (W) directions at different stations for the $Re_b \approx 37000$ case. Red solid lines, equilibrium wall-stress model; blue solid lines, dynamic slip-wall model; unfilled circles, Experiment \cite{rapp2011flow}.}
  \label{fig:Re37k_mean_vel_model_comps}
	\end{figure}

\begin{figure}
		\centering
		\begin{subfigure}{1.0\textwidth}
			\centering
			\includegraphics[width=1.0\linewidth, trim = 0mm 0mm 0mm 0mm]{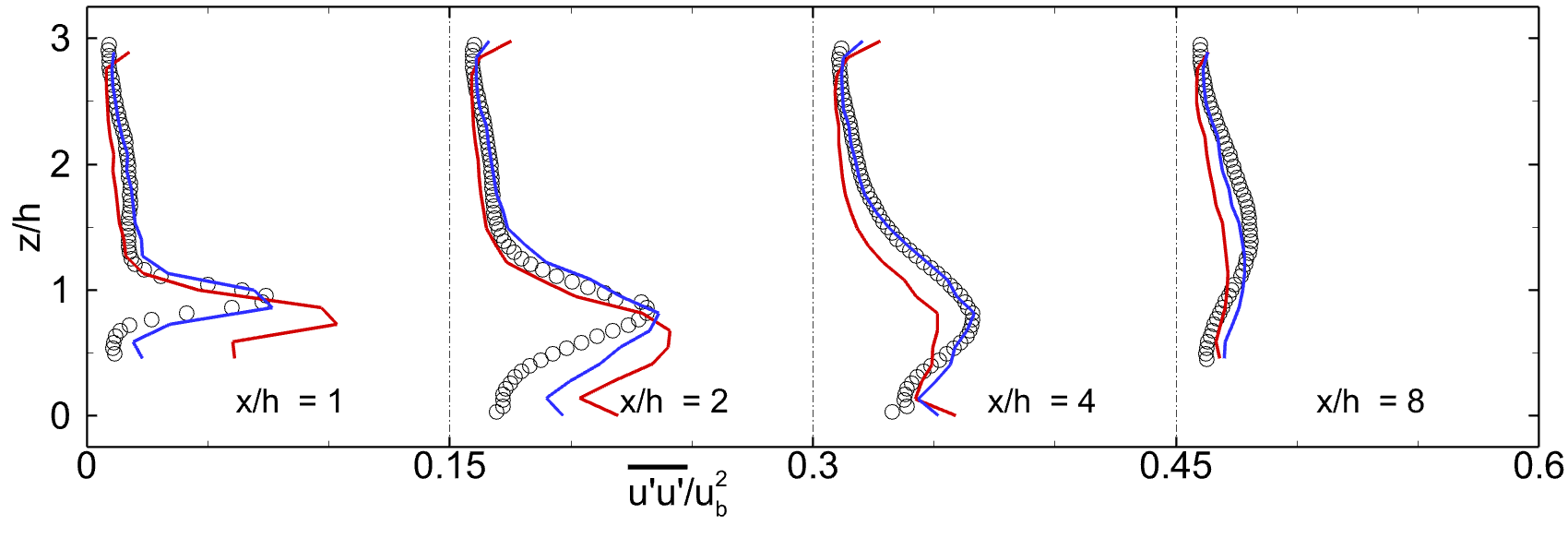}
			\caption{}
		\end{subfigure} \\ 
		\begin{subfigure}{1.0\textwidth}
			\centering
			\includegraphics[width=1.0\linewidth, trim = 0mm 0mm 0mm 0mm]{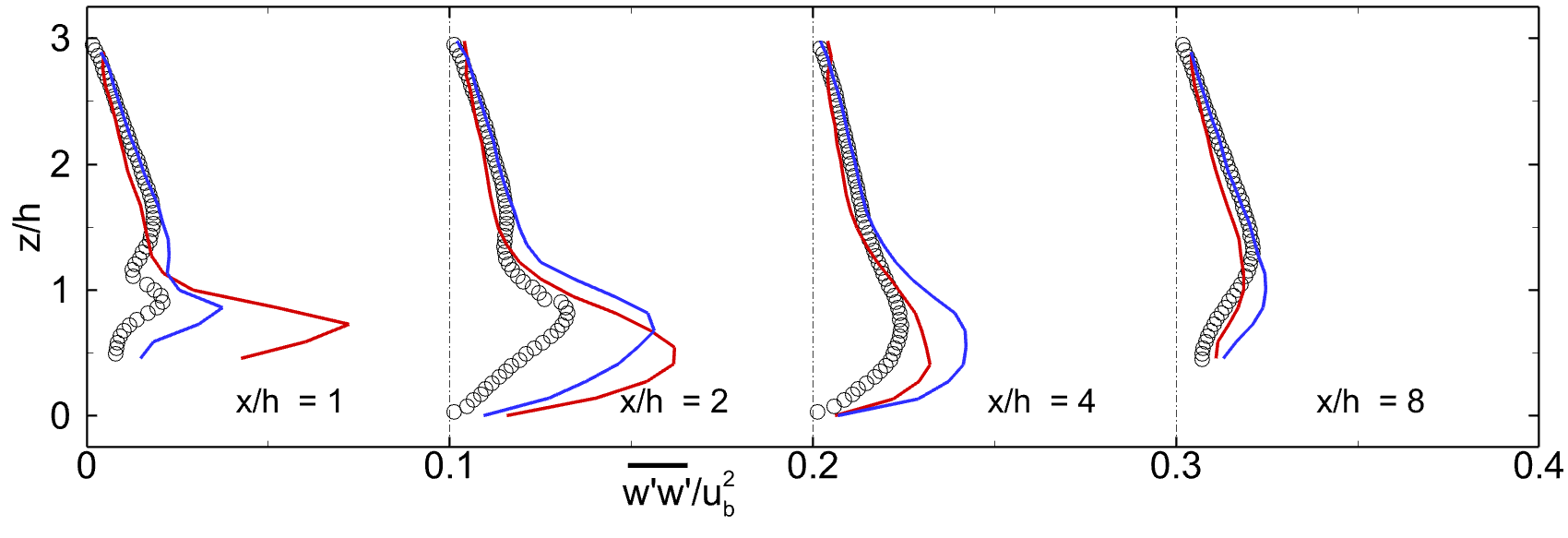}
			\caption{}
		\end{subfigure}	\\
          \begin{subfigure}{1.0\textwidth}
			\centering
			\includegraphics[width=1.0\linewidth, trim = 0mm 0mm 0mm 0mm]{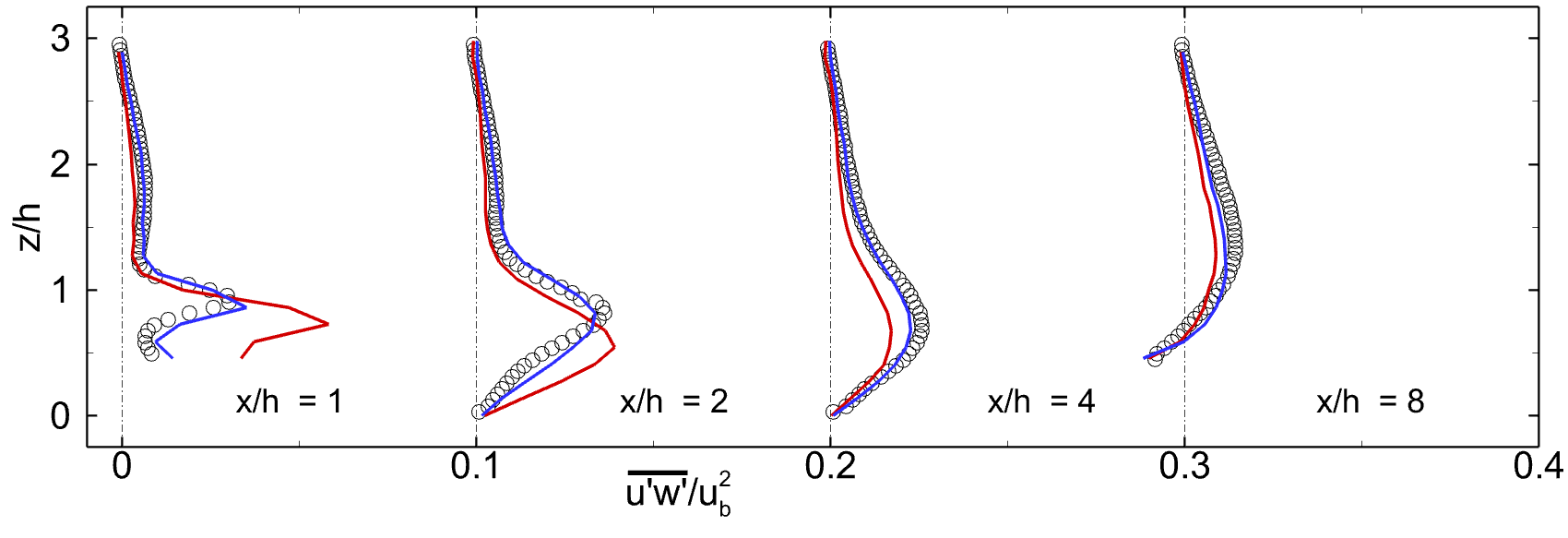}
			\caption{}
		\end{subfigure}
  \caption{Profiles of Reynolds stresses in the streamwise (U) and vertical (W) directions at different stations for the $Re_b \approx 37000$ case. Red solid lines, equilibrium wall-stress model; blue solid lines, dynamic slip-wall model; unfilled circles, Experiment \cite{rapp2011flow}.}
  \label{fig:Re37k-Re_stresses}
	\end{figure}

The periodic hill flow is then computed at a higher Reynolds number of $Re_b = 37000$ using the fine grid and comparisons between the new dynamic slip-wall model and the EQWM predictions for the mean velocities and Reynolds stresses are shown in Figs.~\ref{fig:Re37k_mean_vel_model_comps} and \ref{fig:Re37k-Re_stresses}, respectively. The EQWM significantly underpredicts the separation and shows a faster recovery and the mean stream-wise and wall-normal velocity profiles in the separated regions show a significant mismatch with the experiments. On the other hand, the dynamic slip-wall model accurately captures the separation and shows an excellent match with the experiments for the mean velocities at the four locations in Fig.~\ref{fig:Re37k_mean_vel_model_comps}. A discrepancy is observed for the wall-normal velocity at $x/h = 8$ similar to the $Re_b \approx 10600$ case, which is consistent with the study of \cite{gloerfelt2019large}. The Reynolds shear stress profiles predicted by the dynamic slip-wall model also closely agree with the experiments whereas the EQWM model predictions show a considerable mismatch. This is also the case for the stream-wise Reynolds stress profiles. However, the dynamic slip-wall model predictions for the wall-normal Reynolds stress show some discrepancies and overpredict the levels found in the experiments. This is again consistent with the LES studies of \cite{gloerfelt2019large} which showed dramatic overprediction for the wall-normal Reynolds stresses while getting a good match for the other Reynolds stress components. Overall, the dynamic slip-wall model predictions are considerably better than the EQWM.

%%%%%%%%%%%%%%%%%%%%%%%%%%%%%%%%%%%%%%%%%%%%%%%%%%%%%%%%%%%
	
	%%%%%%%%%%%%%%%%%%%%%%%%%%%%%%%%%%%%%%%%%%%%%%%%%%%%%%%%%%%%%%%%

%%% Reb = 37k results
%%%%%%%%%%%%%%%%%%%%%%%%%%%%%%%%%%%%%%%%%%%%%%%%%%%%%%%%%%%
	
	%%%%%%%%%%%%%%%%%%%%%%%%%%%%%%%%%%%%%%%%%%%%%%%%%%%%%%%%%%%%%%%%

%%%%%%%%%%%%%%%%%%%%%%%%%%%%%%%%%%%%%%%%%%%%%%%%%%%%%%%%%%%
	
	%%%%%%%%%%%%%%%%%%%%%%%%%%%%%%%%%%%%%%%%%%%%%%%%%%%%%%%%%%%%%%%%

%%%%%%%%%%%%%  Cw,lambda %%%%%%%%%%%%%%%%%%%%%%%%%%%%%%%%%%%

\begin{figure}
\centering
\includegraphics[width=0.45\linewidth, trim = 0mm 0mm 0mm 0mm]{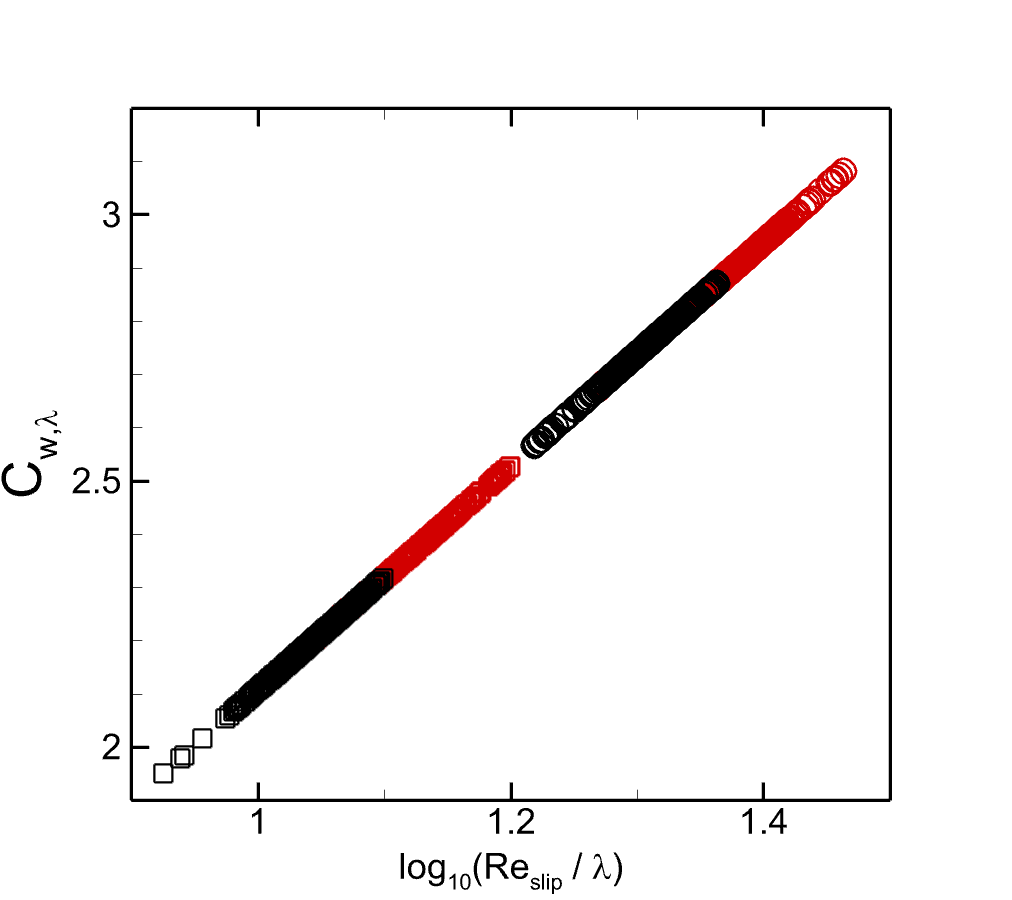}
\caption{$\lambda$-normalized slip parameter $C_w$ i.e. $C_{w,\lambda}$ as a function of $Re_{slip}$ and $\lambda$ at $Re_b \approx 10600$ (squares) and $Re_b \approx 37000$ (circles). Color code : black -- grid G1; red -- grid G2. }	
\label{fig:cwl-hill}
\end{figure} 

%%%%%%%%%%%%%%%%%%%%%%%%%%%%%%%%%%%%%%%%%%%%%%%%%%%%%%%%%%%%%%%%%%%%%%%%%%

\begin{figure}
\centering
\includegraphics[width=0.45\linewidth, trim = 0mm 0mm 0mm 0mm]{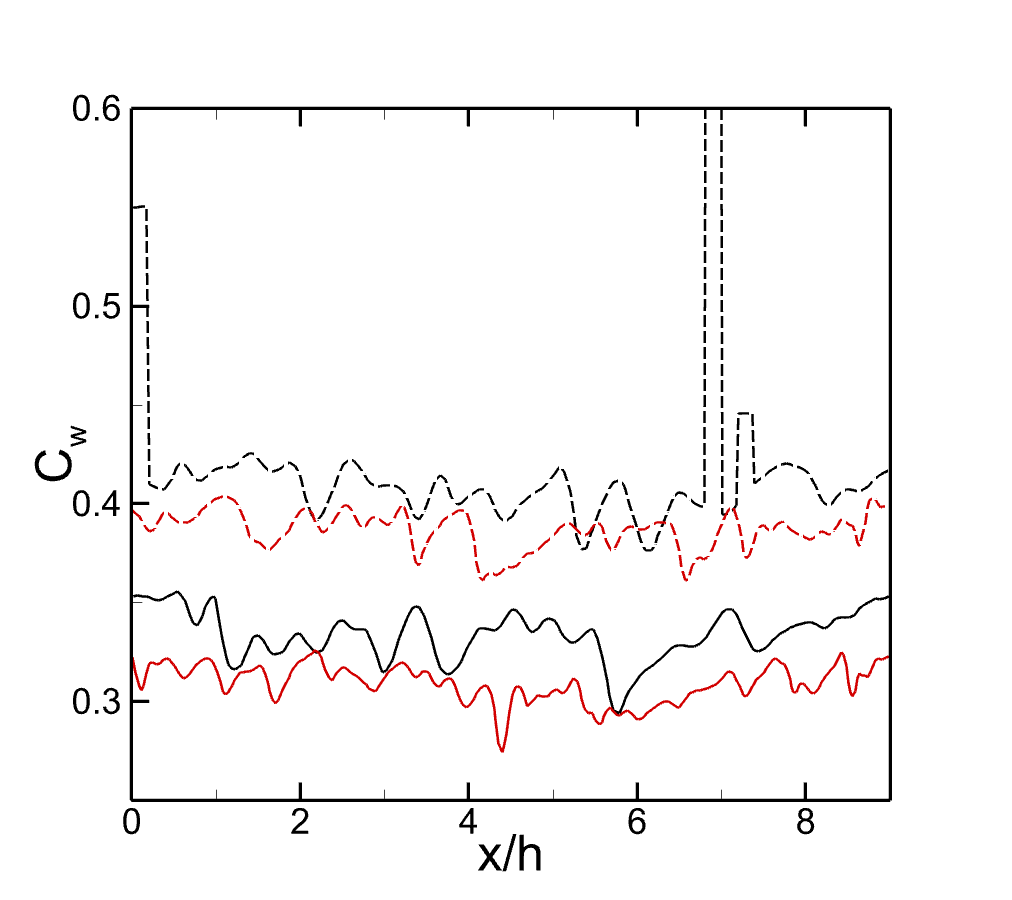}
\caption{Streamwise distribution of the slip parameter $C_{w}$ at $Re_b \approx 10600$ (solid lines) and $Re_b \approx 37000$ (dashed lines) obtained on grid G1 (black lines) and grid G2 (red lines). }	
\label{fig:cw-hill}
\end{figure}

Figure~\ref{fig:cwl-hill} plots the variation of $\lambda$-normalized slip parameter $C_w$ i.e. $C_{w,\lambda}$ as a function of the slip velocity-based Reynolds number $Re_{slip}$ and $\lambda$ for the two Reynolds numbers of $Re_b \approx 10600$ and $37000$ obtained on grids G1 and G2. The curves for $C_{w,\lambda}$ are found to collapse as a function of $Re_{slip}$ for the two grid resolutions, and the universal scaling relationship discovered in \cite{pradhan2023unified} for turbulent channel flows using an \emph{a priori} analysis is found to remain valid for these separated flows as well in the \emph{posteriori} computations. The parameter $C_{w,\lambda}$ follows the expected trend; assumes lower values at the lower Reynolds number and coarser near-wall grid resolutions and decays as the near-wall grid is coarsened or Reynolds number is decreased. The streamwise variation of the slip parameter $C_w$ is shown in Fig.~\ref{fig:cw-hill} and it also follows the expected trend at the two Reynolds numbers and grid resolutions. Variations in $C_w$ along the streamwise direction are also observed for these cases. The sudden peaks at the starting of the hill and at $x/h \approx 7$ for the high Reynolds number case of $Re_b \approx 37000$ are indicative of the unphysical effects due to the coarse grid resolution and these peaks vanish on the fine grid G2. 

%%%%%%%%%%%%%  slip-velocity %%%%%%%%%%%%%%%%%%%%%%%%%%%%%%%%%%%

%%%%%%%%%%%%%%%%%%%%%%%%%%%%%%%%%%%%%%%%%%%%%%%%%%%%%%%%%%%%%%%%%%%%%%%%%%

%%%%%%%%%%%%%  slip-velocity %%%%%%%%%%%%%%%%%%%%%%%%%%%%%%%%%%%

\begin{figure}
\centering
\includegraphics[width=0.45\linewidth, trim = 0mm 0mm 0mm 0mm]{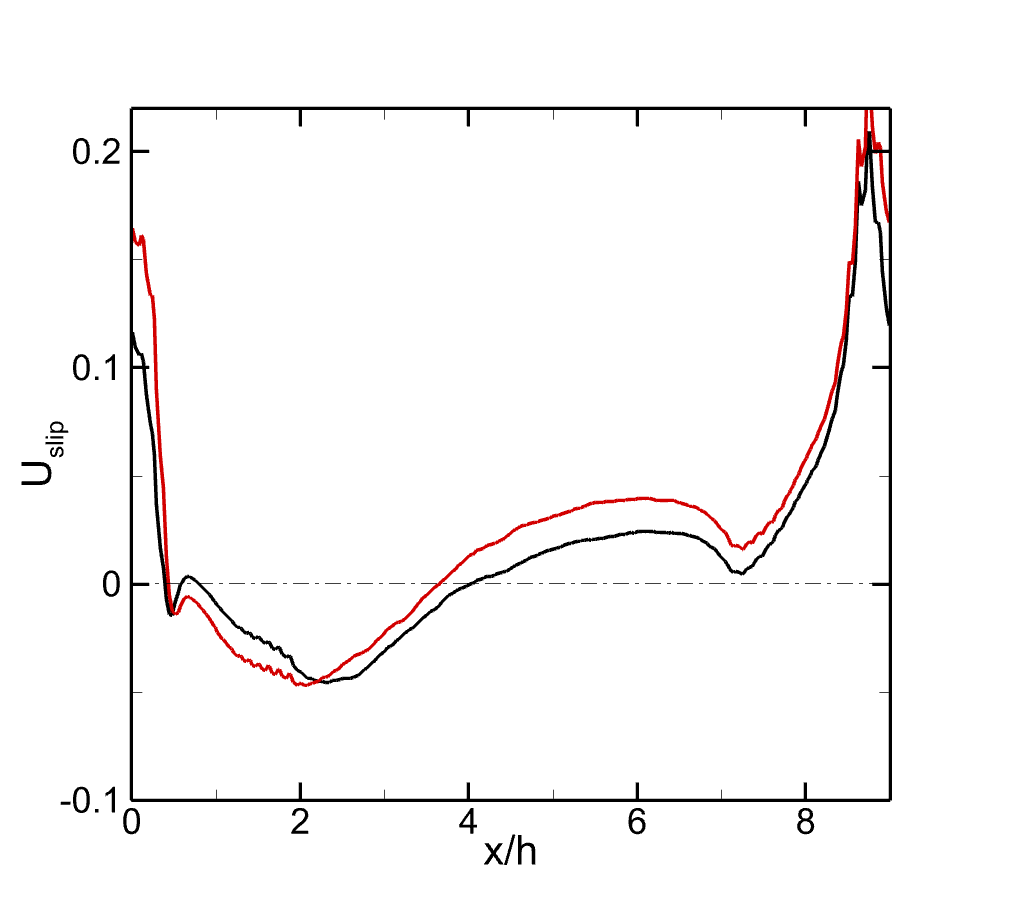}
\caption{Streamwise distribution of slip-velocity $U_{slip}$ at $Re_b \approx 10600$ (black line) and $Re_b \approx 37000$ (red line) obtained on grid G2.} 
\label{fig:uslip-hill}
\end{figure} 

%%%%%%%%%%%%%%%%%%%%%%%%%%%%%%%%%%%%%%%%%%%%%%%%%%%%%%%%%%%%%%%%%%%%%%%%%%

The streamwise distributions of the mean streamwise slip wall velocity at $Re_b \approx 10600$ and $37000$ obtained on the fine grid G2 are shown in Fig.~\ref{fig:uslip-hill}. The slip velocity $U_{slip}$ for the two cases changes the sign near the separation point, remains negative in the separation bubble, and reverses the sign near the reattachment location. For the $Re_b \approx 10600$ case, the locations where the slip velocity changes the sign closely match the experimental measurements and WRLES results for the separation and reattachment points at $x/h \approx 0.2$ and $x/h \approx 4.6$, respectively. As indicated by the negative values of the slip velocity, the separation region slightly reduces in size at the higher Reynolds case which is physically consistent.

%%%%%%%%%%%%%%%%%%%%%%%%%%%%%%%%%%%%%%%%%%%%%%%%%%%%%%%%%%%%%%%%%%%%%%%

%%%%%%%%%%%%%%%%%%%%%%%%%%%%%%%%%%%%%%%%%%%%%%%%%%%%%%%%%%%%%%%%%%%%%%%%%%%%%%

\section{Conclusion}\label{sec:sec7}

 Several strategies have been proposed to bypass the stringent near-wall grid resolution requirement for performing  LES of high Reynolds number flows in the presence of solid walls. In this work, we focus on the slip-wall modeling approach - originally proposed by Bose et al.~\cite{bose2014dynamic} -  and replace the conventional no-slip velocity boundary condition with slip velocities at the wall. The major objective is to accurately capture the mean flow characteristics at Reynolds numbers of practical relevance using a significantly coarse near-wall LES mesh, and do so in a robust manner.

We present a new formulation of a dynamic slip-wall model that is consistent with the Discontinuous-Galerkin (DG) framework and is tightly integrated with DG operators. The model coefficients of the modified slip-wall model of \cite{pradhan2023unified} are based on \textit{a priori} estimates obtained using an optimal finite-element projection framework. Here, we propose a dynamic modeling procedure to compute the scaling parameter $\lambda$ for the slip-wall model coefficient $C_w$. The dynamic part of the model is based on a modified form of Germano identity and coupled with the dynamic Smagorinsky model. The level of under-resolution is represented by a slip Reynolds number and the proposed model attempts to also incorporate the effects of the numerical discretization and the SGS model.

The canonical case of statistically stationary turbulent channel flow is first used to validate the new dynamic slip-wall model. The model predictions are compared with the available DNS data at three Reynolds numbers of $Re_\tau \approx 2000, 5200,$ and $10000$. Grid independence studies are performed at these Reynolds numbers by considering significantly underresolved LES meshes with stream-wise, span-wise, and wall-normal grid resolutions corresponding to $\Delta_{x} \simeq 0.1-0.4\delta$, $\Delta_{y} \simeq 0.05-0.2\delta $, and $\Delta_{z} \simeq 0.03-0.125\delta$, respectively. These mesh resolutions are significantly coarser than the WMLES mesh recommendations of \cite{larsson2016large} corresponding to $\Delta_{x} \simeq 0.08 \delta$, $\Delta_{y} \simeq 0.05 \delta $, and $\Delta_{z} \simeq 0.01-0.05 \delta$. Mean velocity profiles show an excellent match with the DNS at the considered Reynolds on all the grids with $L_2$ error less than $3 \%$ for all the cases. Reynolds shear and normal stress profiles resolved on the significantly coarse grids also show excellent agreement with the DNS. The model performance is shown to be similar to that of the EQWM, which is known to predict the equilibrium wall-bounded flows without separation accurately. This is a considerable improvement over the dynamic slip-wall model of \cite{bae2019dynamic} which shows a significant log-layer mismatch at similar Reynolds numbers but on comparatively finer grid resolutions.

The model performance is evaluated in flow separation and reattachment over periodic hills at Reynolds numbers of $Re_b \approx 10600$ and $37000$ using two different grid resolutions. The meshes used for the computations are significantly coarser than the conventional LES meshes, e.g., the fine mesh used here has about $20$ times fewer degrees of freedom than the implicit LES performed by \cite{krank2018direct}. The stream-wise and wall-normal mean velocity profile predictions obtained using the dynamic slip-wall model on the two grids compare well with the experimental data in the separated and post-reattachment flow regions at $Re_b \approx 10600$. Reynolds shear stress predictions obtained using the two grids also match very well with experiments at different stream-wise locations. However, the Reynolds normal stresses are better predicted on the fine grid. Computations at $Re_b \approx 37000$ using the fine grid show that the dynamic slip-wall model predictions for the mean velocity profiles agree well with the experiments. The Reynolds shear stress profiles are also in excellent agreement with the experiments, with some discrepancies in the Reynolds normal stress predictions. On the other hand, EQWM for this case shows significant discrepancies with the experimental data for the mean velocities as well as Reynolds shear and normal stresses. 

%%%%%%%%%%%%%%

The new model contains parameters like $C_{w,\lambda}$, $ C_{wR}$, and $\Delta_R$ which are empirically established, but the model does not assume the state of a boundary layer. It is important to note that the main purpose of a slip-wall model is similar to that of a traditional wall-stress model i.e. accurate estimation of wall shear stress. Achieving this goal without prior assumptions regarding the state of the boundary layer or embedded empirical parameters is an outstanding challenge. Moreover, the instantaneous velocity field is intertwined with the effects of the LES grid resolution and Reynolds number for a given numerical discretization and SGS model as discussed in \cite{bae2019dynamic}. The modeling choices made in this work are consistent with the observations of \cite{pradhan2023unified} and the works of \cite{bose2014dynamic,bae2019dynamic}. The empirical parameters in the proposed model, especially $C_{w,\lambda}$, provide an explicit reference to how the near-wall flow should behave at different near-wall grid resolutions and Reynolds numbers. The model is found to be somewhat insensitive to the parameters $ C_{wR}$ and $\Delta_R$ in their plausible range for the considered cases, which includes the smooth body-separated flows with separation and reattachment.

The new model can consistently predict mean velocity and Reynolds shear and normal stress profiles for the equilibrium as well as separated flows at high Reynolds numbers using significantly coarse near-wall LES meshes. The model performs at a computational cost similar to the EQWM which is the cheapest state-of-the-art WMLES strategy. In the authors' opinion, the excellent performance of the model may be attributed to the integration of the optimal finite-element projection framework used to obtain the slip-wall parameters with the consistent dynamic procedures for the SGS and slip-wall modeling coupled with the DG framework. 
This work is a step towards making the slip-wall model a viable computing tool for predicting complex engineering flows, and further evaluations are required.

While the present work demonstrates the implementation and validation of the dynamic slip wall model within a DG framework, we emphasize that the fundamental approach is not inherently tied to DG methods. The essential requirement is a rigorous coarse-graining or scale separation operator, which can be provided by various numerical frameworks including finite element methods, variational multi-scale approaches, or projection-based methods. For finite volume and finite difference techniques, while the implementation is less straightforward, agglomeration-based techniques (e.g. ~\cite{gravemeier2006scale}) can provide viable pathways. Changing the numerical framework (for instance, to continuous finite elements) would necessitate recalibration of the base parameters, along with a redefinition of the length scales. This requirement stems from our fundamental observation that subgrid and wall models cannot be decoupled from the underlying numerical method due to the strong interactions between unresolved and coarsely resolved dynamics.

%%%%%%%%%%%%%%%%%%%%%%%%%%%%%%%%%%%%%%%%%%%%%%%%%%%%%%%%%%%%%%%%%%

\section*{Acknowledgement}

This research was funded by NASA under the project "Scale-resolving turbulence simulations through adaptive high-order discretizations and data-enabled model refinements", grant number 80NSSC18M0149 (Technical monitor: Dr. Gary Coleman). We acknowledge Prof. Krzysztof Fidkowski, Dr. Gary Coleman, and Dr. Aniruddhe Pradhan for their valuable discussions.

\section*{Declaration of interests.}

The authors report no conflict of interest.

%%%%%%%%%%%%%%%%%%%%%%%%%%%%%%%%%%%%%%%%%%%%%%%%%%%%%%%%%%%%%%%%%%%%%%%%%%%%%%%%%%%%%%%

\appendix

%%%%%%%%%%%%%%%%%%%%%%%%%%%%%%%%%%%%%%%%%%%%%%%%%%%%%%%%%%%%%%%%%%%%%%%%%%%%%%%%%%%%%%%%
\section{Sharp modal cut-off filter implementation}\label{appendixA}

The implementation of the sharp modal cut-off filter is a three-step process; transforming the nodal solution coefficients to a hierarchical modal representation, applying a filter on the modal coefficients, and then transforming back into the nodal representation. Let us denote the nodal solution coefficients as $u_j$, the nodal basis functions as $\phi_j$, the modal solution coefficients as $b_j$, and the modal basis functions as $\psi_j$. Then, we can write the approximation $u_h$ to any flow variable $u$ in an element as
    \begin{equation}
      u \approx  u_h = \sum_{j=1}^{p+1} u_j \phi_j = \sum_{j=1}^{p+1} b_j \psi_j \hspace{0.02in} .
    \end{equation}
Multiplying the above equation by $ \psi_i $ and integrating over the standard element, we get
    \begin{equation}
       \sum_{j=1}^{p+1} C_{ij} u_j =  \sum_{j=1}^{p+1} M_{ij} b_j
    \end{equation}
 or in Matrix and vector form, we can write
    \begin{equation}
        [C] \Vec{u} = [M] \Vec{b} \hspace{0.02in} .
        \label{eqn:modal-nodal}
    \end{equation}
Here, the Modal Mass Matrix [M] is given by
    \begin{equation}
        [M] = M_{ij} = \int_{\Omega_k} \psi_i \psi_j \,dx \hspace{0.02in} .
    \end{equation}
The Mixed Mass Matrix [C] is given by,
    \begin{equation}
        [C] = C_{ij} = \int_{\Omega_k} \psi_i \phi_j \,dx \hspace{0.02in} .
    \end{equation}
Using Eqn.~(\ref{eqn:modal-nodal}), we can obtain the modal solution coefficients, $\Vec{b}$, from the nodal solution coefficients $\Vec{u}$ by inverting the Modal Mass Matrix [M] i.e.
    \begin{equation}
        \Vec{b} = [M]^{-1} [C] \Vec{u}
        \label{eqn:modal_sol}
    \end{equation}
Now that the hierarchical modal basis coefficients have been obtained, a square filter matrix, [F], can be applied as a matrix-vector product:
\begin{equation}
\widehat{\Vec{b}} = [F] \Vec{b},
\label{eqn:filter_modal}
\end{equation}
where $\widehat{\Vec{b}}$ are the filtered modal solution coefficients.
The sharp cut-off Filter matrix $[F]$ is diagonal with its entries being $0$ or $1$. If all entries are $1$ giving the identity matrix, the filtering operation returns the original solution. To obtain a cut-off filter of order $(p^\star + 1)$, all diagonal entries of the filter matrix are $1$ up to and including the ($p^\star + 1$) diagonal entry with the rest of the entries $0$. The last step in the modal decomposition filtering procedure is to transform the filtered modal coefficients back to nodal basis coefficients to give the filtered nodal solution. This reverse transformation can be performed as follows :
  \begin{align}
      \widehat{\Vec{u}} &= [C]^{-1} [M] \widehat{\Vec{b}} \\
      &= [C]^{-1} [M] [F] \Vec{b} \hspace{0.2in} \text{by : Eqn.(\ref{eqn:filter_modal})}  \\
      &= [C]^{-1} [M] [F] [M]^{-1} [C] \Vec{u} \hspace{0.2in} \text{by : Eqn.(\ref{eqn:modal_sol})} 
  \end{align}
Let 
  \begin{equation}
   [B] = [M]^{-1} [C] \hspace{0.1in} \text{and} \hspace{0.1in}
   \widehat{[F]} = [B]^{-1} [F] [B] \hspace{0.02in} .
  \end{equation}
We can then write the final filtered nodal solution coefficients as follows :
  \begin{align}
      \widehat{\Vec{u}} = \widehat{[F]} \Vec{u}
  \end{align}
The final filter matrix $ \widehat{[F]} $ can be assembled as a pre-processing step as it does not have a dependence on the solution.

%%%%%%%%%%%%%%%%%%%%%%%%%%%%%%%%%%%%%%%%%%%%%%%%%%%%%%%%%%%
	
     \begin{figure}
	\centering
		\begin{subfigure}{0.49\textwidth}
			\centering
			\includegraphics[width=0.9\linewidth, trim = 0mm 0mm 0mm 0mm]{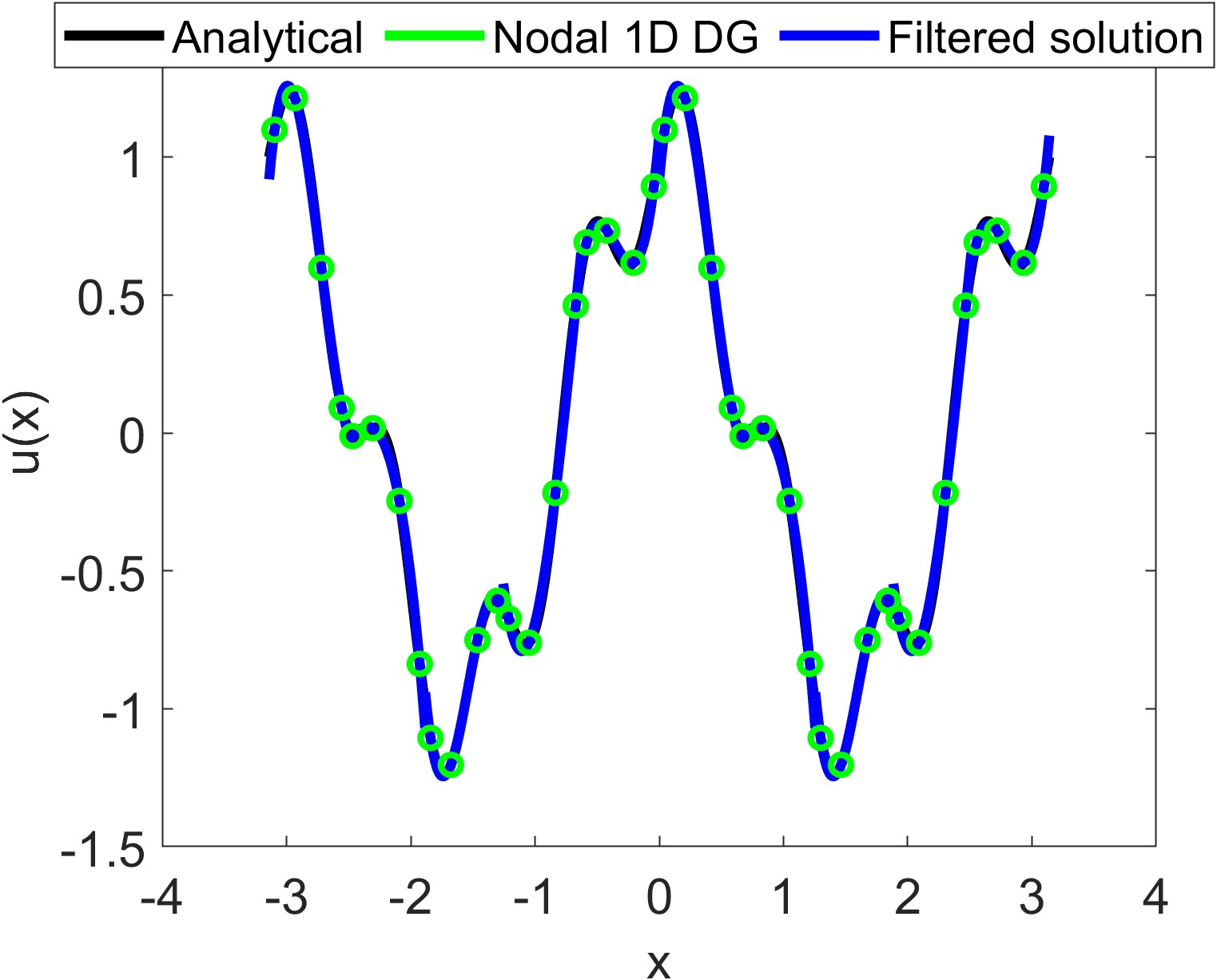}
			\caption{$p = 3, p^\star = 3$ }
		\end{subfigure} 
		\begin{subfigure}{0.49\textwidth}
			\centering
			\includegraphics[width=0.9\linewidth, trim = 0mm 0mm 0mm 0mm]{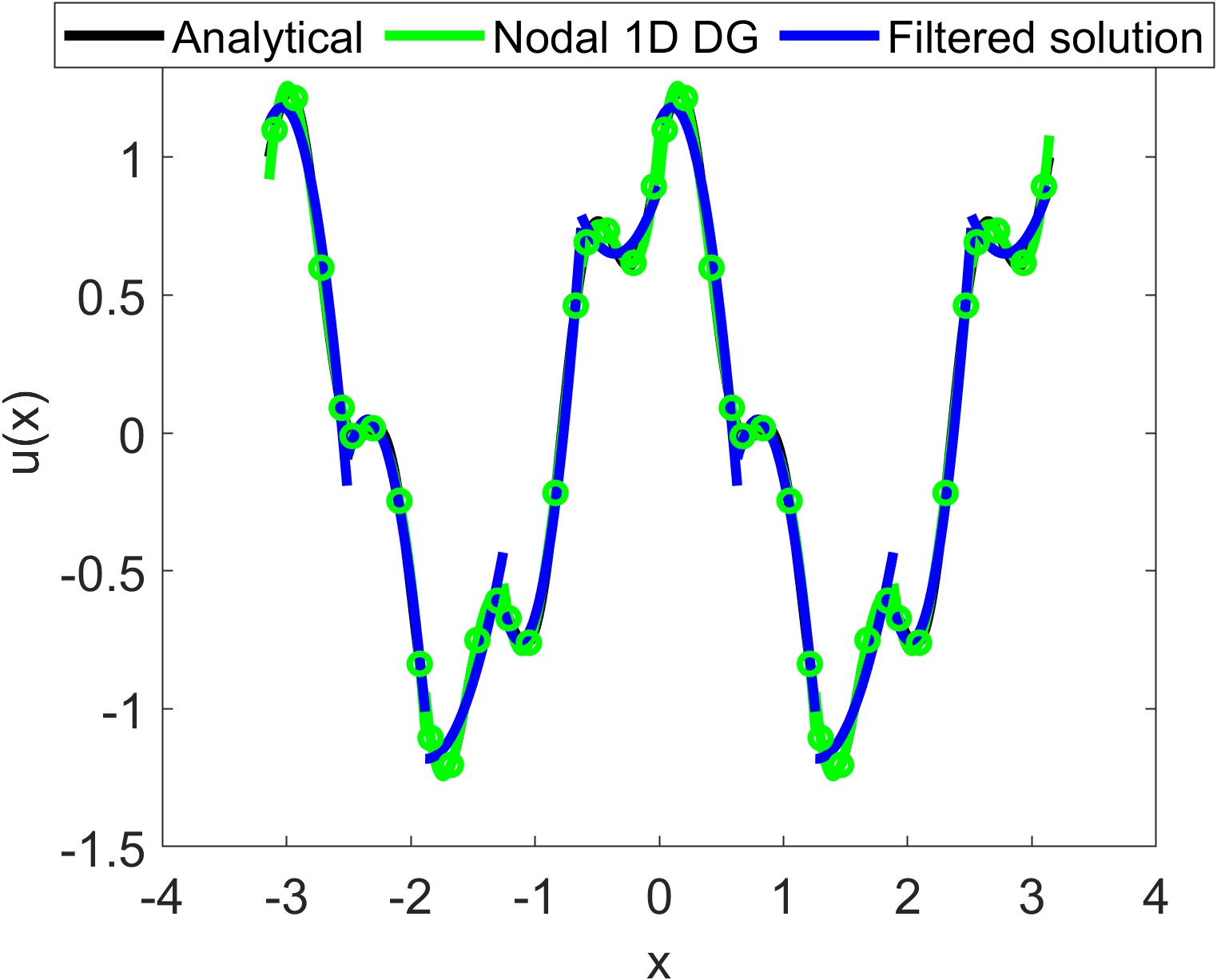}
			\caption{$p = 3, p^\star = 2$ }
		\end{subfigure}	\\
		\begin{subfigure}{0.49\textwidth}
			\centering
			\includegraphics[width=0.9\linewidth, trim = 0mm 0mm 0mm 0mm]{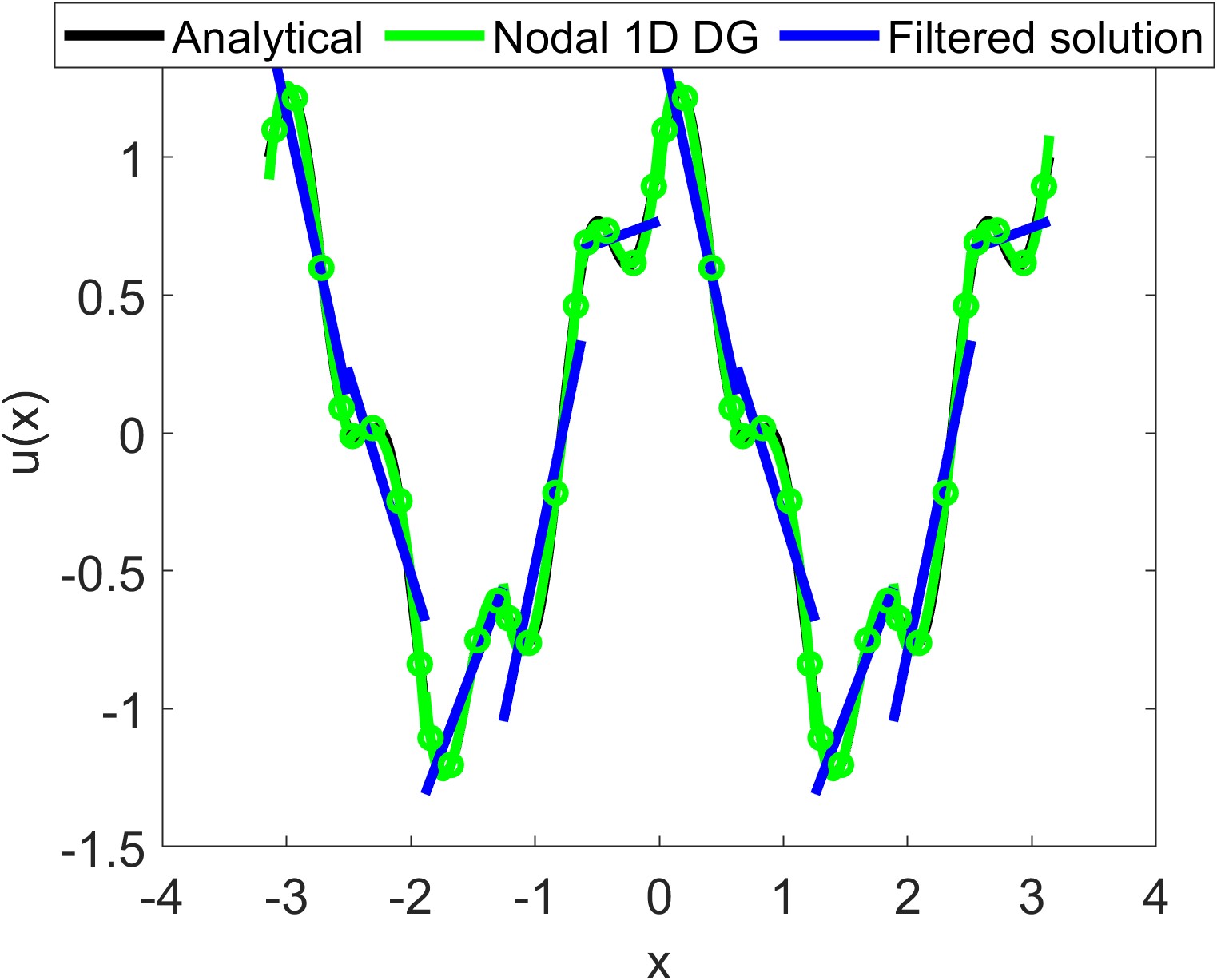}
			\caption{$p = 3, p^\star = 1$ }
		\end{subfigure} 
		\begin{subfigure}{0.49\textwidth}
			\centering
			\includegraphics[width=0.9\linewidth, trim = 0mm 0mm 0mm 0mm]{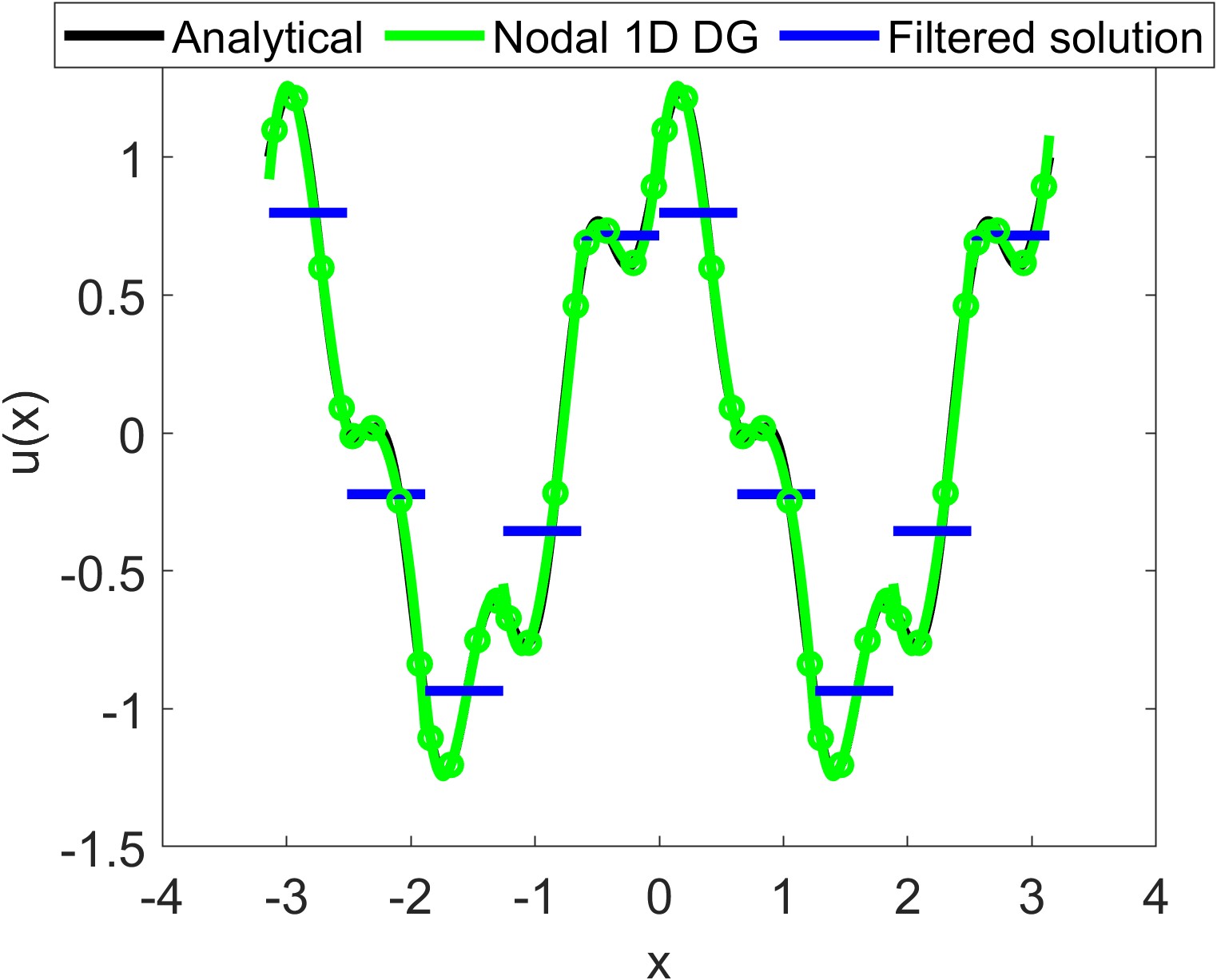}
			\caption{$p = 3, p^\star = 0$ }
		\end{subfigure}	
  \caption{Comparisons of analytical solution for a function $u(x) = cos(2x) + 0.3sin(8x) $ in the range $ [-\pi,\pi] $ with the best fit obtained using one-dimensional nodal DG employing 10 elements and $p = 3$ along with filtered solutions for $p^\star = 0,1,2$, and $3$. Unfilled green circles indicate the quadrature points within each element. }
  \label{fig:1D-filter}
	\end{figure} 
	
	%%%%%%%%%%%%%%%%%%%%%%%%%%%%%%%%%%%%%%%%%%%%%%%%%%%%%%%%%%%%%%%%	

The sharp-modal filter is first tested in a one-dimensional DG set-up. The objective is to find a best-fit for the function $ u(x) = cos(2x) + 0.3sin(8x) $. The degree of the Lagrange basis function is set to $p = 3$, and the sharp modal cut-off filter is tested for $p^\star = 0,1,2$, and $3$. The number of elements used is $10$ with $4$ quadrature points on each element, and the results are presented in Fig.~\ref{fig:1D-filter}. The analytical solution is plotted for the domain $[-\pi,\pi]$. The nodal DG solution with $p = 3$ matches closely with the analytical solution, and the jumps at the element approximation denote the discontinuous nature of the approximation. The sharp modal cut-off filter with $p^\star = 3$ does not affect the solution and the results are identical to the original solution. On the other hand, lower cut-off orders of $p^\star = 2,1$, and $0$ result in a piece-wise quadratic, linear, and constant solution, respectively.

\begin{figure}
		\centering
		\begin{subfigure}{0.49\textwidth}
			\centering
			\includegraphics[width=1.\linewidth, trim = 0mm 0mm 0mm 0mm]{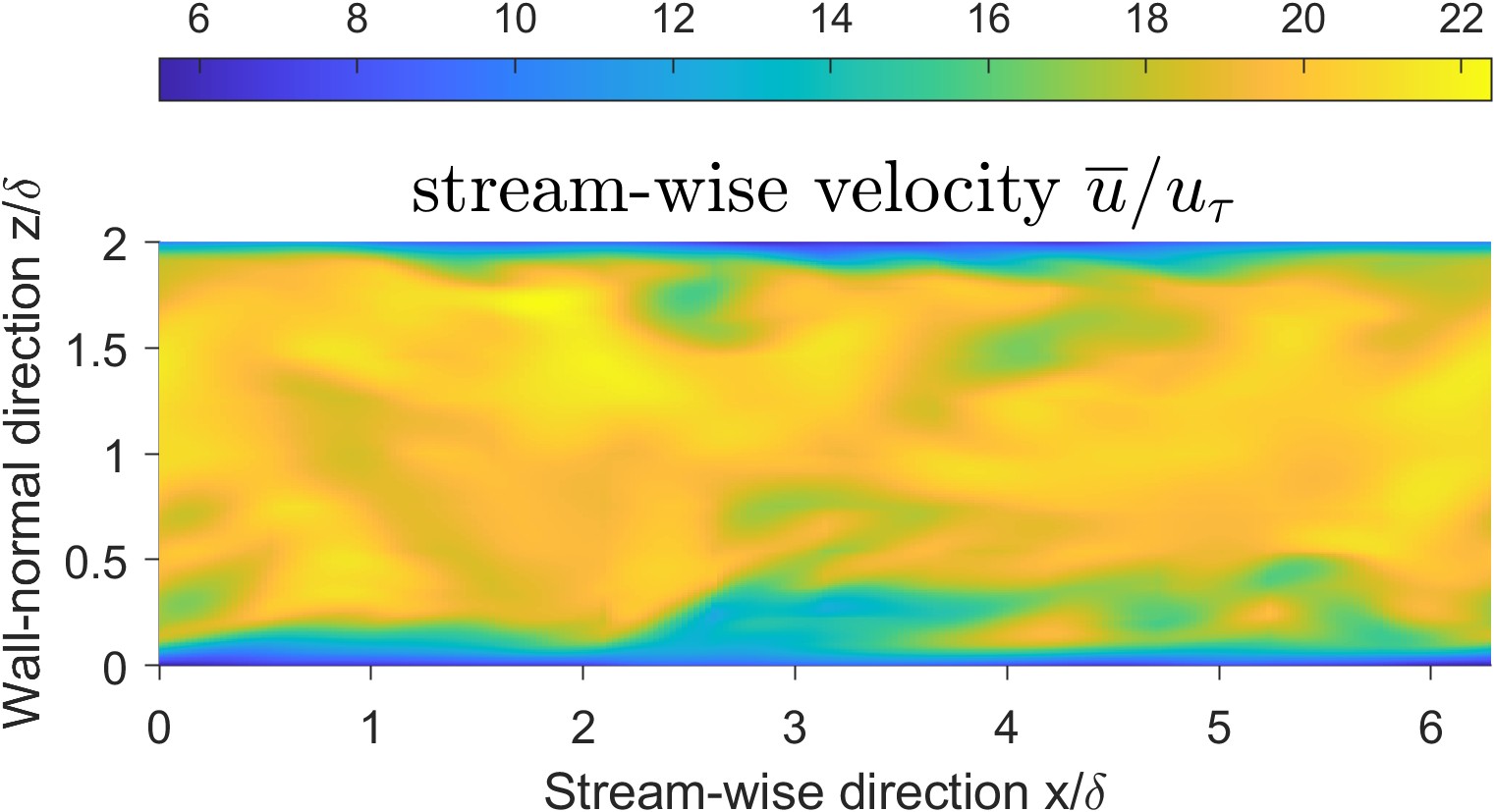}
			\caption{Filtered solution : $p = 3$, $p^\star = 3$ }
		\end{subfigure} 
		\begin{subfigure}{0.49\textwidth}
			\centering
			\includegraphics[width=1.\linewidth, trim = 0mm 0mm 0mm 0mm]{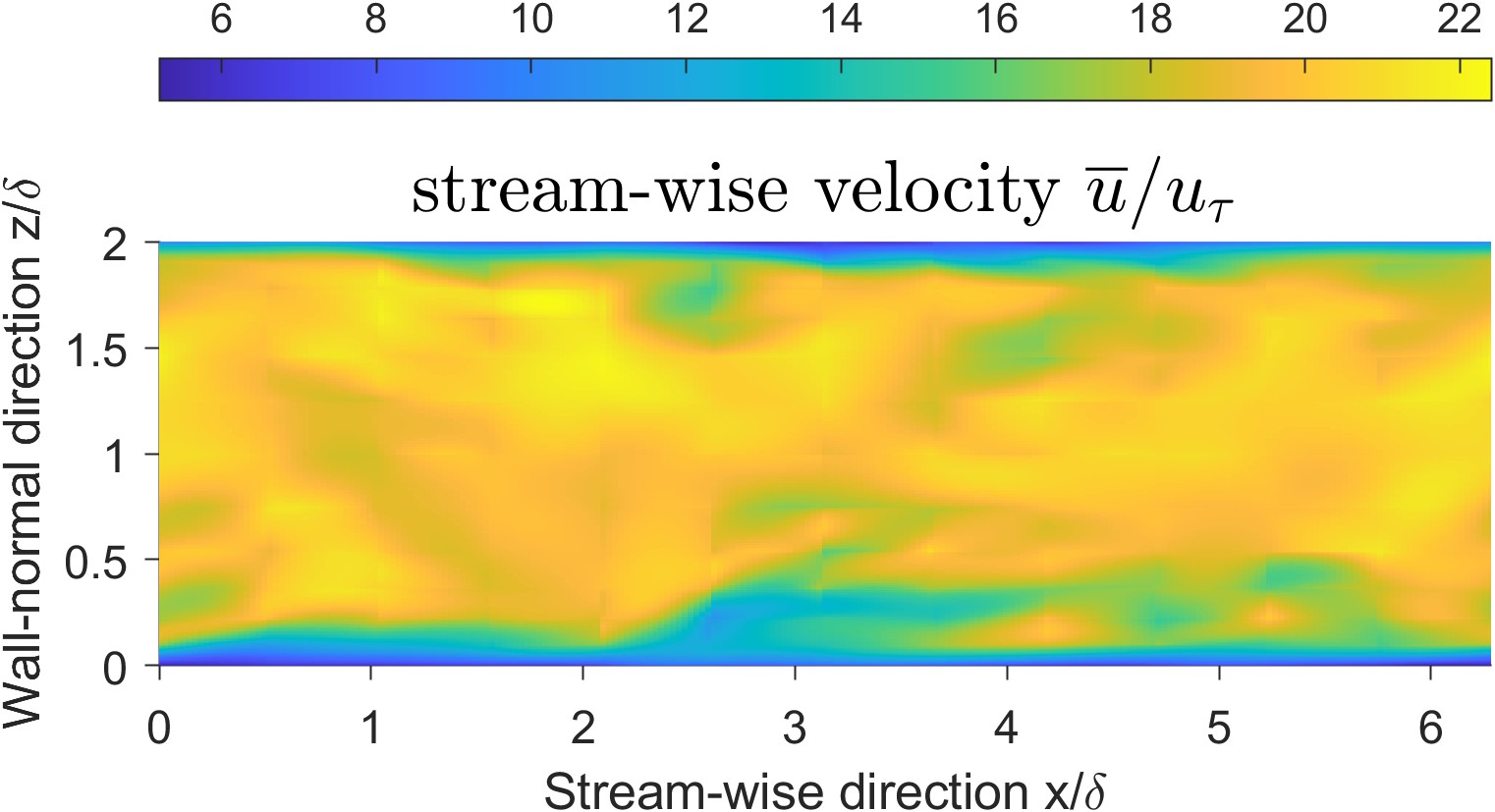}
			\caption{Filtered solution : $p = 3$, $p^\star = 2$ }
		\end{subfigure}	\\
		\begin{subfigure}{0.49\textwidth}
			\centering
			\includegraphics[width=1.\linewidth, trim = 0mm 0mm 0mm 0mm]{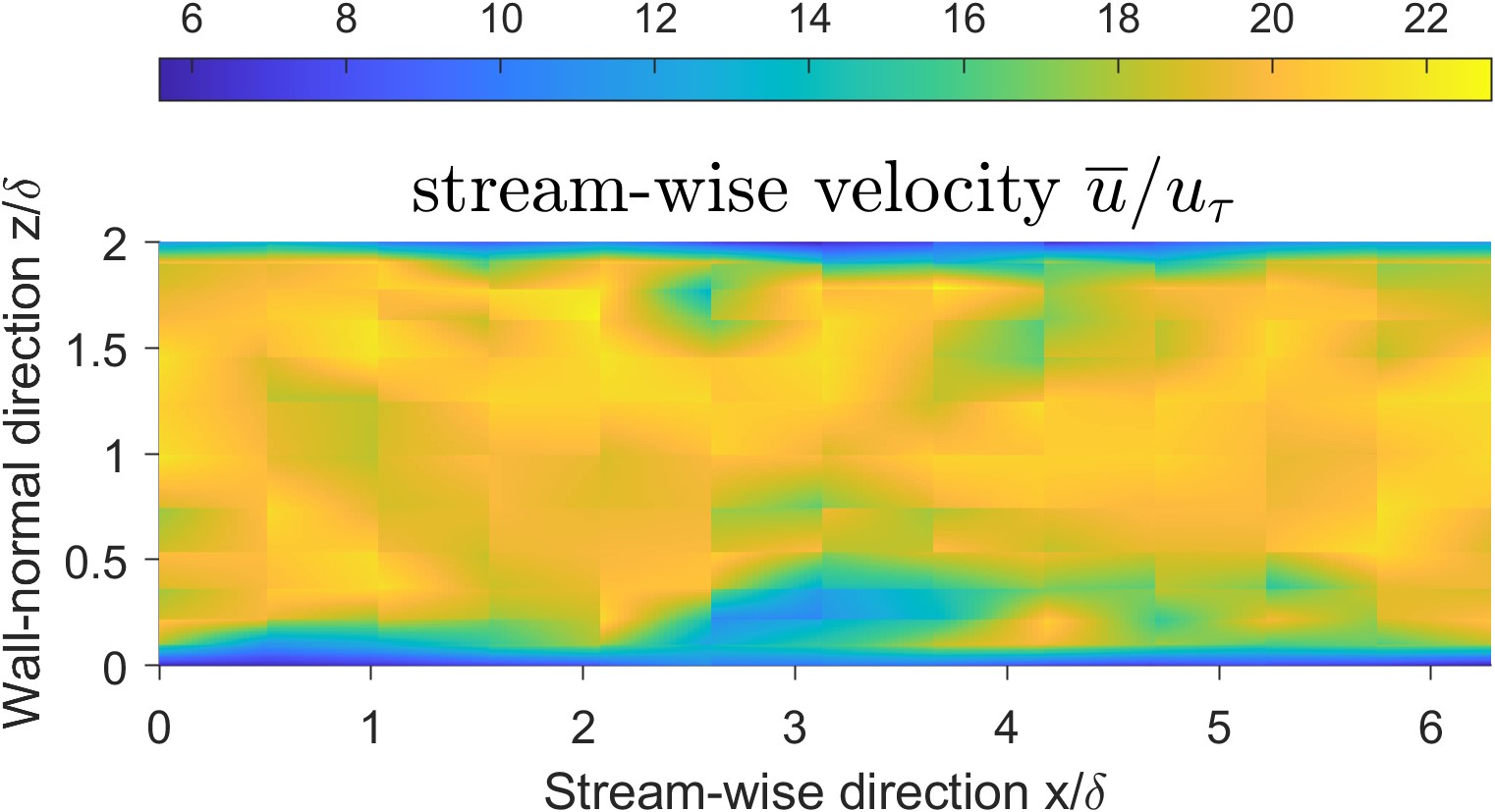}
			\caption{Filtered solution : $p = 3$, $p^\star = 1$ }
		\end{subfigure} 
		\begin{subfigure}{0.49\textwidth}
			\centering
			\includegraphics[width=1.\linewidth, trim = 0mm 0mm 0mm 0mm]{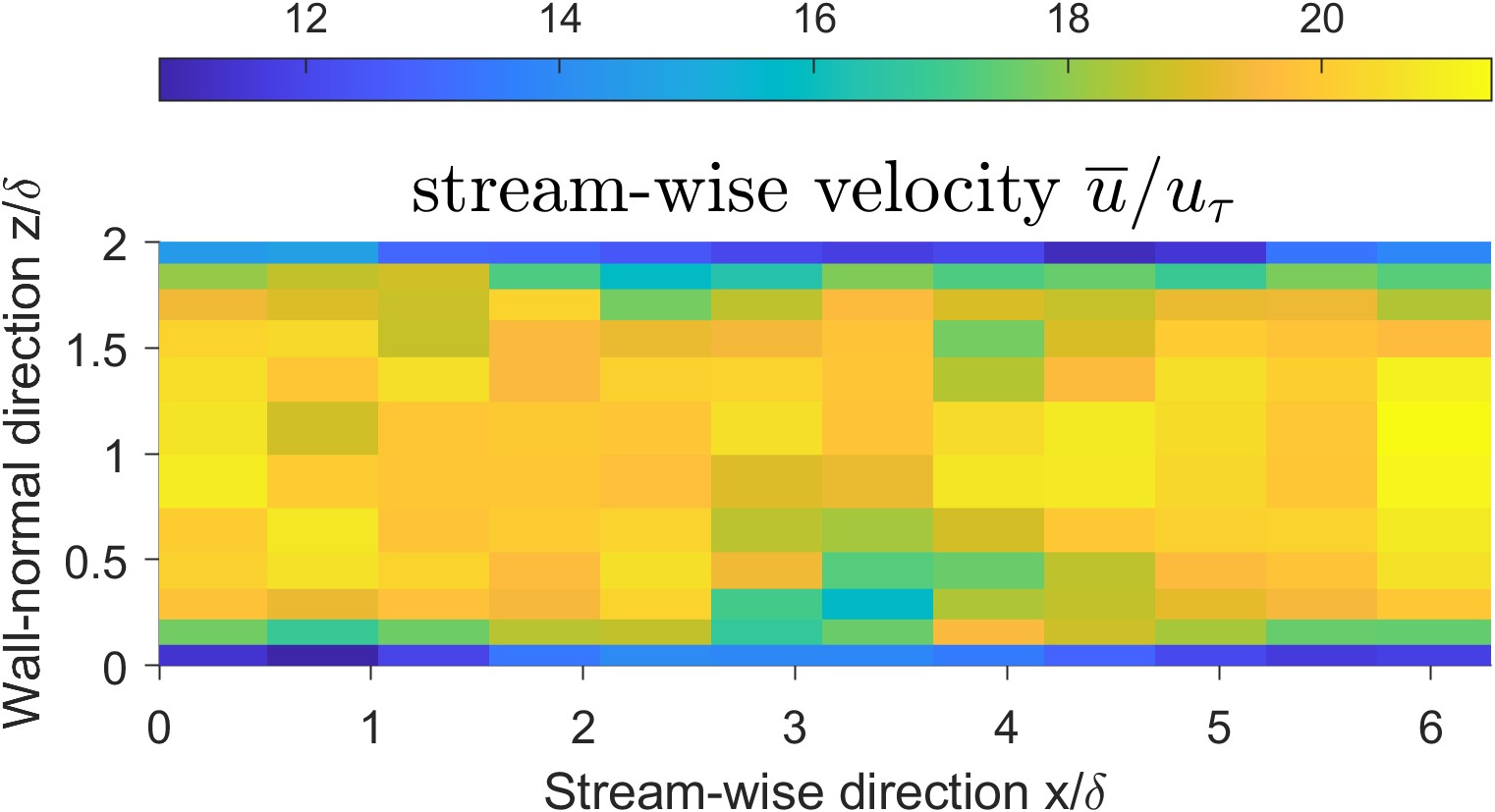}
			\caption{Filtered solution : $p = 3$, $p^\star = 0$ }
		\end{subfigure}	
  \caption{Snapshots of normalized stream-wise velocity in $xz$ plane passing midway through the span-wise domain dimension with $p^\star = 3, 2, 1,$ and $0$ showing the effects of test filtering operation at $Re_\tau \approx 544$. }
  \label{fig:2D-filter-velocity}
	\end{figure}

The modal sharp cut-off filter in its 1D form discussed above is extended to 3D in a tensor product fashion and applied before every RK3-TVD step in our in-house DG code. The effect of test filtering operation using the modal sharp cut-off filter on the normalized instantaneous stream-wise velocity $\overline{u}/ u_\delta$ is shown in Fig.~\ref{fig:2D-filter-velocity}. The degree of polynomial used is $p = 3$ and results are shown for $Re_\tau \approx 544$ case with filter orders of $p^\star = 3,2,1$, and $0$. The snapshots of $\overline{u}/ u_\tau$ show the loss of information and decrease in resolution of the flow-field as the filter cut-off order is reduced. 

%%%%%%%%%%%%%%%%%%%%%%%%%%%%%%%%%%%%%%%%%%%%%%%%%%%%%%%%%%%%%%%%%%%%%%%%%%

 %%%%%%%%%%%%%%%%%%%%%%%%%%%%%%%%%%%%%%%%%%%%%%%%%%%%%%%%%%%
	
	%%%%%%%%%%%%%%%%%%%%%%%%%%%%%%%%%%%%%%%%%%%%%%%%%%%%%%%%%%%%%%%%

%%%%%%%%%%%%%%%%%%%%%%%%%%%%%%%%%%%%%%%%%%%%%%%%%%%%%%%%%%%%%%%%%%%%%%%%%%%%%%%%%%%%

\section{Wall-resolved LES at $Re_\tau \approx 544$}

A wall-resolved large eddy simulation (WRLES) of the turbulent channel flow is performed at $Re_\tau \approx 544$ using the dynamic Smagorinsky model (DSM) as the SGS model to verify our in-house DG solver. The mesh size is $36 \times 30 \times 24$ elements in the stream-wise, span-wise, and wall-normal directions, respectively. The grid is uniform in the stream-wise and span-wise directions and it is geometrically stretched in the wall-normal direction with a stretching ratio of $1.2$. The effective grid sizes in each direction in wall units are $\Delta_x^+ \approx 38 $,  $\Delta_y^+ \approx 19$, and $\Delta_{z}^+ $ at the wall is $ \Delta_{z_w}^+ \approx 4.5$ and at the channel center is $ \Delta_{z_c}^+ \approx 32.2$. The grid resolution is based on the recommendation of \cite{bose2014dynamic} for a wall-resolved LES, i.e. $\Delta_x^+ \lesssim 50 $,  $\Delta_y^+ \lesssim 30$, and $\Delta_{z_w}^+ \sim O(1)$. In comparison, the grid resolution of the available DNS \citep{lee2015direct} is $\Delta_x^+ \approx 8.9 $,  $\Delta_y^+ \approx 5$, $ \Delta_{z_w}^+ \approx 0.019$, and $ \Delta_{z_c}^+ \approx 4.5$. Please note that, the effective grid sizes $\Delta_x$, $\Delta_y$ and $\Delta_z$ for the finite element grid are defined as $\Delta_x = \Delta^e_x/p$, $\Delta^e_y/p$, and $\Delta^e_z/p$, respectively. The quantities $\Delta^e_x$, $\Delta^e_y$, and $\Delta^e_z$ represent the actual element sizes in the finite element mesh.

 %%%%%%%%%%%%%%%%%%%%%%%%%%%%%%%%%%%%%%%%%%%%%%%%%%%%%%%%%%%
	
	\begin{figure}
		\centering
        		\begin{subfigure}{0.49\textwidth}
			\centering
			\includegraphics[width=1.\linewidth, trim = 0mm 0mm 0mm 0mm]{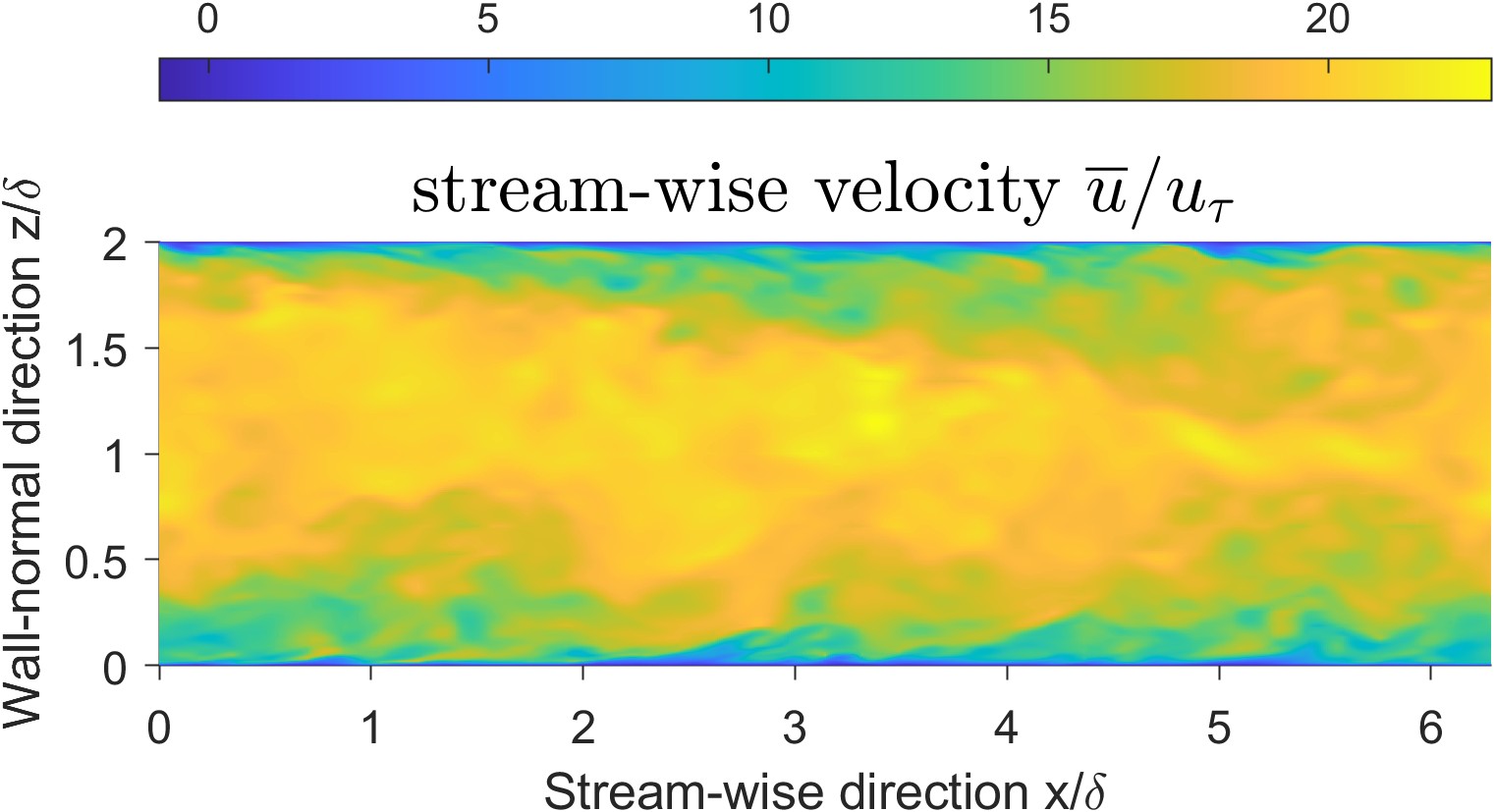}
			\caption{}
		\end{subfigure} \\
		\begin{subfigure}{0.49\textwidth}
			\centering
			\includegraphics[width=1.\linewidth, trim = 0mm 0mm 0mm 0mm]{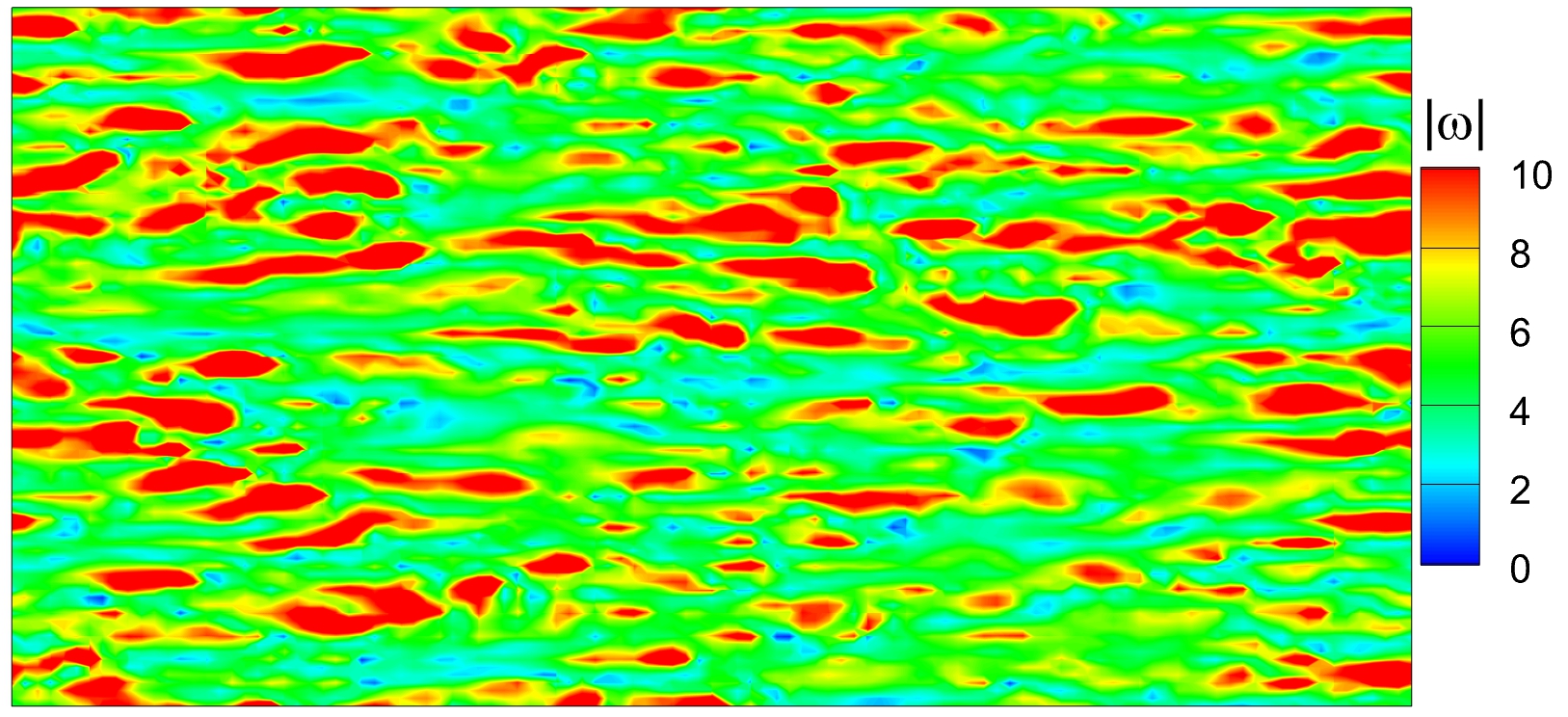}
			\caption{}
		\end{subfigure}
		\begin{subfigure}{0.49\textwidth}
			\centering
			\includegraphics[width=1.\linewidth, trim = 0mm 0mm 0mm 0mm]{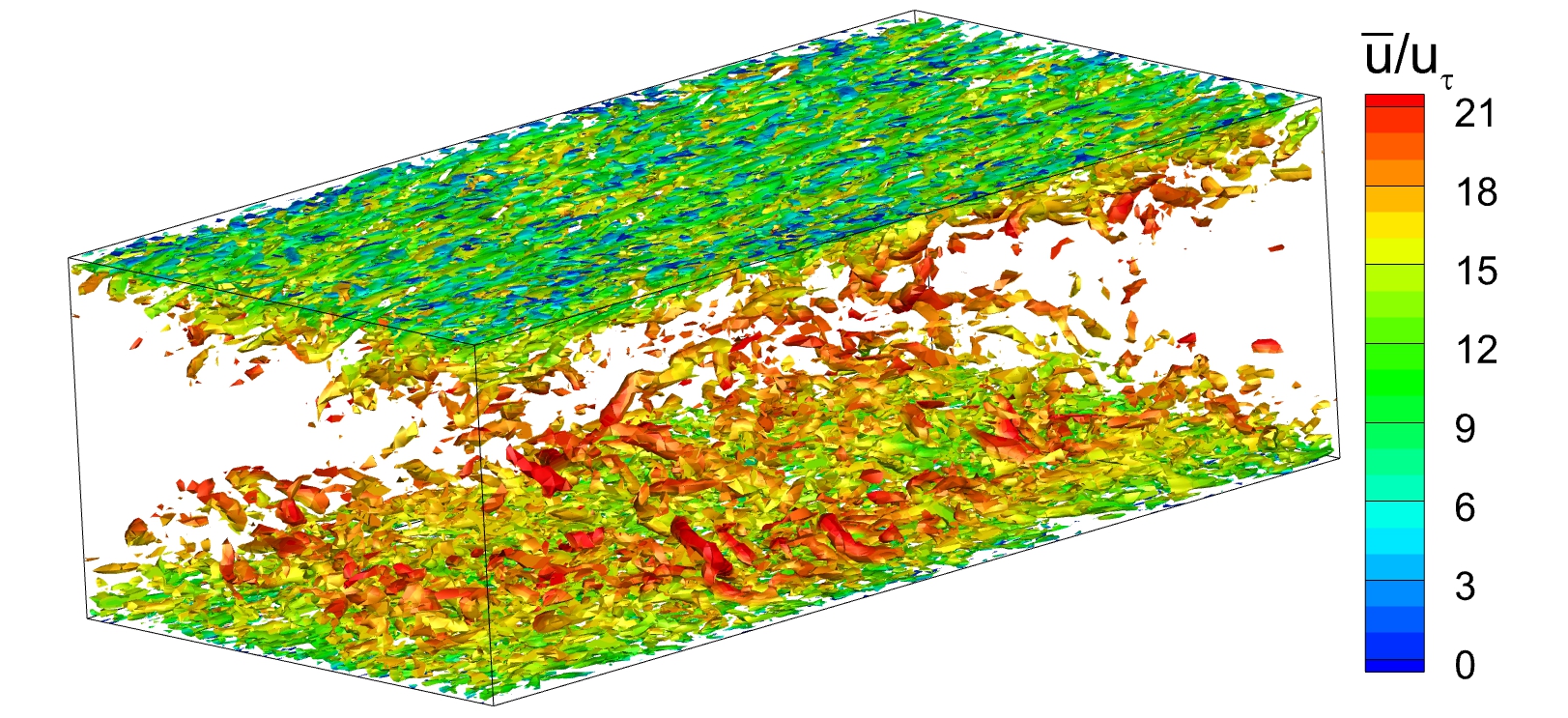}
			\caption{}
		\end{subfigure}	
  \caption{(a) Snapshot of normalized stream-wise velocity in a $xz$ plane passing midway through the span-wise dimension, (b) snapshot of vorticity magnitude on the bottom wall, and (b) iso-surface of Q-criterion colored with normalized stream-wise velocity $\overline{u}/u_\tau$ for a WRLES at $Re_\tau \approx 544$ obtained using the DSM SGS model.} 
  \label{fig:WRLES-vel-vort-iso}
	\end{figure} 
	
	%%%%%%%%%%%%%%%%%%%%%%%%%%%%%%%%%%%%%%%%%%%%%%%%%%%%%%%%%%%%%%%%

%%%%%%%%%%%%%%%%%%%%%%%%%%%%%%%%%%%%%%%%%%%%%%%%%%%%%%%%%%%%%%%%%%%%%%%%%
	
\begin{figure}
	\centering
 \begin{subfigure}{0.49\textwidth}
		\centering
		\includegraphics[width=0.95\linewidth, trim = 0mm 0mm 0mm 0mm]{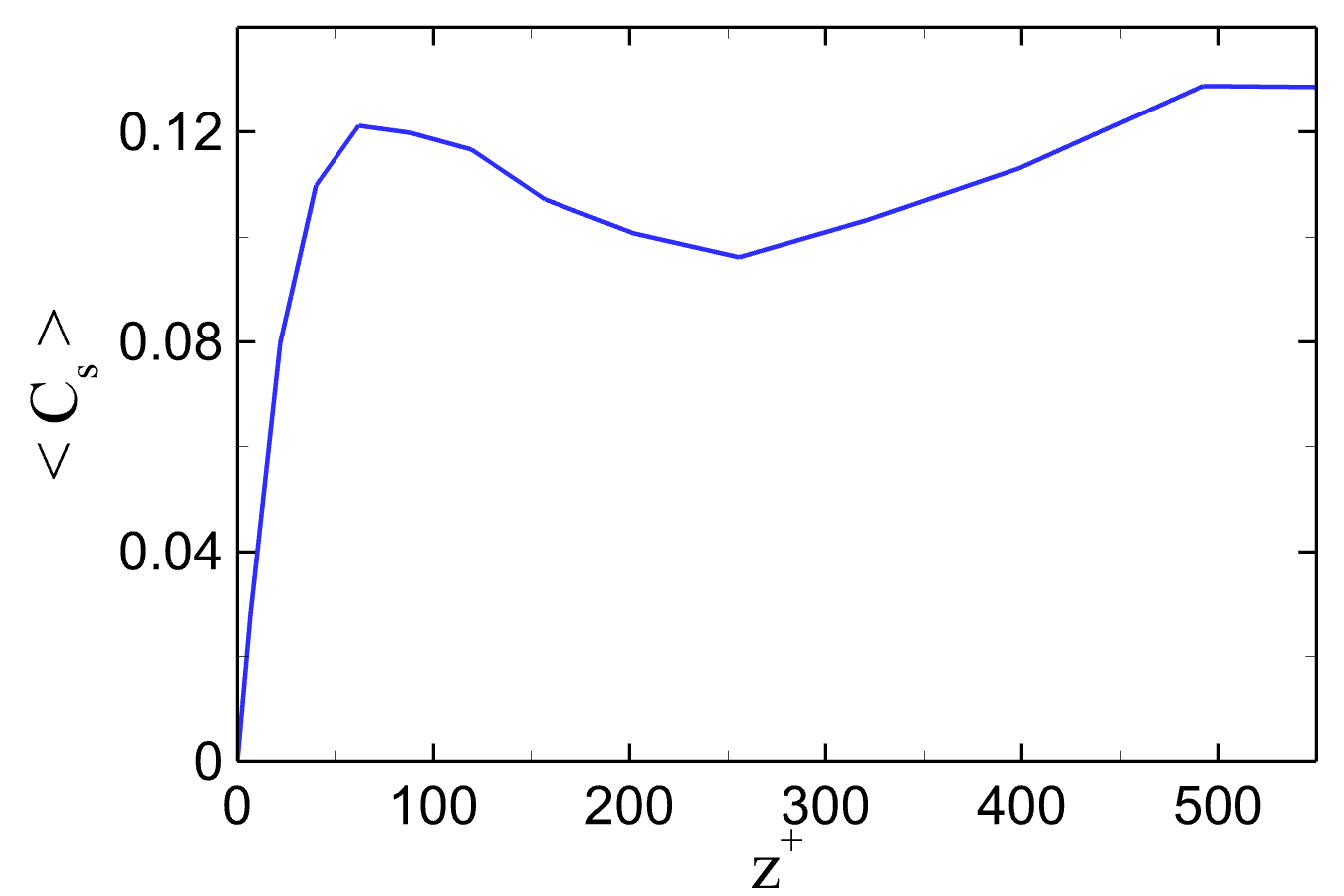}
		\caption{}
	\end{subfigure}
	\begin{subfigure}{0.49\textwidth}
		\centering
		\includegraphics[width=0.95\linewidth, trim = 0mm 0mm 0mm 0mm]{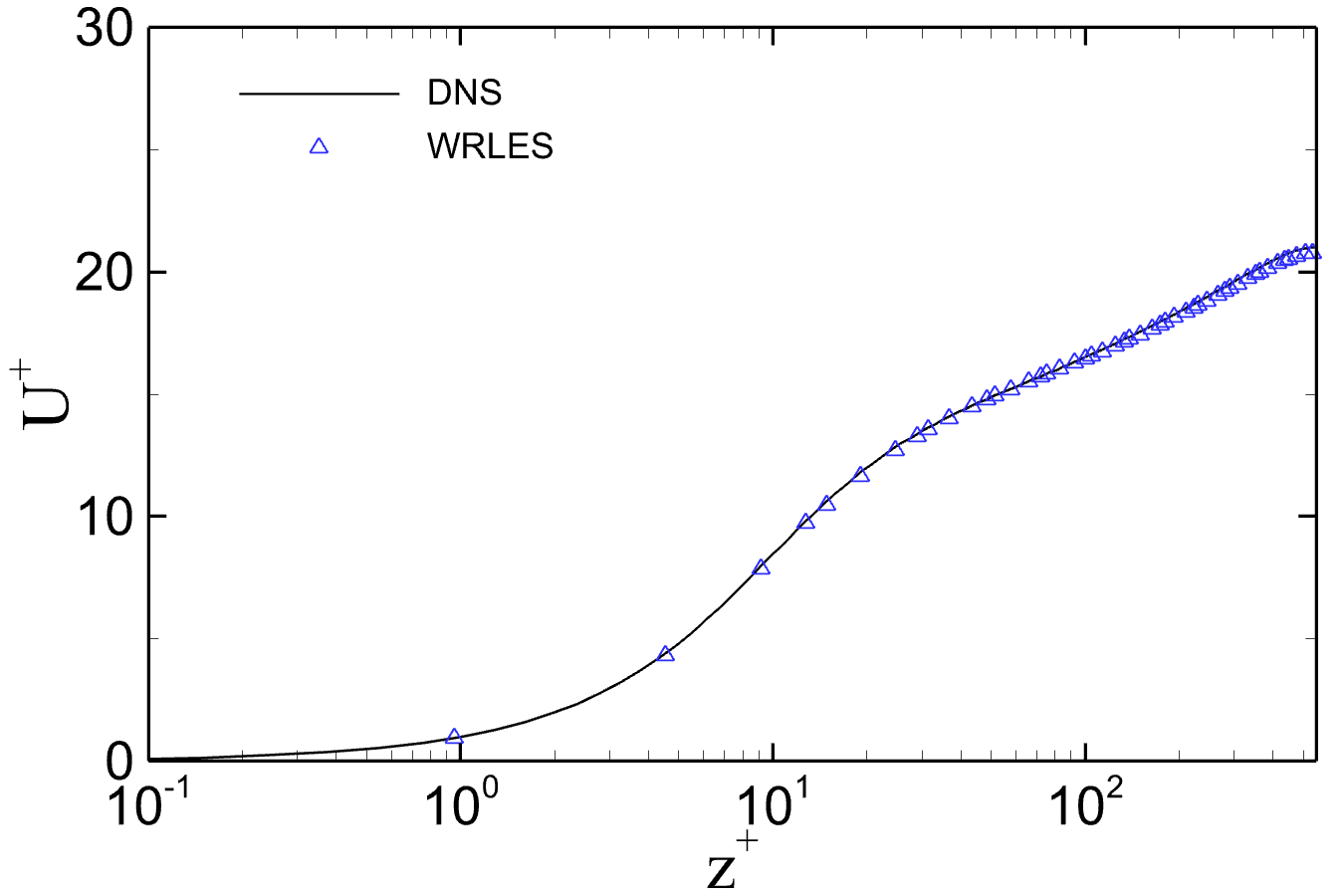}
		\caption{}
	\end{subfigure} 
	\begin{subfigure}{0.49\textwidth}
		\centering
		\includegraphics[width=0.95\linewidth, trim = 0mm 0mm 0mm 0mm]{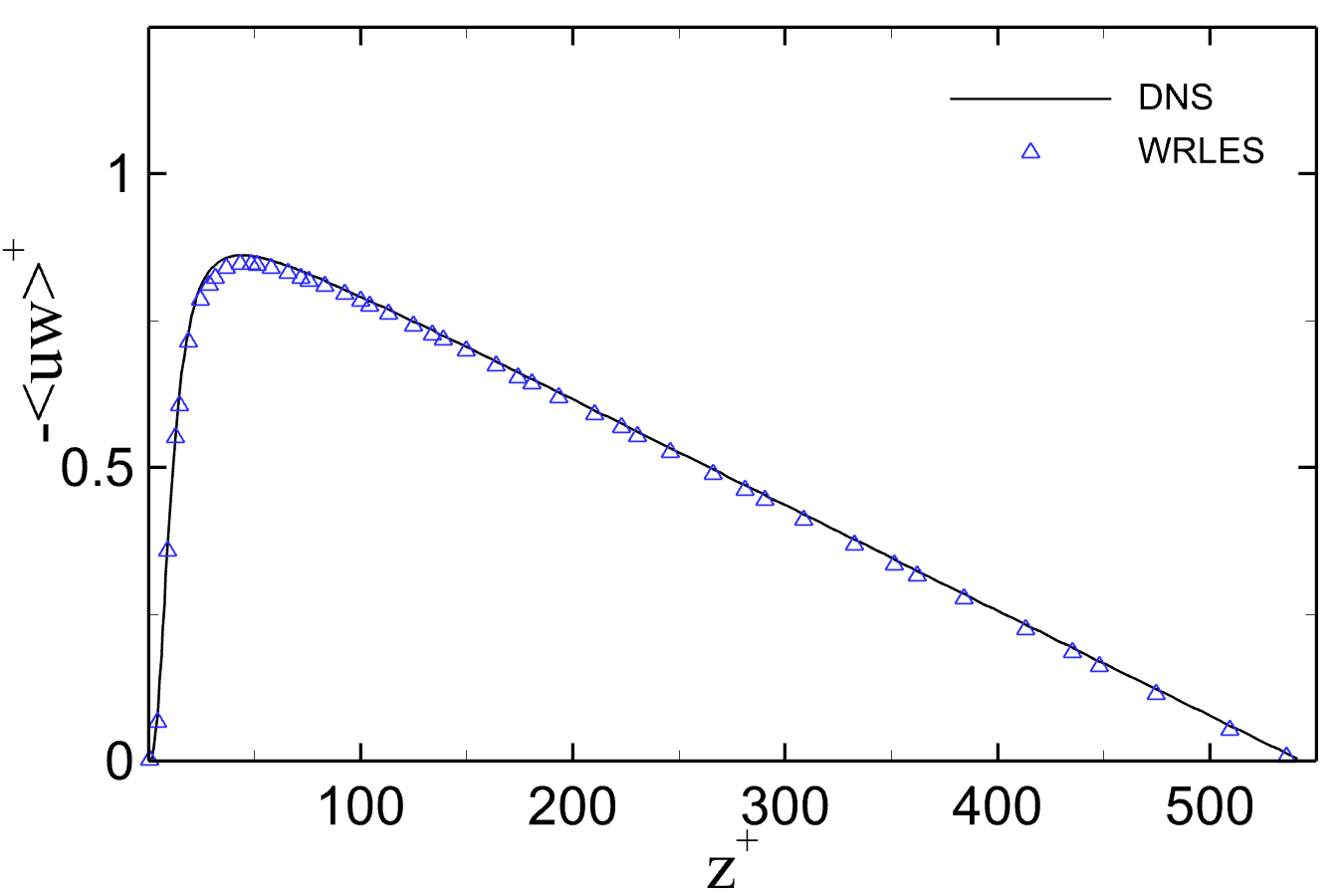}
		\caption{}
	\end{subfigure} 
	\begin{subfigure}{0.49\textwidth}
		\centering
		\includegraphics[width=0.95\linewidth, trim = 0mm 0mm 0mm 0mm]{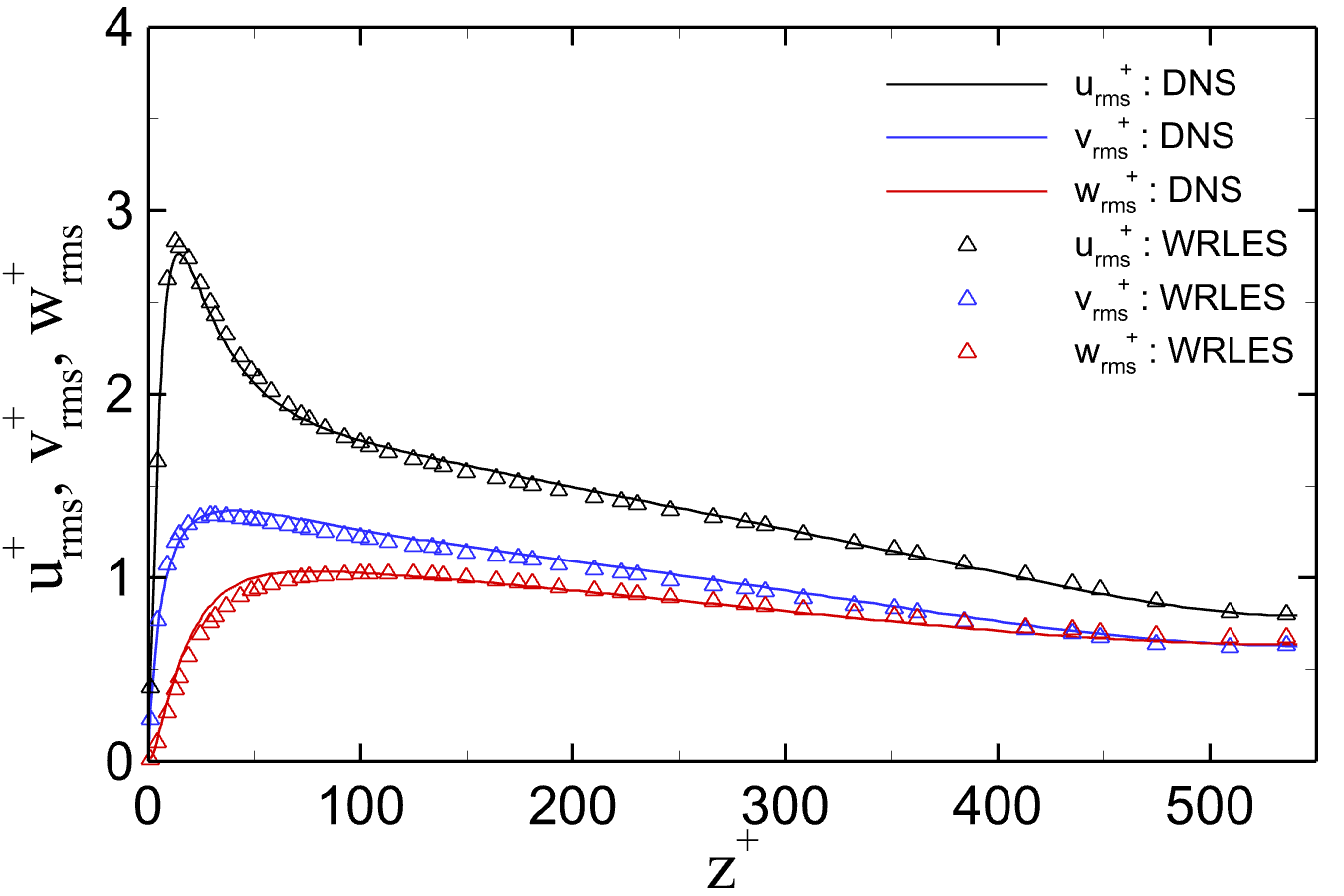}
		\caption{}
	\end{subfigure}	
     \caption{Wall-normal variation of (a) mean Smagorinsky coefficient $\langle C_s \rangle$, (b) mean velocity, (c) Reynolds shear stress, and (d) r.m.s. velocity fluctuations for a wall-resolved LES employing DSM as the SGS models compared with the DNS at $Re_\tau \approx 544$.}
     \label{fig:WRLES-Re550}
\end{figure} 	
	
%%%%%%%%%%%%%%%%%%%%%%%%%%%%%%%%%%%%%%%%%%%%%%%%%%%%%%%%%%%%%%%%%%%%%%%%%

The instantaneous stream-wise velocity $\overline{u}$ normalized with $u_\tau$ for the WRLES in a $xz$-plane are shown in Fig.~\ref{fig:WRLES-vel-vort-iso}(a). The solution is reasonably resolved. Snapshots of the vorticity magnitude on the bottom wall are shown in Fig.~\ref{fig:WRLES-vel-vort-iso}(b). On the other hand, Fig.~\ref{fig:WRLES-vel-vort-iso}(c) shows the isometric view of the iso-surfaces of the Q-criterion to visualize the near-wall eddies.

%%%%%%%%%%%%%%%%%%%%%%%%%%%%%%%%%%%%%%%%%%%%%%%%%%%%%%%%%%%%%%%5

The wall-normal variation of the mean Smagorinsky coefficient $\langle C_s \rangle$ along with mean velocity and Reynolds shear and normal stresses for the WRLES compared with the DNS is shown in Fig.~\ref{fig:WRLES-Re550}. The Smagorinsky coefficient $C_s$ for the DSM assumes a value of zero at the wall as $L_{ij}^{d}$ in Eqn.~(\ref{eqn:dyn-smag}) is equivalently zero at the wall owing to the no-slip velocity boundary condition. It gradually increases in the viscous-sublayer before reaching a value of about $0.12$ in the log-layer at about $ z^+ \approx 60$ after which it remains close to $0.1-0.12$ till the half channel height $\delta^+$. The mean velocity and Reynolds shear and normal stress profiles obtained using the WRLES are nearly identical to the DNS data.

%%%%%%%%%%%%%%%%%%%%%%%%%%%%%%%%%%%%%%%%%%%%%%%%%%%%%%%%%%%%%%%%%%%%%%%%%%%%%%%

\section{Sensitivity analysis}\label{appendixC}

\begin{figure}
		\centering
		\begin{subfigure}{1.0\textwidth}
			\centering
			\includegraphics[width=0.55\linewidth, trim = 0mm 0mm 0mm 0mm]{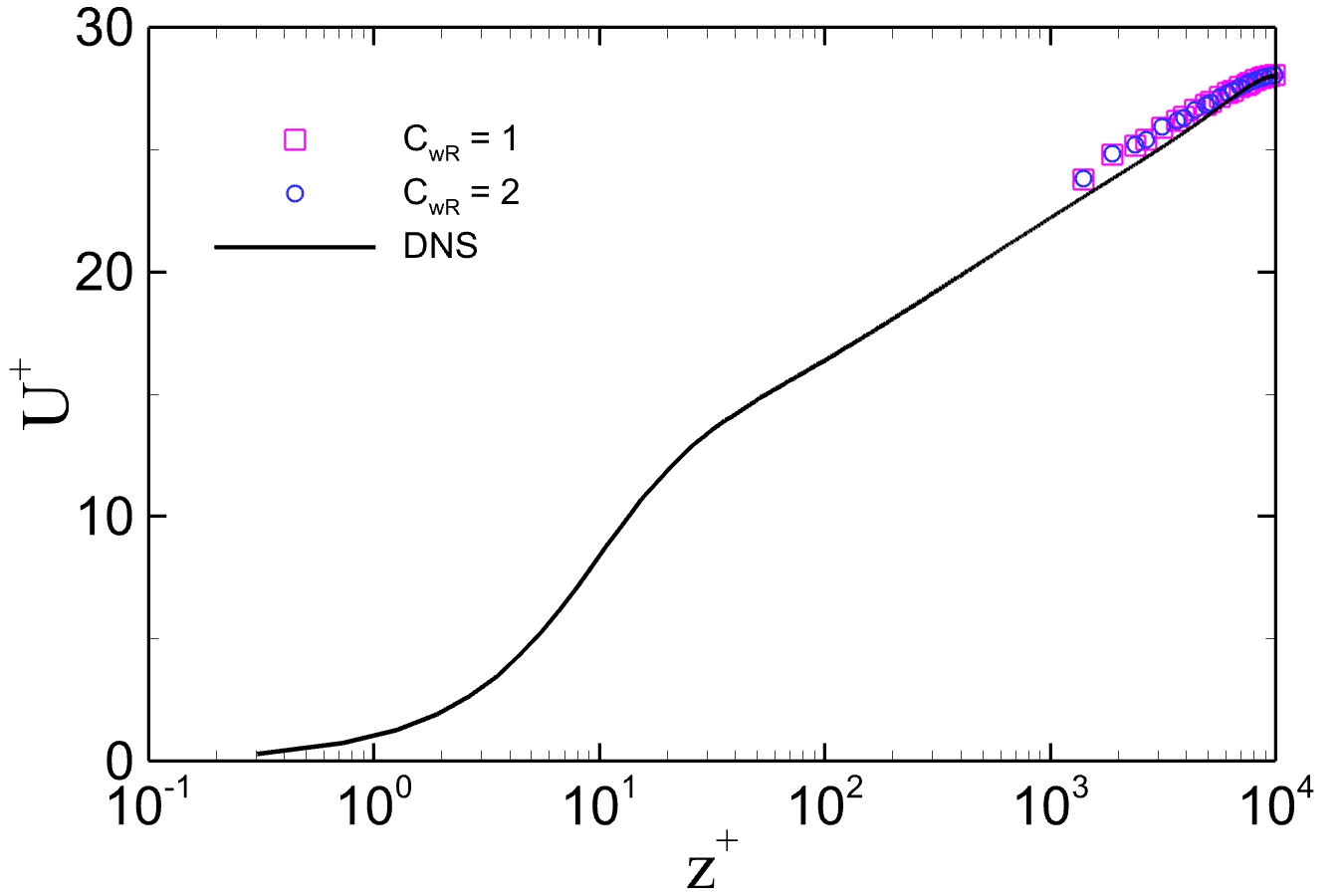}
			\caption{}
		\end{subfigure} \\
		\begin{subfigure}{0.49\textwidth}
			\centering
			\includegraphics[width=0.99\linewidth, trim = 0mm 0mm 0mm 0mm]{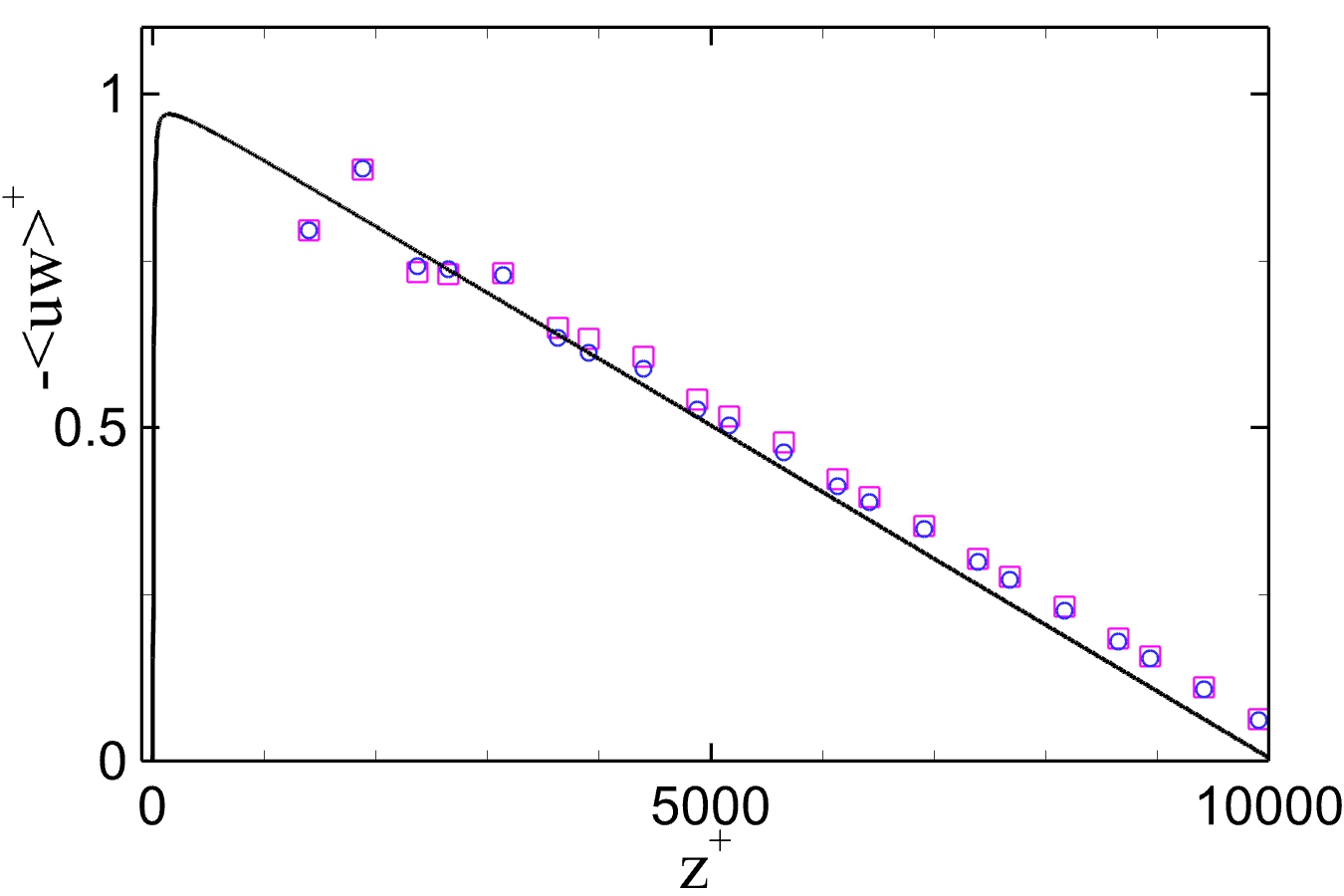}
			\caption{}
		\end{subfigure} 
		\begin{subfigure}{0.49\textwidth}
			\centering
			\includegraphics[width=0.99\linewidth, trim = 0mm 0mm 0mm 0mm]{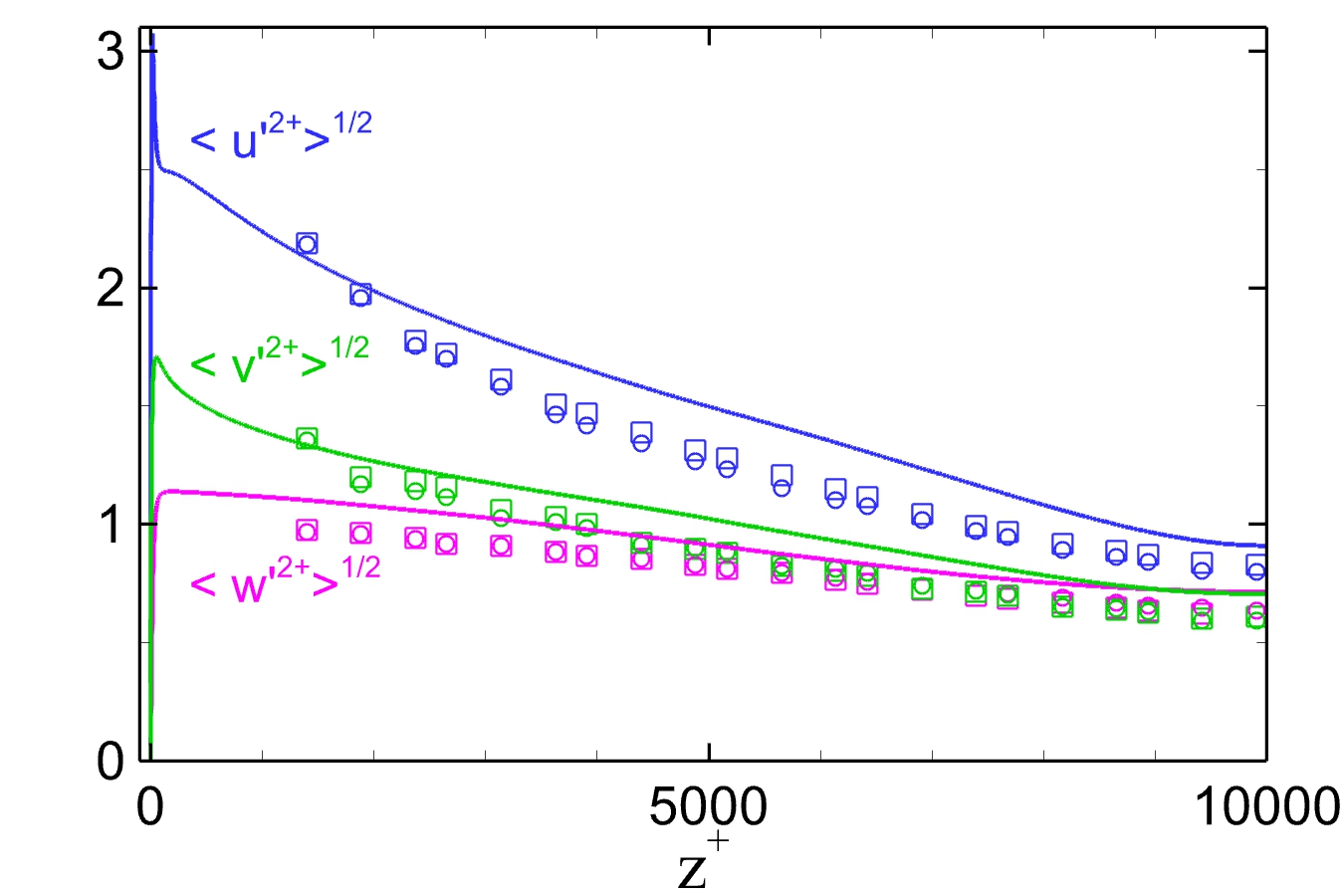}
			\caption{}
		\end{subfigure}		
            \caption{Effect of the parameter $ C_{wR}$ values on the proposed dynamic wall model predictions for the case DSW-10000-G2 along with DNS comparisons for (a) mean velocity, (b) Reynolds shear stress, and (c) r.m.s. velocity fluctuations.}
            \label{fig:sensitivity}
	\end{figure}

The proposed dynamic slip wall model involves two model parameters, namely $ C_{wR}$ and $\Delta_R$. The definition of the filter operation fixes the value of $\Delta_R$, which is the ratio of the test filter width to the grid filter width. We have used a value for $\Delta_R$ as recommended by \cite{brazell2015using}. Numerical experiments using different values for $\Delta_R$ in the plausible range of $ \Delta_{R} = [1,2]$ for $p = 2$ and $p^\star = 1$ resulted in negligible differences in the results and these observations are similar to those made by \cite{bae2019dynamic} for their dynamic slip wall model. On the other hand, the parameter $ C_{wR}$ comes into the picture because of the use of different values of the model coefficient $C_{w}$ at the test filter and grid filter levels. The sensitivity to $ C_{wR}$ is tested for values in the plausible range $ C_{wR} = [1,2]$. Results for the two extreme values in this range i.e. $ C_{wR} = 1$ and $2$ are shown in Fig.~\ref{fig:sensitivity} for $Re_\tau \approx 10000$ case obtained using grid G2. The effect of $ C_{wR}$ on the mean velocity and Reynolds stress predictions is also found to be negligible and the results are almost identical. This suggests that the model coefficient $ C_{w}$ can be taken to be the same at the test and grid filtered levels, which is the general practice \citep{bose2014dynamic,bae2019dynamic}

%%%%%%%%%%%%%%%%%%%%%%%%%%%%%%%%%%%%%%%%%%%%%%%%%%%%%%%%%%%%%%%%

%%%%%%%%%%%%%%%%%%%%%%%%%%%%%%%%%%%%%%%%%%%%%%%%%%%%%%%%%%%%%%%%%%%%%%%%%%%%%%%

\section{Computational cost}\label{appendixD}

The simulations were performed on NASA's Pleiades Supercomputer on the Broadwell compute nodes consisting of E5-2680v4 Intel Xeon processors at $2.4$ GHz. For the channel flow computations on the finest mesh G3 consisting of $32 \times 32 \times 32 $ elements with about $0.885$ million degrees of freedom, the dynamic slip-wall model takes about $0.135$s of wall time per time-step on $512$ processors. For the $Re_\tau \approx 10000$ case, the dynamic slip-wall model requires about wall time of $11.8$ mins for a single flow-through ($= L_x /U_b$) on the grid G3. On the other hand, for the periodic hill cases using the fine grid consisting of $ 75 \times 36 \times 15$ elements with about $1.1$ million degrees of freedom, the wall time required by the dynamic slip-wall model per time-step is approximately $ 0.088$s on $ 3330 $ processors. For the $Re_b = 37000$ case, the wall time required for a single flow-through is about $22$ mins. The equilibrium wall-stress model requires a similar time per time-step as that of the dynamic slip-wall model for the channel flow and periodic hill cases on identical grids. On the other hand, a static slip-wall model using an arbitrary constant value of the slip-length takes about $1$\% less time per time-step for the channel flow on the G3 mesh and about $7$\% less time per time-step for the periodic hill case on the G2 grid in comparison to the dynamic slip-wall model on an identical number of processors.

\bibliographystyle{jfm}
\bibliography{jfm}

\end{document}